\pdfoutput=1
\documentclass[12pt]{article}
\usepackage{amsmath,amssymb}
\usepackage{graphicx}

\newcommand{\be}{\begin{eqnarray}}
\newcommand{\ee}{\end{eqnarray}}
\def\G{{\cal G}^{+++}}

\def\D{H}

\def\d{\delta}
\def\r{\rho}
\def\l{\lambda}
\def\s{\sigma}

\def\eps{\epsilon}

\def\p{\partial}
\def\nn{\nonumber}
\def\cE{{\cal{E}}}

\def\z{\zeta}
\def\mmp{{\widehat B}}
\def\ep{B}
\def\cW{{\cal{W}}}
\def\ints{\mathbb{Z}}
\def\nn{\nonumber}

\begin{document}

\thispagestyle{empty}
\setcounter{page}{0}
\renewcommand{\theequation}{\thesection.\arabic{equation}}

{\hfill{\tt hep-th/0703285}}

{\hfill{ULB-TH/07-09}}

{\hfill{AEI-2007-014}}

\begin{center} {\bf \Large An $E_9$ multiplet of BPS states}

\vspace{.2cm}

Fran\c cois Englert${}^{a}$,  Laurent
Houart${}^{b}$\footnote{Research Associate F.N.R.S.}, Axel
Kleinschmidt${}^{c}$, Hermann Nicolai${}^{c}$ , Nassiba
Tabti${}^{b}\footnote{ F.R.I.A. Researcher}$

\footnotesize
\vspace{.2 cm}

${}^a${\em Service de Physique Th\'eorique, Universit\'e Libre de
Bruxelles\\Campus Plaine C.P.225, Boulevard du Triomphe, B-1050
Bruxelles, 
Belgium\\and\\ The International Solvay Institutes \\Campus Plaine
C.P.231, Boulevard du Triomphe, B-1050 Bruxelles, 
Belgium}

\vspace{.2cm}

${}^b${\em Service de Physique Th\'eorique et Math\'ematique,
Universit\'e Libre de Bruxelles\\ Campus Plaine C.P. 231, Boulevard du
Triomphe, B-1050 Bruxelles, 
Belgium\\and\\ The International Solvay Institutes \\Campus Plaine
C.P.231, Boulevard du Triomphe, B-1050 Bruxelles, 
Belgium}

\vspace{.2cm}

${}^c${\em Max Planck Institute for Gravitational Physics, Albert
  Einstein Institute\\ 
  Am M\"uhlenberg 1, 14476 Potsdam,
Germany}

 {\tt
fenglert, lhouart, ntabti@ulb.ac.be \\ axel.kleinschmidt,
hermann.nicolai@aei.mpg.de} 

\end{center}

\vspace {.7cm}

\begin{center}
{\bf Abstract} 

\begin{tabular}{p{15cm}}
{\small
\noindent
We construct an infinite $E_9$ multiplet of BPS states for 11D supergravity.
For each positive real root of $E_9$ we obtain a BPS solution of 11D
supergravity, or of its exotic counterparts, depending on two non-compact
transverse space variables. All these solutions are related by U-dualities
realised via $E_9$ Weyl transformations in the regular embedding $E_9\subset
E_{10} \subset E_{11}$. In this way we recover the basic BPS solutions,
namely the KK-wave, the M2 brane, the M5 brane and the KK6-monopole,
as well as other solutions admitting eight longitudinal space
dimensions. A novel technique of combining Weyl reflexions with compensating
transformations allows the construction of many
new BPS solutions, each of which can be mapped to a solution of a dual
effective action of gravity coupled to a certain higher rank tensor field.
For real roots of $E_{10}$ which are not roots of $E_9$, we obtain
additional BPS solutions transcending 11D supergravity (as exemplified by
the lowest level solution corresponding to the M9 brane). The relation
between the dual formulation and the one in terms of the original 11D
supergravity fields has significance beyond the realm of BPS solutions. We
establish the link with the Geroch group of general relativity, and explain
how the $E_9$ duality transformations generalize the standard Hodge
dualities to an infinite set of `non-closing dualities'.}
\end{tabular}
\end{center}

\newpage
\tableofcontents
\newpage
\setcounter{equation}{0}
\addtocounter{footnote}{-2}

\section{Introduction}

String theories, and  particularly superstrings and their   
possible merging at the non-perturbative level in an elusive M-theory,
are often 
viewed in the   double perspective of  a consistent quantum gravity
theory  and   of 
fundamental interactions unification. It is of interest to inquire
into the symmetries which would underlie the M-theory project, using
as a guide symmetries rooted in  its conjectured classical low energy
limit, namely 11-dimensional supergravity whose bosonic action is

\begin{eqnarray}
\label{Mth}
{\cal S}^{(11)} &=&\frac{1}{16\pi G_{11}}\,\int d^{11}x
\sqrt{-g^{(11)}}\bigg(R^{(11)}- \frac{1}{2  \cdot 4!} 
F_{\mu\nu\sigma\tau}F^{\mu\nu\sigma\tau} \nn\\
&&\quad\quad\quad\quad\quad\quad+ \frac1{(144)^2}
\eps^{\mu_1\ldots \mu_{11}} F_{\mu_1\ldots
  \mu_4}F_{\mu_5\ldots\mu_8}A_{\mu_9\mu_{10}\mu_{11}} \bigg).
\end{eqnarray}

Scalars in the dimensional reduction of the action Eq.(\ref{Mth}) to
three space-time dimensions realise non-linearly  the maximal
non-compact form of the Lie group $E_8$  as a coset $E_8/SO(16)$ where
$SO(16)$ is its maximal compact subgroup. Here, the symmetry of the
(2+1) dimensionally reduced action has been enlarged from the $GL(8)$
deformation group of the compact torus $T^8$ to the simple Lie group
$E_8$. This symmetry enhancement stems from the detailed structure of
the action Eq.(\ref{Mth}).  

Coset symmetries were first found in  the dimensional reduction of
11-dimensional supergravity \cite{Cremmer:1978km}  to four space-time
dimensions \cite{Cremmer:1979up} but appeared also in other
theories. They have been the subject of much study, and some classic
examples are given in \cite{Ferrara:1976iq, Cremmer:1977tt,
  Julia:1980gr, Julia:1981wc, Cremmer:1978ds, Schwarz:1983wa}. In
fact, all simple maximally non-compact Lie groups $\cal G$   can be
generated from the reduction down to three  dimensions of suitably
chosen actions \cite{Cremmer:1999du}. In particular the effective
action of the 26-dimensional bosonic string without tachyonic term
yields $D_{24}$ and pure gravity in $D$ space-time dimensions yields
$A_{D-3}$. 

It has been suggested that such actions, or possibly some unknown
extensions of them, possess a much larger symmetry than the one
revealed by their dimensional reduction to three space-time dimensions
in which all fields, except  $(2+1)$-dimensional gravity itself,  are
scalars.  Such hidden symmetries would be, for each simple Lie group
$\cal G$, the Lorentzian  `overextended' $\cal
G^{++}$ \cite{Damour:2002fz}  or the `very-extended' $\cal G^{+++}$
\cite{Gaberdiel:2002db,Lambert:2001gk,Englert:2003zs} Kac--Moody
algebras generated 
respectively by  adding 2 or 3 nodes to the Dynkin diagram defining
$\cal G$. One first adds the affine node, then a second node connected
to it by a single line to get the $\cal G^{++}$ Dynkin diagram and
then similarly a third one connected to the second to generate $\cal
G^{+++}$. In particular, the $E_8$ invariance of the 
dimensional reduction to   three dimensions of 11-dimensional supergravity
would be enlarged to  $E_8^{++} \equiv E_{10}$
\cite{Julia:1980gr,Damour:2002cu} or to
$E_8^{+++} \equiv E_{11}$ \cite{West:2001as}.  In our quest for the
symmetries of M-theory we 
shall restrict here our considerations to $ E_{10}$ and $ E_{11}$ and
their gravity subalgebras $A_8^{++}$ and $A_8^{+++}$. The extension of
the Dynkin diagram of $E_8$ to $E_{11}$ is depicted in Fig.1.

To explore the possible fundamental significance of these huge
symmetries a Lagrangian formulation \cite{Damour:2002cu}  {\it
  explicitly} invariant under $E_{10}$ has been proposed. It was
constructed as a reparametrisation invariant $\sigma$-model of fields
depending on one parameter $t$, identified as a time parameter, living
on the coset space $E_{10}/K_{10}^+$. Here $K_{10}^+$ is the
subalgebra of $E_{10}$ invariant  under the Chevalley involution.  The
$\sigma$-model contains an infinite number of fields and is built in a 
recursive way  by a level expansion of $E_{10}$ with respect to its
subalgebra $A_9$ \cite{Damour:2002cu, Nicolai:2003fw}  whose Dynkin
diagram is the `gravity line' defined in Fig.1, with the node 1
deleted\footnote{ Level expansion of $\G $ algebras in terms of a
  subalgebra $A_{D-1}$ have been considered in
  \cite{West:2002jj,Kleinschmidt:2003mf}.}.  The level of an
irreducible representation of $A_9$ occurring in the decomposition of
the adjoint  representation of $E_{10}$  counts the number of  times
the simple root $\alpha_{11}$ not pertaining to the gravity line
appears in 
the decomposition.  The $\sigma$-model, limited to the roots up to
level 3 and height 29, reveals  a perfect  match with the bosonic
equations of motion of 11-dimensional supergravity in the vicinity of
the  spacelike singularity of the cosmological billiards
\cite{Damour:2000hv,Damour:2001sa,Damour:2002et}, where fields depend
only on time.  It was conjectured that   space derivatives are hidden
in some higher level fields of the
$\sigma$-model~\cite{Damour:2002cu}. We shall label  this
one-dimensional $\sigma$-model $S^{cosmo}$. 

\begin{figure}[h]
   \centering
   \includegraphics[width=7cm]{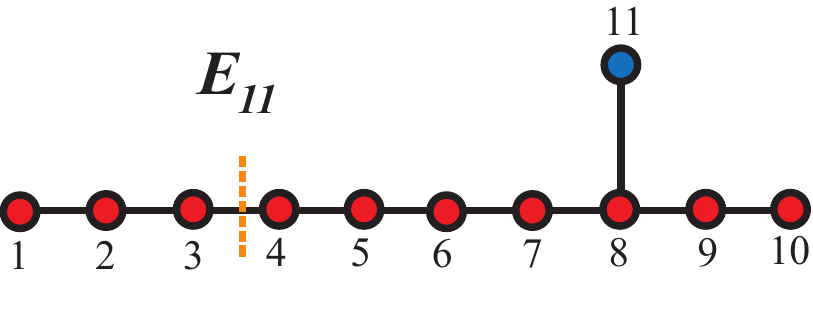} 
 \caption { \sl \small The Dynkin diagram of $E_{11}$ and its gravity
 line. The roots 3 and 2 extend $E_8$ to $E_{10}$ and the additional
 root 1 to $E_{11}$. The Dynkin diagram of the  $A_{10}$  subalgebra
 of $E_{11}$,  represented in the figure by the horizontal line is its
 `gravity line'.} 
   \label{ffirst}
\end{figure}

An alternate $E_{10}$ $\sigma$-model parametrised by a space variable
$x^1$ can be formulated on a coset space $E_{10}/K_{10}^-$, where
$K_{10}^-$ is invariant under a `temporal' involution ensuring the
Lorentz invariance $SO(1,9)$ (rather than $SO(10)$ in $S^{cosmo}$) at
each  level in the $A_9$ decomposition of $E_{10}$.  This
$\sigma$-model provides a natural framework for 
studying static solutions. It yields all the basic BPS solutions of
11D supergravity~\cite{Englert:2003py}, namely the KK-wave, the M2
brane, the M5 brane and 
the KK6-monopole, smeared in all space dimensions but one, as well as
their exotic counterparts. We shall label the action of this
$\sigma$-model $S^{brane}$~\cite{Englert:2004ph}.  The algebras
$K_{10}^+$  and $K_{10}^-$ 
are both subalgebras of  the algebra $K_{11}^-$ invariant under the
temporal involution defined on $E_{11}$, which selects the Lorentz
group  $SO(1,10)=K_{11}^-\cap A_{10}$ in the $A_{10}$ decomposition of
$E_{11}$~\cite{Englert:2003py, Englert:2004ph}.  

Elucidating the role of the huge number of  $E_{10}$ and $E_{11}$
generators is an important problem in the Kac--Moody approach to
M-theory. In this paper we focus on the  real roots of $E_{10}$, and
in particular to those belonging to its regular affine subalgebra
$E_9$.  We consider fields parametrising  the Borel representatives of
the coset space $E_{10}/K_{10}^-$, that is the Cartan and the positive 
generators of $E_{10}$. $E_{10}$ is taken to be embedded regularly
in $E_{11}$. The Dynkin diagrams of $E_{10}$ and $E_9$ are obtained
from the Dynkin diagram of Fig.1 by deleting successively the nodes 1
and~2. We find that each positive real root determines a BPS state in
space-time\footnote{This is in line with the analysis of
  \cite{West:2004st} where BPS states are associated with $E_{11}$
  roots.}, where a BPS state is defined by the no-force condition
allowing superposition of configurations centred at different
space-time points. We obtain explicitly an  infinite $E_9$ multiplet
of BPS static solutions of 11D supergravity depending on two
non-compact  space variables. They are all related  by U-dualities
realised  by the $E_9$ Weyl transformations. 

An obvious question at this point concerns the relation between our results
and the conjecture of~\cite{Hull:1994ys}, according to which the BPS solitons
of the toroidally compactified theory should transform under the arithmetic
group $E_9({\mathbb Z})$. Like~\cite{Obers:1998rn}, we here consider
only the Weyl group of $E_9$, realized as a subgroup of $E_9({\mathbb
  Z})$. More specifically, 
we consider the Weyl groups of various affine $A_1^+ \equiv A_1^{(1)}$
subgroups of $E_9$. Each one of these is then found to act via inversions
and integer shifts on a certain analytic function characterising the given
BPS solution, cf. Eq.(\ref{Moebact}) ---  much like the modular group
$SL(2,{\mathbb Z})$ (see Appendix G for details on the embedding of
the Weyl group into the Kac--Moody group itself). There is also an
action on the conformal factor, which is, however, more complicated
and cannot be interpreted in this simple way (and which can be locally
undone by a conformal coordinate transformation). The remaining Weyl
reflections `outside' the given $A_1^+$, and associated to the simple roots of the $E_9$, are realized as permutations.  However, we should like
to stress that the full $E_9({\mathbb Z})$ contains many more transformations
than those considered here.

We show that the BPS states we find admit an equivalent description as
solutions of effective actions, where the infinite set of $E_9$ fields
in the Borel representative of $E_{10}/K_{10}^-$ play the role of
matter fields in a `dual' metric. We label the description in terms of
11D supergravity fields the `direct' one and the description in terms
of Borel fields the `dual' one. Comparing the direct with the dual
description sheds some light on the significance of the $E_9$ real
roots. We see indeed that the Borel fields corresponding to these
roots are related to the supergravity fields by an infinite set of
`non-closing dualities' generalizing the Hodge duality.  This is a
feature which is well-known in the context of the Geroch symmetry of
standard $D=4$ gravity reduced to two
dimensions~\cite{Geroch:1970nt,Geroch:1972yt,Breitenlohner:1986um,Nicolai:2001cc}. 
There one also defines an infinite set of dual potentials from the
standard fields and uses the infinite Geroch group to generate new
solutions. The Geroch group is the affine $A_1^+$ extension of $SL(2)$
and we will make use of various subgroups of $E_9\subset E_{10}$
isomorphic to this basic affine group. 

One of the original motivations for the present work was to get a
better understanding of the significance of the higher level fields
in $E_{10}$ and maybe expand the known dictionary beyond level three (rather
height $29$). Since the role of some of these higher levels is
understood in the two-dimensional $E_9$ context in terms of the
generalised dual potentials one can 
anticipate that they will play a similar role in 
the one-dimensional $\s$-models $S^{brane}$ and $S^{cosmo}$. Indeed
this is what we find in the BPS case but we
have not obtained an analytic form for these non-closing dualitites in
the form of an extended dictionary.
Still it is clear that their definition is not restricted to BPS
states.   

Borel fields attached to roots of $E_{10}$ that are not roots of $E_9$
define BPS states depending on one non-compact space variable. These
may not admit a direct description in terms of 11D supergravity fields
but their dual description is well defined. The explicit BPS solution
attached to a level 4 root of $E_{10}$ is obtained, in agreement with
reference \cite{West:2004st}, and describes the M9 solution in 11
dimensions which is the `uplifting' of the D8-brane of Type IIA string
theory.  

The paper is organised as follows.

In Section 2 we review the construction of the $\sigma$-model
$S^{brane}$ and its relation to the basic BPS solutions of 11D
supergravity and of its exotic counterparts.   
 
 In Section 3, we classify {\em all} $E_9$ generators in $A_1^+$
 subgroups with central charge in $E_{10}$. We select  particular
 $A_1^+$ subgroups containing two infinite `brane' towers of
 generators, or one infinite `gravity' tower. The tower generators are
 recurrences of the generators defining the basic M2 and M5,  or the
 KK-wave and the KK6-monopole.  The other $A_1^+$ subgroups needed to
 span all $E_9$ generators  are obtained from the chosen ones by  Weyl
 transformations in the $A_8\subset E_9$ gravity line.               
 The fields characterising the basic BPS solutions smeared to two
 space dimensions are encoded as parameters in Borel
 representatives of $E_{10}/K_{10}^-$: each basic solution is fully
 determined by a specific positive generator associated to a specific
 positive real root.                                        All
 $E_9\subset E_{10}$ real roots are related by Weyl
 transformations. We use sequences of Weyl reflexions to reach any
 positive real root from roots corresponding to basic BPS solutions.
 We then express through dualities and compensations the fields
 defined by a given root  in terms of the 11-dimensional metric and
 the 3-form potential. We verify 
 that these  fields  yield  a new solution of 11D supergravity or of
 its exotic counterparts.  In this way we generate an infinite
 multiplet of $E_9$  BPS solutions  depending on two space
 variables. 
 In the string theory context this constitutes an infinite
 sequence of U-dualities realised  as Weyl transformations of
 $E_9$. 
 It is shown that the full BPS multiplet of states is
 characterised by group transformations preserving the analyticity of
 the Ernst potential originally introduced in the context of the
 Geroch symmetry of 4-dimensional gravity with one time and one space
 Killing vectors. 
 
Section 4 discusses the nature of the different BPS states. One
introduces the dual formalism which proves a convenient tool to
analyse the charge and mass content of the $E_9$ BPS states. The
masses are defined and computed in the string theory context. We show
that the $E_9$ multiplet can be split into three different classes
according to the $A_9$ level $l$. For $0\le l \le 3$ one gets the
basic BPS states smeared to two non-compact space dimensions. For
levels 4, 5 and 6  the BPS states depending on two non-compact space
variables can not be `unsmeared' in higher space dimensions. We
qualify the eight remaining space dimensions, and the time dimension,
as longitudinal ones.  For $l>6$ all BPS states admit nine
longitudinal dimensions, including time, and we argue that they are
all compact. 

Section 5 shows that $E_{10}$ fields associated to real roots which
are not in $E_9$ are BPS solutions of  $S^{brane}$ and admit thus a
space-time description with one non-compact transverse space
dimension. They may not admit a direct description but the dual
description is still well defined. These facts are exemplified by a
level 4 field which yields the M9 brane. 

We summarize the results in Section 6 and discuss the emergence of
non-closing dualities. 

Several appendices complement arguments in the main part of the
paper.

\setcounter{equation}{0}

\section{Basic BPS states in $E_{10}\subset E_{11}$}
\subsection{From $E_{11}$ to $E_{10}$ and the coset space $E_{10}/K^-_{10}$}

We recall that the Kac-Moody algebra $E_{11}$ is
entirely defined by the commutation relations of its Chevalley generators and
by  the Serre relations~\cite{Kac:book}. The Chevalley presentation
consists of the generators $e_m,f_m$ and $h_m, \ m=1,2,\dots 11$~with
commutation relations 
\begin{eqnarray}
   [h_m,h_n]=0~~, ~~[h_m,e_n]&=&a_{mn}e_n \, ,\nonumber \\
\label{chevalley}
   [h_m,f_n]&=&-a_{mn} f_n~~,~~[e_m,f_n]=\delta_{mn} h_m\, ,
\end{eqnarray}  where $a_{mn}$ is the Cartan matrix which can be
expressed in terms of scalar products of the simple roots $\alpha_m$
as  
\begin{equation}
\label{cartan}  
a_{mn}=2\frac{\langle\alpha_m,\alpha_n\rangle}{\langle\alpha_m,
\alpha_m \rangle}\, .
\end{equation}
The Cartan subalgebra is generated by $h_m$, while  the positive
(negative) step operators are  the  $e_m$ ($f_n$) and their
multi-commutators, subject to the Serre relations 
\begin{equation}
\label{serre} [e_m,[e_m,\dots,[e_m,e_n]\dots]]=0,\quad
[f_m,[f_m,\dots,[f_m,f_n]\dots]]=0\, ,
\end{equation} 
where the number of $e_m$ ($f_m$) acting on $e_n$     ($f_n$) is
given by $1-a_{mn}$. The Cartan matrix of $E_{11}$ is encoded in its
Dynkin diagram depicted in  Fig.1. Erasing the node 1 defines the
regular embedding of its $E_{10}$ hyperbolic subalgebra and erasing
the nodes 1 and 2 yields the regular embedding of the affine
$E_9$.\footnote{We will usually not distinguish in notation between
  group and algebra since it should be clear from the context which is
  meant.}

$E_{11}$ contains a  subgroup $GL(11)$ such that $SL(11)\equiv A_{10}
\subset GL(11) \subset
E_{11}$. The generators of the $GL(11)$ subalgebra are taken to be
$K^a_{~b}\ (a,b=1,2,\ldots ,11)$   with commutation relations
\begin{equation}
\label{Kcom} [K^a_{~b},K^c_{~d}]   =\delta^c_b
K^a_{~d}-\delta^a_dK^c_{~b}\,  .
\end{equation}
The relation  
between the
commuting generators $K^a{}_a$ of 
$GL(11)$ and   the Cartan generators $h_m$ of $E_{11}$ in the   
Chevalley basis
follows from comparing the commutation relations Eqs.(\ref{chevalley})  
and
(\ref{Kcom}) and from the identification of the simple roots of
$E_{11}$. These are
$e_m=\delta_m^{a} K^a{}_{a+1},\ m=1,\ldots ,10$ and  $e_{11}= R^{\, 9\,  
10\,11}$
where
$R^{abc}$ is a generator in
$E_{11}$ that is a third rank anti-symmetric tensor under $A_{10}$.   
One gets
\begin{eqnarray}
\label{aa} h_m&=&\delta_m^{a}(K^a{}_a-K^{a+1}{}_{a+1})~~~~  
m=1,\dots,10\\
\label{el} h_{11}&=&-\frac{1}{3}(K^1{}_1+\ldots +K^8{}_8) +\frac{2} 
{3}(K^9{}_9+
K^{10}{}_{10}+K^{11}{}_{11})\,.
\end{eqnarray}
The positive (negative) step operators in the $A_{10}$ subalgebra are
the $K^a{}_b$ with $b>a$ ($b<a$). The adjoint representation of
$E_{11}$ can be written as an infinite direct sum of representations
of the $GL(11)$ generated by the $K^a{}_b$. This is known as the
$A_{10}$ level decomposition of
$E_{11}$~\cite{Damour:2002cu,West:2002jj,Nicolai:2003fw}.
The $K^a{}_b$ define the level zero positive
(negative) step  operators.  The  positive (negative) level $l$  step
operators are defined by the number  of times the root $\alpha_{11}$
appears in the decomposition of the adjoint representation of $E_{11}$  into  
irreducible representations of  $A_{10}$ .  At level 1, one has the
single representation spanned by the anti-symmetric tensor
$R^{abc}$. At each level the number of irreducible representations of
$A_{10}$ is finite and the symmetry properties of the irreducible
tensors $R^{c_1\dots c_r}$ ($ R_{c_1\dots c_r}$) are fixed by the
Young tableaux of the representations. 
In what follows, positive level step operators will always be denoted
with upper indices and negative level ones with lower ones. For
positive level $l$ the number of indices on a generator $R^{c_1\dots
  c_r}$ is $r=3l$.

The Borel group formed by the Cartan generators and the positive level
0 generators can be taken as representative of the coset space
$GL(11)/ SO(1,10)$ and hence the parameters of the Borel group can be
used to define in a particular gauge the 11-dimensional metric
$g_{\mu\nu}$  which spans this coset space at a given space-time
point. We note that the subgroup 
$SO(1,10)$ of $GL(11)$ is the subgroup invariant under a temporal
involution $\Omega^0$  which generalizes the Chevalley involution by
allowing the identification of the tensor index  1 to be the time
index. Namely we define $\Omega^{0}$ by the map 
\begin{equation}
\label{map0} K^a_{~b}\stackrel{\Omega^0}{\mapsto}
-\epsilon_a\epsilon_b K^b_{~a}\, ,
\end{equation}
with $\epsilon_a =-1$ if $a=1$ and
$\epsilon_a=+1$ otherwise. This suggests to define in general  all
$E_{11}$ fields as parameters of the coset space $E_{11}/K_{11}^-$
where $K_{11}^-$ is invariant under the more general temporal
involution $\Omega$ \cite{Englert:2003py} with map 
\begin{equation}
\label{map} K^a_{~b}\stackrel{\Omega}{\mapsto}
-\epsilon_a\epsilon_b K^b_{~a}\quad
, \quad R^{c_1\dots c_r}
\stackrel{\Omega}{\mapsto}
-\epsilon_{c_1}\dots\epsilon_{c_r}   R_{ c_1\dots c_r}\, .
\end{equation}
Here $ R_{ c_1\dots c_r}$ is the negative step operator 
 corresponding to the positive one $R^{c_1\dots c_r}$.
One sees that $K_{11}^-\cap A_{10}=SO(1,10)$. We shall henceforth
 label the $A_{10}$ Dynkin subdiagram of $E_{11}$, the `gravity line'
 depicted in Fig.1. 

For the regular embedding of $E_{10}$ in $E_{11}$ obtained by deleting
the node 1 in Fig.1, the description in terms of $GL(10)$ follows from
the description of $E_{11}$ in terms of $GL(11)$. Omitting the
generators $K^1{}_2,K^2{}_1$ and $K^1{}_1$, the relations
Eqs.(\ref{Kcom}), (\ref{aa}) remain valid and  Eq.(\ref{el}) becomes  
\begin{equation}
\label{el10}
 h_{11}\to -\frac{1} {3}(K^2{}_2+\ldots +K^8{}_8) +\frac{2}
{3}(K^9{}_9+
K^{10}{}_{10}+K^{11}{}_{11})\,.
\end{equation}
The temporal involution $\Omega$, $\varepsilon_a=+1$ for
$a=2,3\dots,11$ reduces to the Chevalley involution acting on the
$E_{10}$ generators.  It leaves invariant a subalgebra 
$K^+_{10}$ of $E_{10}$. The Borel representative of the coset space
$E_{10}/ K^+_{10}$ is now parametrized by $A_9$ tensor fields in the
Euclidean metric $GL(10)/ SO(10)$.  Taking these fields to be
functions of the remaining time coordinate~1, one can built a
$\sigma$-model on this coset space.  This model has been used mainly
to study cosmological solutions
\cite{Damour:2002et,Kleinschmidt:2005gz} and we shall label the action
of this $\sigma$-model~\cite{Damour:2002cu} as $S^{cosmo}$.  
 

We could of course have chosen 2 instead of 1 as time coordinate in
$GL(11) \subset E_{11}$.  This change of time coordinate can be
obtained by performing  the $E_{11}$ Weyl reflexion $ W_{\alpha_1}$
sending $\alpha_1\to -\alpha_1$ and $\alpha_2 \to \alpha_1 +\alpha_2$.
Choosing the gravity line of Fig.1 to be the reflected one, one finds
that its time coordinate has switched from 1 to  2.  This results from
the fact that the temporal involution does not in general commute
with Weyl reflexions
\cite{Keurentjes:2004bv,Keurentjes:2005jw,Englert:2004ph}, a property
that has far reaching consequences, as reviewed in
Appendix~\ref{appw}. Deleting the node 1 in Fig.1 we obtain the Dynkin
diagram of $E_{10}$ endowed with the temporal involution
$\Omega_{(\lambda)},\lambda =2$,  with $\varepsilon_a =-1$ for $a=2$
and $+1$ for $a= 3,4\dots 11$ in Eq.(\ref{map}). This involution
leaves invariant a subalgebra $K_{10}^-$ of $E_{10}$.  The coset space
$E_{10}/K^-_{10}$ accommodates the Lorentzian metric  $GL(10)/
SO(1,9)$. Performing  products of $E_{10}$ Weyl reflexions on the
gravity line $W_{\alpha_i},\, i=2,\dots, 10$, one obtains ten possible
different identifications of the time coordinate from
$\Omega_{(\lambda)}, \, \lambda=2,3\dots 11$.   The
$\sigma$-model build upon the coset $E_{10}/K^-_{10}$  can be
constructed for any choice of $\lambda$ in $\Omega_{(\lambda)}$.
These formulations of the $\sigma$-model are all equivalent up to the
field redefinitions by $E_{10}$ Weyl transformations and we shall
label them by the generic notation $S^{brane}$, leaving implicit the
choice of the time coordinate $\lambda$. 

For sake of completeness, we recall the construction of $S^{brane}$
\cite{Englert:2003py,Englert:2004ph}. 
We take as representatives of $E_{10}/K^-_{10}$ the elements of
the Borel group of $E_{10}$ which we write
as\footnote{\label{borelfn}As a warning 
  to the reader we note that this type of Borel gauge for the coset
  space is not always accessible since the denominator $K^-_{10}$ is
  not the compact subgroup of the split $E_{10}$ and therefore the
  Iwasawa decomposition theorem fails. A simple finite-dimensional
  example is the coset space $SL(2)/SO(1,1)$ which will also play a
  role below. In this space  one cannot find  upper triangular
  representatives for matrices of the form
  \begin{equation*}
  {\cal V} = \left[\begin{array}{cc}a&b\\a&c\end{array}\right]
  \quad\quad\quad \text{with}\quad\big[a(c-b)=1\big],
  \end{equation*}
  since the lightlike first column vector cannot be Lorentz boosted
  to a spacelike one.}
\begin{equation}
\label{positive} {\cal V}(x^1)= \exp \left[\sum_{a\ge b}
h_b^{~a}(x^1)K^b_{~a}\right]\exp \left[\sum
\frac{1}{r!} A_{ a_1\dots a_r}(x^1) R^{
a_1\dots   a_r} +\cdots\right]\, ,
\end{equation}
where from now on all indices run from 2 to 11. The first exponential
contains only  level zero  operators  and the second one the positive
step operators of $E_{10}$ of levels strictly greater 
than zero. Define
\begin{equation}
\label{sym}
v(x^1)= \frac{d{\cal V}}{dx^1} {\cal V}^{-1}\quad \widetilde
v(x^1)=   -\Omega_{(\lambda)}
\, v(x^1)
\qquad\quad v_{sym}=\frac{1}{2} (v+\widetilde v)\, ,
\end{equation}
with $\lambda$ equal to the  chosen time coordinate. 
Using the invariant scalar product $\langle\cdot|\cdot\rangle$ for
$E_{10}$ one obtains a $\sigma$-model constructed on the coset
$E_{10}/K^-_{10}$  
\begin{equation}
\label{action} S^{brane}=\int dx^1 \frac{1}{n(x^1)}\langle
v_{sym}(x^1)|v_{sym}(x^1)\rangle\, ,
\end{equation} where
$n(x^1)$ is an arbitrary lapse function ensuring reparametrisation
invariance on the world-line. Explicitly, defining 
\begin{equation}
\label{vielbein}
e_\mu^{~m}=(e^{-h})_\mu^{~m} \qquad g_{\mu\nu}
=e_\mu^{~m}e_\nu^{~n}\eta_{mn} \, , 
\end{equation}
where $\eta_{mn}$ is the Lorentz metric with
$\eta_{\lambda\lambda}=-1$, one writes Eq.(\ref{action}) as 
\begin{equation}
\label{full} S^{brane} =S_0+\sum_A
S_A\, ,
\end{equation} with
\begin{equation}
\label{fullzero} S_0 =\frac{1}{4}\int dx^1
\frac{1}{n(x^1)}(g^{\mu\nu}g^{\sigma\tau}-
g^{\mu\sigma}g^{\nu\tau})\frac{dg_{\mu\sigma}}{dx^1}
\frac{dg_{\nu\tau}}{dx^1},
\end{equation}
\begin{equation}
\label{fulla} S_A=\frac{1}{2 r!}\int dx^1
\frac{ 1}{n(x^1)}\left[
\frac{DA_{\mu_1\dots \mu_r}}{dx^1} g^{\mu_1{\mu}^\prime_1}...\,
g^{\mu_r{\mu}^\prime_r}
\frac{DA_{{\mu}^\prime_1\dots {\mu}^\prime_r}}{dx^1}\right].
\end{equation} 
Here $D/dx^1$ is a group covariant derivative and all indices run from
2 to 11. Note that in Eq.(\ref{vielbein}) one may extend the range of
indices to include $\mu,\nu,m,n =1$, using the embedding relation
$E_{10}$ in $E_{11}$ which reads~\cite{Englert:2003zs} 
\begin{equation}
\label{embedding}
h_1^{~1}=\sum_{a=2}^{11} h_a^{~a}\qquad, \qquad  h_1^{~a}=0\qquad
a=2,3,\dots 11\, . 
\end{equation}

Up to level 3 and height 29, the fields in Eqs.(\ref{fullzero}) and
(\ref{fulla}) can be identified with fields of 11D
supergravity~\cite{Damour:2002cu}. They
can be used, as shall now be recalled, to characterise its basic BPS
solutions depending on one non-compact space variable. 

 \subsection{The basic BPS solutions in 1 non-compact dimension}
 
\subsubsection{Generalities and Hodge duality}
The basic BPS solutions of 11D supergravity are the 2-brane (M2) and
its magnetic counterpart the 5-brane (M5), and in the pure gravity
sector the Kaluza--Klein wave (KK-wave) whose magnetic counterpart is
Kaluza-Klein monopole (KK6-monopole). These are static solutions
which, wrapped on tori, leave respectively 8, 5, 9 and 3 non-compact
space dimensions.  
It is convenient to express the magnetic solutions in terms of an
`electric' potential with a time index. This is done  by trading {\em
  on the equations of motion} the field strength for its Hodge
dual. For the M5 the field
$F_{\mu{}_1\mu{}_2\mu{}_3\mu{}_4}=4\,\partial_{[\mu{}_1}
  A_{\mu{}_2\mu{}_3\mu{}_4]}$ has as dual the field 
 $\widetilde
F_{\nu{}_1\nu{}_2\nu{}_3\nu{}_4\nu{}_5\nu{}_6\nu{}_7}=7\,\partial_{[\nu{}_1}
  A_{\nu{}_2\nu{}_3\nu{}_4\nu{}_5\nu{}_6\nu{}_7]}$ defined by 
\begin{equation}
\label{Pdual1}
\sqrt{- g}\widetilde
F^{\nu{}_1\nu{}_2\nu{}_3\nu{}_4\nu{}_5\nu{}_6\nu{}_7}=\frac{1}{4!}\epsilon^{\nu{}_1\nu{}_2\nu{}_3\nu{}_4\nu{}_5\nu{}_6\nu{}_7\mu{}_1\mu{}_2\mu{}_3\mu{}_4}
F_{\mu{}_1\mu{}_2\mu{}_3\mu{}_4}\, , 
\end{equation}
For the KK6-monopole the KK-potential $A_\mu^{(\nu)}$ in terms of the
vielbein $e_\mu{}^n$ is given by  $A_\mu^{(\nu)} = -e_\mu{}^n 
(e^{-1})_n{}^\nu$ where $\mu$ labels the non-compact directions,
$\nu$ is the Taub-NUT direction in coordinate indices and $n$ is
the Taub-NUT direction in flat frame indices and there is no summation
on $n$. The  field strength is
$F_{\mu{}_1\mu{}_2}^{(\nu)}=\,\partial_{\mu{}_1} A_{\mu{}_2}^{(\nu)}
-\partial_{\mu{}_2} A_{\mu{}_1}^{(\nu)}$. Its dual is $\widetilde
F_{\nu{}_1\nu{}_2\nu{}_3\nu{}_4\nu{}_5\nu{}_6\nu{}_7\nu{}_8
  \nu{}_9\vert\nu{}_9}=9\,
\partial_{[\nu{}_1}A_{\nu{}_2\nu{}_3\nu{}_4\nu{}_5\nu{}_6\nu{}_7\nu{}_8
    \nu{}_9]\vert\nu{}_9}$ where 
\begin{equation}
\label{Pdual2}
\sqrt{- g }\widetilde
F^{\nu{}_1\nu{}_2\nu{}_3\nu{}_4\nu{}_5\nu{}_6\nu{}_7\nu{}_8
  \nu{}_9\vert\nu{}_9}=\frac{1}{2}\epsilon^{\nu{}_1\nu{}_2\nu{}_3\nu{}_4\nu{}_5\nu{}_6\nu{}_7\nu{}_8 \nu{}_9 \mu{}_1\mu{}_2} F_{\mu{}_1\mu{}_2}^{(\nu{}_9)}\, .   
\end{equation}
These BPS solutions depend on space variables in the non-compact
dimensions only. One may further compactify on tori some of these
`transverse' directions. This increases accordingly the number of
Killing vectors and one obtains in this way new `smeared' solutions
depending only on the space variables in the remaining non-compact
dimensions\footnote{In the string language, the smearing process
  amounts to introducing image branes in the compact dimensions and
  averaging them over the torus radii (or equivalently considering in
  the non-compact dimensions distances large compared to these
  radii). Compact dimensions which cannot be `unsmeared'  are labelled
  `longitudinal'. It will be convenient  in what follows to take the
  time dimension as compact (and longitudinal). Decompactifying
  longitudinal space-time dimensions does not affect the field
  dependence of the solutions. However longitudinal dimensions cannot
  always be decompactified, as exemplified by the Taub-NUT direction
  of the KK6-monopole. This feature will be studied in detail in
  Section~\ref{chargesec}.}. 
As we shall see the smearing process is straightforward, except for
the smearing of magnetic solutions to one non-compact dimensions,
which can only be performed in the electric language, hinting on the
fundamental significance of the dual formulation as will indeed be
later confirmed.  

As recalled below, all BPS solutions, smeared up to one non-compact
dimensions are solutions of the    $\sigma$-model  $S^{brane}$ given
by Eq.(\ref{full}) where $x^1$ is identified with the non-compact
space dimension \cite{Englert:2003py,Englert:2004ph}. 

\subsubsection{Levels 1 and 2: The 2-brane and the 5-brane}

For the 2-brane (M2) solution of 11D supergravity wrapped in the
directions 10 and 11, we choose 9 as the time coordinate, so that the
only non vanishing component  of the 3-form potential is
$A_{9\,10\,11}$. For the 5-brane (M5) wrapped in the directions
4,5,6,7,8, we choose 3 as time coordinate so that the 3-form potential
is still $A_{9\,10\,11}$ and the Hodge dual Eq.(\ref{Pdual1}) is
$A_{3\,4\,5\,6\,7\,8}$. One gets for these BPS solutions the following
metric and fields  
\begin{eqnarray}
 \label{M2metric}
{\rm M2} &:& g_{11}=g_{22}=H^{1/3}\,,\quad g_{33}=g_{44}=\dots
 =g_{88}=H^{1/3}\,,\quad -g_{99}=g_{10\,10}=g_{11\,11}=
 H^{-2/3}\,,\nonumber\\ 
 &&A_{9\,10\, 11} =\frac{1}{H}\, ,\\
\nonumber \\
\label{M5metric}
 {\rm M5}&:& g_{11}=g_{22}=H^{2/3}\,,\quad -g_{33}=g_{44}=\dots
 =g_{88}=H^{-1/3}\,,\quad g_{99}=g_{10\,10}=g_{11\,11}=
 H^{2/3}\,,\nonumber\\ 
 &&A_{3\,4\, 5\,6\,7\,8} =\frac{1}{H}\, .
\end{eqnarray}
Here $H$ is a harmonic function of the non-compact space dimensions
[(1,2,3,4,5,6,7,8) for M2 and  (1,2,9,10,11) for M5] with
$\delta$-function singularities at the location of the
branes. Smearing simply reduces the number of variables in the
harmonic functions to those labelling the remaining non-compact
dimensions. For instance a single M2 (M5) brane smeared to two
non-compact  dimensions, located at the origin of the coordinates
$(x^1,x^2)$, is described by $H= (q /2\pi)\ln r =(q
/2\pi)\ln\sqrt{(x^1)^2 +(x^2)^2}$ where $q$ is an electric (magnetic)
charge density. When smeared to one non-compact dimension one gets $H=
(q /2)\, \vert x^1\vert$. We see that in the one-dimensional case only
the electric dual description of the magnetic M5 brane is available,
as the duality relating the 6-form Eq.(\ref{Pdual1}) to the 3-form
supergravity potential   requires at least two transverse
dimensions. Note that this one-dimensional solution is still a
solution of 11D supergravity because the replacement in the equations
of motion  of the 4-form field strength by the 7-form dual is valid as
long as the Chern-Simons term contributions vanish, as is indeed the
case for the above brane solutions.  Actually the electric description
of the M5, Eq.(\ref{M5metric}), is a solution of the following
effective action (in any number of transverse non-compact dimensions)  
\begin{equation}
\label{M5dual}
{\cal S}^{(11)}_{M5} =\frac{1}{16\pi G_{11}}\,\int d^{11}x
\sqrt{-g^{(11)}}\left[R^{(11)}-  \frac{1}{2\cdot7!  
}\widetilde F_{\mu_1\mu_2\mu_3\mu_4\mu_5\mu_6\mu_7}\widetilde
  F^{\mu_1\mu_2\mu_3\mu_4\mu_5\mu_6\mu_7}\right]\, . 
\end{equation}
 
We recall \cite{Englert:2003py} how the M2 solution of 11D
supergravity smeared over all dimensions but one can be obtained as a
solution of the  the $\sigma$-model  by putting in $S^{brane}$
(Eq.(\ref{full})) all non-Cartan fields to zero except  the level 1
3-form component $A_{9\,10\,11}$ with time in 9. Similarly the M5
solution smeared in all directions but one solves the  equations of
motion of this  $\sigma$-model  by retaining for the non-Cartan fields
only the component $A_{3\,4\,5\,6\,7\,8}$ of the level 2 6-form  with
time in 3. These are respectively parameters of the Borel generators  
\begin{eqnarray}
\label{3ge}
 R^{[3]}_1&\stackrel {def}{=}&R^{9\,10\,11}\\
 \label{6ge}
 R^{[6]}_2&\stackrel {def}{=}& R^{345678}\, ,
\end{eqnarray}
where the subscripts denote the $A_{9}$ level in the decomposition of
the adjoint representation of~$E_{10}$. As we will see in more detail
below, the roots corresponding to the elements $R^{[3]}_1$ and
$R^{[6]}_2$ have scalar product $-2$ and thus give rise to an
infinite-dimensional affine $A_1^+$ subalgebra.

These two solutions of the $\sigma$-model are characterised by Cartan fields
 $ h_a^{~a}, a=2,3,\dots,11$. 
One has\footnote{\label{intconsfn}For simplicity we have chosen zero for the
 integration constants in the solutions of the equations of motion of
 the  fields $A_{9\,10\, 11}$ and $A_{3\,4\, 5\,6\,7\,8}$.}
 \cite{Englert:2003py} 
 \begin{eqnarray}
 \label{M2}
{\rm M2}&:& h_a^{~a} =\left\{\frac{-1}{6}\,\Big|\,
\frac{-1}{6},\frac{-1}{6},\frac{-1}{6},\frac{-1}{6},\frac{-1}{6},\frac{-1}{6},\frac{-1}{6},\frac{1}{3},\frac{1}{3},\frac{1}{3}\right\}
  \ln H\,,\qquad h_a^{~b}=0 ~\hbox {for} ~a\neq b \nonumber\\ 
&&h_a^{~a}K^a_{~a}=\frac{1}{2}\ln H \,\cdot\,h_{11}\,,\nonumber\\
&&  A_{9\,10\, 11} =\frac{1}{H}\, ,\\
\nonumber\\
\label{M5}
{\rm M5}&:& h_a^{~a}=\left\{\frac{-1}{3}\,\Big|\,
\frac{-1}{3},\frac{1}{6},\frac{1}{6},\frac{1}{6},\frac{1}{6},\frac{1}{6},\frac{1}{6},\frac{-1}{3},\frac{-1}{3},\frac{-1}{3}\right\}\ln
H\,,\qquad h_a^{~b}=0 ~\hbox {for} ~a\neq b \nonumber \\ 
&&h_a^{~a}K^a_{~a}=\frac{1}{2}\ln H \,\cdot\,(-h_{11}-K^2{}_2\,)\,,\nonumber\\
&& A_{3\,4\, 5\,6\,7\,8} =\frac{1}{H}\, ,
\end{eqnarray}
where we used Eq.(\ref{el10}) and $H(x^1)$ is a harmonic function.  To
the left of the $\vert$ symbol in the first line of Eqs.(\ref{M2}) and
(\ref{M5}), we have added the field $h_1^{~1}$, evaluated from the
embedding of $E_{10}$ in $E_{11}$ Eq.(\ref{embedding}). All other
quantities in these equations are defined in $E_{10}$. From
Eq.(\ref{vielbein}), the $\sigma$-model results Eqs.(\ref{M2}) and
(\ref{M5}) are equivalent to the supergravity results
Eqs.(\ref{M2metric}) and (\ref{M5metric}) for branes smeared to one
space dimension.

\subsubsection{Levels 0 and 3: The Kaluza-Klein wave and the
  Kaluza-Klein monopole} 

We now examine the BPS solutions involving only gravity. 

First consider  the KK-wave solution. The supergravity solution with
time in 3 and torus compactification in the 11 direction, is  
\begin{equation}
 \label{G0metric1}
 ds^2= -H^{-1}(dx^3)^2+(dx^1)^2 +(dx^2)^2+(dx^4)^2+\dots +(dx^{10})^2+
 H[dx^{11}- A_{3}^{(11)}dx^3]^2\, , 
\end{equation}
where the electric potential  $A_{3}^{(11)}$ is related to the harmonic
function $H$ in nine space dimensions with suitable source
$\delta$-function singularities by   
\begin{equation}
\label{potential}
A_{3}^{(11)} = (1/H)-1\, .
\end{equation}
Smearing over any number of space dimensions results in taking  $H$ as
a harmonic function of only the remaining non-compact space
variables. For non-compact space dimension $d>2$, the constant $-1$ in
Eq.(\ref{potential}) makes the potential vanish in the asymptotic
Minkowskian space-time if the limit of $H$ at spatial infinity is
chosen to be one. For $d=2$ or 1 space is not asymptotically flat and
we  keep for convenience the non vanishing constant in
Eq.(\ref{potential}) to be one. 

Smearing the KK-wave to  one non-compact dimension, the above
supergravity solution is recovered from the  $\sigma$-model
Eq.(\ref{full}) by putting to zero all fields parametrising the
positive roots in the Borel representative Eq.(\ref{positive})  except
the level 0 field $h_{3}^{~11}(x^1)$, taking 3 as the time
coordinate. To see this, it is convenient to rewrite
Eq.(\ref{positive}) by disentangling the Cartan generators and the
level zero positive step operators in two separate exponentials. One
writes 
\begin{equation}
\label{wave}
{\cal V}(x^1)= \exp \left[\sum_{a=2}^{11}
h_a^{~a}(x^1)K^a_{~a}\right]\exp \left[A_{3}^{~(11)}(x^1)K^3_{~11}\right]\, .
\end{equation}
The expression of $A_{3}^{(11)}$  in terms of the vielbein defined by
Eq.(\ref{vielbein}) is given in Appendix~\ref{appcg} by
Eq.(\ref{kka}), namely 
\begin{equation}
\label{potential0}
A_{3}^{(11)}=-e_{3}^{~11}(e^{-1})_{11}^{~11}\, .
\end{equation}
The solution is \cite{Englert:2003py}, taking into account the
embedding relation Eq.(\ref{embedding}), 
 \begin{eqnarray}
 \label{GO}
{\rm KKW}&:& h_a^{~a} =\left\{0\Big|
0,\frac{1}{2},0,0,0,0,0,0,0,\frac{-1}{2}\right\}\ln H\,, \nonumber\\ 
&&h_a^{~a}K^a_{~a}=\frac{1}{2}\ln H \,\cdot\,
(K^3{}_3-K^{11}{}_{11})\,,\nonumber\\ 
&& A_{3}^{(11)} =\frac{1}{H}-1\, ,
\end{eqnarray}
which, using Eqs.(\ref{vielbein}) and (\ref{potential0}), is
equivalent to the KK-wave solution Eqs.(\ref{G0metric1}) and
(\ref{potential}) of general relativity.

Consider now the KK6-monopole solution. In 11 dimensions, it has 7
longitudinal dimensions (see footnote 5). Taking 11 as the Taub-NUT direction and 4 as
the timelike direction, the general relativity solution reads 
\begin{equation}
\label{G3metric1}
ds^2= H\left[(dx^1)^2+(dx^2)^2+(dx^3)^2\right]
   -(dx^4)^2+(dx^5)^2+\dots+(dx^{10})^2
  +H^{-1}\left[ dx^{11} -\sum_{i=1}^3   A_i^{(11)} dx^i\right]^2 
\end{equation}
and
\begin{equation}
\label{F}
F_{ij}^{(11)}\equiv \partial_i A_j^{(11)}-\partial_j
A_i^{(11)}=-\varepsilon_{ijk}\, \partial_k H\, , 
\end{equation}
where $H(x^1,x^2,x^3)$ is the harmonic function. It can be smeared to
2 spatial dimensions by taking the index $j$ in Eq.(\ref{F})  to label
a compact dimension, say 3,  
\begin{equation}
\label{F2}
\partial_i A_3^{(11)}=\varepsilon_{ik}\, \partial_k H\qquad i,k = 1,2
\end{equation}
and $H(x^1,x^2)$ is now  harmonic in two dimensions. The metric
Eq.(\ref{G3metric1}) becomes 
\begin{equation}
\label{G3metric2}
ds^2= H\left[(dx^1)^2+(dx^2)^2+(dx^3)^2\right]
  -(dx^4)^2+(dx^5)^2+\dots+(dx^{10})^2
  +H^{-1}\left[ dx^{11} - A_3^{(11)}  dx^3\right]^2\, 
\end{equation}
and is a solution of Einstein's equations.

The smearing to one space dimension is more subtle. As for the
magnetic 5-brane, it requires a dual formulation which in this case is
defined by the duality relation Eq.(\ref{Pdual2}). However the
reformulation of the supergravity action is now less
straightforward. To understand the dual formulation we first show how
to use it for the unsmeared KK6-monopole given by
Eqs.(\ref{G3metric1}) and (\ref{F}).  
We rewrite the metric by setting $e_3^{~11}$, hence $A_3^{(11)}$, to
zero and substitute for it the field dual to $A_3^{(11)}$, defined
with field strength
$\widetilde
F_{\nu{}_1\nu{}_2\nu{}_3\nu{}_4\nu{}_5\nu{}_6\nu{}_7\nu{}_8
  \nu{}_9\vert\nu{}_9}=9\,
\partial_{[\nu{}_1}A_{\nu{}_2\nu{}_3\nu{}_4\nu{}_5\nu{}_6\nu{}_7\nu{}_8
    \nu{}_9]\vert\nu{}_9}$, where the dual field strength $\widetilde
F$ is defined by Eq.(\ref{Pdual2}). The dual (diagonal) metric and the
dual potential read 
\begin{eqnarray}
 \label{KKMmetric}
{\rm KK6M} &:& g_{11}=g_{22}= g_{33}= H\,,\quad  -g_{44}=g_{55}\dots
 =g_{10\,10}=1\,,\quad  g_{11\,11}= H^{-1}\nonumber\\ 
 &&A_{4\,5\,6\,7\,8\,9\,10\, 11\vert 11} =\frac{1}{H}\,  .
\end{eqnarray}
One verifies that the dual description  of the KK6-monopole given by
Eq.(\ref{KKMmetric}) can be derived from an effective action  in
analogy with  the action Eq.(\ref{M5dual}) for the M5. Here the dual
field plays the role of a matter field. In the gauge considered here
one takes as effective action action 
\begin{equation}
\label{KKMdual}
{\cal S}^{(11)}_{KK6} =\frac{1}{16\pi G_{11}}\,\int d^{11}x \sqrt{-g^{(11)}}\left[R^{(11)}- {1\over 2  
}\widetilde F_{i\,4\,5\,6\,7\,8\,9\,10\, 11\vert 11}\widetilde F^{i\,4\,5\,6\,7\,8\,9\,10\, 11\vert 11}\right] \, ,
\end{equation}
where $i$ runs over the three non-compact dimensions $1,2,3$. In this
dual description we may trivially smear the KK monopole to two or to
one non-compact space dimensions by letting the index $i$  in
Eq.(\ref{KKMdual}) run over the remaining non-compact dimensions. In
two non-compact space dimensions, one obtains the dual of the
description Eqs.(\ref{F2}) and (\ref{G3metric2}) and in one dimension
one gets  in this way  a definition of the smeared KK6-monopole which
inherits its charge and mass from the parent one with 3 non-compact
dimensions.  

The charge carried by the KK6-monopoles in three or less non-compact
dimensions can be obtained in the dual formulation  from the equations
of motion of the field
$A_{\nu{}_2\nu{}_3\nu{}_4\nu{}_5\nu{}_6\nu{}_7\nu{}_8
  \nu{}_9\vert\nu{}_9}$. From Eq.(\ref{KKMdual}), one gets 
\begin{equation}
\label{current}
\sum_{i,j=1}^2
\partial_i\big(\sqrt {-g}
g^{ij}g^{44}g^{55}g^{66}g^{77}g^{88}g^{99}g^{10\,10}(g^{11\,11})^2\partial_j 
A_{4\,5\,6\,7\,8\,9\,10\, 11\vert 11}\big)=0 \, , 
\end{equation}
and using Eq.(\ref {KKMmetric}) one finds
\begin{equation}
\label{currentH}
\sum_{i=1}^2 \partial_i\partial_i H =0\, ,
\end{equation}
outside the source singularities of the harmonic function $H$. If the
latter yields in Eq.(\ref{currentH}) only $\delta$-function
singularities $\sum_k q_k\delta(\vec r-\vec r_k)$ located in
non-compact space points $\vec r_k$, one may extend
Eq.(\ref{currentH}) to the whole non-compact space. We write  
\begin{equation}
\label{h}
\sum_{i=1}^2
\partial_i\partial_i H\propto \sum_k q_k \delta(\vec r-\vec r_k)\, ,
 \end{equation}
where $q_k$ is the charge of the monopole located  at $\vec r_k$. For
instance a single KK6-monopole located at the origin in 2 non-compact
space is described by $H= (q/2\pi)\ln r$.  
 
For 3 or 2 non-compact dimensions, writing Eq.(\ref{F}) or
Eq.(\ref{F2}) as $F_{\mu\nu}^{(11)}$ one recovers from Eq.(\ref{h}) the
charge of the monopoles from the conventional surface integral  
\begin{equation}
\label{chargeKKM}
\int F \equiv\int (1/2) F_{\mu\nu}^{(11)}\,dx^\mu \wedge dx^\nu \propto
\sum_k q_k\, , 
\end{equation}
where the surface integral enclosed the charges
$q_k$. Eq.(\ref{chargeKKM}) is equivalent to Eq.(\ref{h}). For
KK6-monopoles smeared to one non-compact dimension, the surface
integral loses its meaning but the direct definition of charge
Eq.(\ref{h}) is still valid. In this way the magnetic KK6-monopole in
any number of non-compact transverse dimensions is suitably described
(as  the magnetic 5-brane by Eq.(\ref{M5dual})) by a dual effective
action  Eq.(\ref{KKMdual}). As expected for a BPS solution, the charge
of the KK6-monopole, smeared or not, is equal to its tension evaluated
in string theory,  as recalled in Section 4 and in
Appendix~\ref{appm3} where the mass of the KK6-monopole is derived
from T-duality.

The above KK6-monopole solution smeared over all dimensions but one
can again be obtained as a solution of the $\sigma$-model  by putting
in $S^{brane}$,  with time in 4, all non-Cartan fields to zero except
the level 3 component $A_{4\,5\,6\,7\,8\,9\,10\,11\vert
  11}$\cite{Englert:2003py}. This is the parameter of the Borel
generator\footnote {We use the bar symbol to distinguish within the
  same irreducible level 3 $A_9$ representation the generator $\bar
  R^{[8,1]}_3$ corresponding to a real $E_{10}$ root from the
  generator  $R^{[8,1]}_3=[R^{[3]}_1,R^{[6]}_2]$ pertaining to the
  degenerate   null root, see also below in Section~\ref{m2m5sec}.}  
\begin{equation}
\label{81ge}
\bar R^{[8,1]}_3\stackrel {def}{=}R^{4\,5\,6\,7\,8\,9\,10\,11\vert 11}\, ,
\end{equation}
where the subscript labels the $A_{9}$ level in the decomposition of
the adjoint representation of $E_{10}$. The solution is, taking into
account the embedding relation Eq.(\ref{embedding}) 
\begin{eqnarray}
 \label{G3}
{\rm KK6M}&:& h_a^{~a} =\left\{\frac{-1}{2}\,\Big|\,
  \frac{-1}{2},\frac{-1}{2},0,0,0,0,0,0,0,\frac{1}{2}\right\}\ln H\,,\\ 
&&h_a^{~a}K^a_{~a} =\frac{1}{2}\ln H
  \,\cdot\,(-K^2{}_2-K^3{}_3+K^{11}{}_{11})\,,\nonumber\\ 
&& A_{4\,5\dots 10 \, 11\vert 11} =\frac{1}{H}\,,\nonumber
\end{eqnarray}
with $H(x^1)$ harmonic. From Eqs.(\ref{vielbein}) one indeed recovers
the KK6-monopole solution of the dual action Eq.(\ref{KKMdual})
displayed in Eq.(\ref{KKMmetric}).

\subsubsection{The exotic BPS solutions}

As a consequence of the non-commutativity of Weyl reflexions with the
temporal involution, the $E_{10}$ $\sigma$-model $S^{brane}$ living on
$E_{10}/K_{10}^-$ was expressed in 10 different ways according to the
choice of the time coordinate in Eq.(\ref{full}) in the global
signature (1,9). These are related through Weyl transformations of
$E_{10}$ from roots of the gravity line. Adding the Weyl reflexion
$W_{\alpha_{11}}$ one gets in addition equivalent expressions for
$S^{brane}$ where the signatures \cite{Englert:2004ph} in
Eq.(\ref{full}) are globally different.  This equivalence realises in
the action formalism  the general analysis of Weyl transformations by
Keurentjes \cite{Keurentjes:2004bv, Keurentjes:2004xx}. Starting with
the global signature (1,9) in 10 dimensions, or (1,10) in 11
dimensions, one reaches  different  signatures $(t,s,\pm)$ in 11
dimensions where $t$ is the number of timelike directions, $s$ is the
number of spacelike directions and $\pm$ encodes the sign of the
kinetic energy term of the level 1 field in the action
Eq.(\ref{fulla}), $+$ being the usual one and $-$  the `wrong'
one. These are \cite{Englert:2004ph}\footnote{ The analysis of the
  different possible signatures related by Weyl reflexions has been
  extended to all ${\cal G}^{++}$ in \cite{deBuyl:2005it,
    Keurentjes:2005jw}.}: $(1,10,+)$, $(2,9,-)$, $(5,6,+)$, $(6,5,-)$
and $(9,2,+)$. The signature changes  under the Weyl transformations
used in the following sections are presented in Appendix~\ref{appw}.

The results obtained in the   $\sigma$-model Eq.(\ref{full}) are in
complete agreement with the interpretation of the Weyl reflexion  
$W_{\alpha_{11}}$ as a double T-duality in the direction $9$ and $10$
plus exchange of the two directions   
\cite{Elitzur:1997zn, Obers:1998rn, Banks:1998vs,
  Englert:2003zs}. Indeed, it has been shown that T-duality involving
a timelike direction changes 
the signature of space-time leading to the exotic phases of M-theory
\cite{Hull:1998vg, Hull:1998ym}.  The signatures found by Weyl
reflexions are thus in perfect agreement with the analysis  of
timelike T-dualities.  

The brane scan of the exotic phases has been studied
\cite{Hull:1998fh, Argurio:1998ad}. Their different BPS branes depend
on the signature and the sign of the kinetic term. The number of
longitudinal timelike directions for a given brane is constrained. As
an example if we consider the so-called ${\rm M}^*$ 
phase characterised by the signature $(2,9, -)$, the wrong  sign of
the kinetic energy term implies that the exotic M2 brane must have
even number of timelike directions. There are thus two different M2
branes in ${\rm M}^*$ theory denoted $(0,3)$ and $(2,1)$ where the
first entry is the number of timelike longitudinal directions and the
second one the number of spacelike longitudinal directions. 
For instance, the metric of a $(2,1)$ exotic M2 brane with timelike
directions $10$ , $11$, and spacelike direction $9$ is  
\begin{eqnarray}
 \label{M2metricE}
{\rm M2}^*&:&g_{11}=g_{22}=H^{1/3}\qquad g_{33}=g_{44}=\dots
=g_{88}=H^{1/3}\,,\nonumber\\  
&&g_{99}=-g_{10\,10}=-g_{11\,11}= H^{-2/3}, 
\end{eqnarray}
where $H$ is  the harmonic function in the transverse non-compact dimensions. 

When smeared in all directions but one this metric 
is also a solution of the  $\sigma$-model  $S^{brane}$ with the
correct identification in Eq.(\ref{full}) of time components and sign
shifts in kinetic energy terms. More generally, all the exotic branes
smeared to one non-compact  space dimension are solutions of this
$\sigma$-model living on the coset $E_{10}/K^-_{10}$
\cite{Englert:2004ph}.

\setcounter{equation}{0}

\section{$E_9$-branes and the infinite U-duality group}

In this section we construct an infinite set of BPS solutions  of 11D
supergravity  and of its exotic counterparts depending on two
non-compact space variables. They are  related by the Weyl group of
$E_9$ to the basic ones reviewed in Section 2 and constitute an
infinite multiplet of U-dualities viewed as Weyl transformations.

\subsection{The working hypothesis}

Our working hypothesis is that the fields describing  BPS solutions of
11D supergravity  depending on two non-compact space variables $(x^1,
x^2)$ are coordinates in  the coset $E_{10}/K_{10}^-$, in the regular
embedding $E_{10}\subset E_{11}$. The coset representatives are taken
in the Borel gauge, subject of course  to the remark in
footnote~\ref{borelfn}.   

We first express in this way the basic solutions of Section 2, smeared
to two space dimensions. 

Consider the M2 and M5 branes.  Their Borel representatives are
\begin{eqnarray}
\label{2M2} {\rm M2}&:& {\cal V}_1= \exp \left [\frac{1}{2}\ln H\,
  h_{11}\right]\, \exp\left [\frac{1}{H}\, R^{[3]}_1\right]\\ 
\label{2M5} {\rm M5} &:&{\cal V}_2= \exp \left[\frac{1}{2}\ln H\,
  (-h_{11} -K^2{}_2)\right]\, \exp \left[\frac{1}{H}\,
  R^{[6]}_2\right]\, . 
\end{eqnarray}
Here  $R^{[3]}_1$ and  $R^{[6]}_2$ are defined in Eqs.(\ref {3ge}) and
(\ref {6ge}), respectively, and $h_{11}$ was defined in
Eq.(\ref{el10}). The Cartan fields and the potentials 
$A_{9\,10\,11}(x_1,x_2)$ for the M2 and
$A_{3\,4\,5\,6\,7\,8}(x_1,x_2)$ for the M5 are given by
Eqs.(\ref{M2}), (\ref{M5}) with $H$ now a function of the two
variables $x^1,x^2$. Their metric Eqs.(\ref{M2metric}) and
(\ref{M5metric})  are encoded in Eq.(\ref{vielbein}) giving the
relation  of the Cartan fields to the vielbein  and in
Eq.(\ref{embedding}) expressing the embedding of $E_{10}$ in
$E_{11}$. The Hodge duality relations 
\begin{eqnarray}
\label{dual1}
\sqrt{\vert g \vert}g^{11}g^{99}g^{10\,10} g^{11\, 11}\partial_1
A_{9\,10\,11}&=&\partial_2  A_{3\,4\,5\,6\,7\,8} \\ 
\label{dual2}
\sqrt{\vert g \vert}g^{22}g^{99}g^{10\,10} g^{11\, 11}\partial_2
A_{9\,10\,11}&=&-\partial_1  A_{3\,4\,5\,6\,7\,8}\, , 
\end{eqnarray}
 reads both for M2 and M5, using Eqs.(\ref{vielbein}), (\ref{M2}) and
 (\ref{M5}),  
\begin{eqnarray}
\label{duality1}
\partial_1 H&=& \partial_2 B\\
\label{duality2}
\partial_2 H&=&-\partial_1 B \, ,
\end{eqnarray}
where $B=A_{3\,4\,5\,6\,7\,8}$ for the M2 and $B=A_{9\,10\,11}$ for
 the M5. In this way, due to the particular choice we made for the
 tensor components defining the branes, the fields $A_{9\,10\,11}$ and
 $A_{3\,4\,5\,6\,7\,8}$ are interchanged  between the M2 and the M5
 when their common value switches from $1/H(x^1,x^2)$ to\footnote{We
 chose zero for the  integration constants  of the dual  fields
 $A_{9\,10\, 11}$ and $A_{3\,4\, 5\,6\,7\,8}$ (cf footnote~\ref{intconsfn}).} 
 $B(x^1,x^2)$.  Note however that for the M2 (M5) the time in
 $A_{3\,4\,5\,6\,7\,8}$ ($A_{9\,10\,11}$) is still 9 (3).  Eqs.(\ref
 {duality1}) and (\ref{duality2}) are the Cauchy-Riemann relations for
 the analytic function
\begin{equation}
\label{analytic}
{\cal E}_{(1)}= H + iB\, ,
\end{equation}
and  $H$ and $B$ are thus conjugate harmonic functions. The duality
relations Eqs.(\ref{duality1}) and (\ref{duality2})  allow for the
replacement of the Borel representatives ${\cal V}_1$ and ${\cal V}_2$
by 
\begin{eqnarray}
\label{boreln1}
{\rm M2}&:& {\cal V}^\prime_1= \exp \left [\frac{1}{2}\ln H\,
  h_{11}\right]\, \exp \left[B \,R^{[6]}_2\right]\\ 
\label{boreln2}
{\rm M5}&:& {\cal V}^\prime_2= \exp \left [\frac{1}{2} \ln H
  \,(-h_{11}-K^2{}_2)\right ] \, \exp \left[B\, R^{[3]}_1\right]\, . 
\end{eqnarray}
Note that the representative of the M5 in Eq.(\ref{boreln2}) is, as
the representative of the M2 in Eq.(\ref{2M2}), expressed in terms of
the supergravity metric and 3-form potential in two non-compact
dimensions.  

Consider now the purely gravitational BPS solutions. According to
Eqs.(\ref{GO}) and (\ref{G3})  the Borel representatives are 
\begin{eqnarray}
\label{bgr1}
&&{\rm KKW}: {\cal V}_0= \exp \left[\frac{1}{2}\ln H \,
  (K^3{}_3-K^{11}{}_{11})\right]\, \exp \left[(H^{-1}-1)\,
  K^3_{~11}\right]\\ 
\label{bgr2}
&&{\rm KK6M}: {\cal V}_3= \exp \left[\frac{1}{2}\ln H
  \,(-K^2{}_2-K^3{}_3+K^{11}{}_{11})\right ] \, \exp \left[H^{-1}\,
  R^{4\,5\,6\,7\,8\,9\,10\,11\vert \,11}\right]\, , 
\end{eqnarray}
with $H=H(x^1,x^2)$. Using the duality relations Eq.(\ref{Pdual2})
between the [8,1]-form and  $A_3^{(11)}$, on may express the Borel
representative of the KK6-monopole, as the representative of the
KK-wave Eq.(\ref{bgr1}), in two space dimensions in terms of the
11-dimensional metric.  

Transforming  by $E_9$ Weyl transformations  the Borel representatives
of the basic solutions given here, we shall obtain for all  $E_9$
generators associated to its real positive roots,  representatives
expressed in terms of the harmonic functions $H=H(x^1,x^2)$. These
will be transformed through dualities and compensations to Borel
representatives expressed in terms of new level 0 fields and 3-form
potentials $A_{9\,10\,11}$. This potential and the metric encoded in
the level 0 fields through the embedding relation Eq.(\ref{embedding})
and Eq.(\ref{vielbein}) will be shown to solve the equations of
motions of 11D-supergravity. In this way we shall find an infinite set
of $E_9$ BPS solutions related to the M2, M5, KK-waves and
KK6-monopoles  by U-duality, viewed as $E_9$ Weyl transformations.

\subsection{The M2 - M5 system}
\label{m2m5sec}

\subsubsection{The group-theoretical setting}

Generalizing the previous notation for generators to all levels by
using subscripts denoting the $A_9$ level, we write $R^{[3]}_1\equiv
R^{9\,10\,11}$ , $R^{[3]}_{-1}\equiv R_{9\,10\,11}$ and
$R^{[6]}_2\equiv R^{3\,4\,5\,6\,7\,8} $ , $R^{[6]}_{-2}\equiv
R_{3\,4\,5\,6\,7\,8}$. One has 

\begin{eqnarray} &&[R^{[3]}_1,
    R^{[3]}_{-1}]=h_{11}\qquad,\qquad[h_{11},
    R^{[3]}_1]=2R^{[3]}_1\nonumber\\ 
\label{geroch}
&&[R^{[6]}_2, R^{[6]}_{-2}]=-h_{11}-K^2_{~2}\qquad,\qquad[h_{11}, R^{[6]}_2]=-2R^{[6]}_2\\
&&[R^{[3]}_1, R^{[6]}_{-2}]=0\, .\nonumber
\end{eqnarray}
These commutation relations form a Chevalley presentation of a group
with Cartan  matrix 
\begin{equation}
\label{gerochC}
{\bf A}=\left[\begin{array}{cc}
2&-2\\-2&2 \\
\end{array}\right]\, ,
\end{equation}
and one verifies that the Serre relations are satisfied. This group is
the affine $A_1^+$ and hence isomorphic to the standard
Geroch group which is also the affine extension of $SL(2)$. The
central charge  is $k$ and derivation $d$, whose eigenvalues define 
the affine level, are here given by  the embedding of the $A_1^+$ in
$E_{10} $ as 
\begin{equation}
\label{brol}
k=-K^2_{\ 2}\quad , \quad d= -\frac {1}{3} K^2_{\
  2}+\frac{2}{9}(K^4_{\ 4} +\dots +K^9_{\ 9}) - \frac{1}{9}(K^3_{\ 3}
+K^{10}_{\ 10} +K^{11}_{\ 11})\, . 
\end{equation}
The level counting operator $d$ is not fixed uniquely by the present
embedding. 

The multicommutators satisfying the Serre relations form three
towers. The  positive  generators, normalized to one, are 
\begin{eqnarray}
\label{tower1}
R^{[3]}_{1+3n}&=& 2^{-n}\,\,
\left[R^{[3]}_1\left[R^{[3]}_1\left[R^{[6]}_2\left[R^{[3]}_1\dots\left[R^{[6]}_2\left[R^{[3]}_1,R^{[6]}_2\right]\right]\dots\right]\right.\right.\right.\qquad 
      n\ge 0\\ 
\label{tower2}
R^{[8,1]}_{3n}
&=&2^{-(n-1/2)}\,\,\left[R^{[3]}_1\left[R^{[6]}_2\left[R^{[3]}_1\dots\left[R^{[6]}_2\left[R^{[3]}_1,R^{[6]}_2\right]\right]\dots\right]\right.\right.
    \qquad n>0\\ 
\label{tower3}
R^{[6]}_{-1+3n}&= &2^{-(n-1)}\,\,
\left[R^{[6]}_2\left[R^{[3]}_1\dots\left[R^{[6]}_2\left[R^{[3]}_1,R^{[6]}_2\right]\right]\dots\right]\right.\qquad 
  n>0\, , 
\end{eqnarray} 
where the affine level $n$ is equal to the number of $R^{[6]}_2$ in
the tower\footnote{Shifts in the affine level by one unit corresponds
  to shifts in $A_9$ levels by three units, see
  also~\cite{Kleinschmidt:2006dy}. In what follows, when the 
  term level is left unspecified, we always mean the $A_9$
  level.}. The  $R^{[8,1]}_{3n} $ tower correspond to the null
roots $n\,\delta$ where 
\begin{equation}
\label{delta}
\delta = \alpha_3 + 2\alpha_4 +3\alpha_5 +4\alpha_6 +5\alpha_7
+6\alpha_8 +4\alpha_9 +2\alpha_{10} +3\alpha_{11}\, , 
\end{equation}
which has the properties
\begin{equation}
\label {products}
\langle \delta,\delta\rangle =0\qquad \langle\delta, \alpha_i
\rangle=0\quad i=3,\dots,11\, . 
\end{equation}
In particular $R^{[8,1]}_{3} $ is a linear combination
$R^{3\,4\,5\,6\,7\,8\,[9\,10\, , 11]}$ of level  3 tensors with all
indices distinct. Its height is 30 and thus exceeds the `classical'
limit 29 of \cite{Damour:2002cu}.  

Substituting $h_{11} =-h_{11}^\prime-K^2{}_2$ in Eq.(\ref{geroch}) one
obtains a presentation with $R^{[3]}_1$ and $R^{[6]}_2$
interchanged. While the $A_1^+$ group in the presentation
Eq.(\ref{geroch}) appears associated with the M2 brane, one could
associate the alternate presentation with the M5 brane. The two
presentations differ by  shifts in the affine level but not by the
$A_9$ level. To avoid complicated notations we keep for the complete
M2-M5 system the description given by Eq.(\ref{geroch}) which is
labelled explicitly in terms of  the $A_9$ level. The generators of
the $A_1^+$ group pertaining to the real roots of the M2-M5 system
appear in Fig.2a and in Fig.2b. 

All  the real roots of $E_9$  can be reached by $E_9$ Weyl
transformations acting on (say) $\alpha_{11}$ defining the generator
$R^{[3]}_1$. We shall find convenient for our construction of the
infinite set of the 
$E_9$ BPS-branes to generate all the real roots from two different
real roots, namely $\alpha_{11}$ and $-\alpha_{11}+\delta$
characterising respectively the generators $R^{[3]}_1$ and
$R^{[6]}_2$.\footnote{We note that in $A_1^+$ not all real roots are
  Weyl equivalent but there are two distinct orbits as we will see in
  more detail below.}
 In this section we obtain the generators of the
$R^{[3]}_{1+3n}$ and $R^{[6]}_{-1+3n}$ towers Eqs.(\ref{tower1}) and
(\ref{tower3}) and their negative counterparts  by the Weyl reflexions
$W_{\alpha_{11}}$ and $W_{-\alpha_{11}+ \delta}$ acting in alternating
sequences, starting  from their action on the generators $R^{[3]}_1$
and $R^{[6]}_2$. This is depicted in Fig.2a.  
The Weyl reflexions $W_{\alpha_{11}}$ and $W_{-\alpha_{11}+ \delta}$
generate the Weyl group of the affine subgroup
$A_1^+$ of $E_9$ depicted in Fig.6.\footnote{This
  is a Coxeter group whose presentation is $ \langle
  \,W_{\alpha_{11}}, W_{-\alpha_{11}+ \delta}\,\vert \,
  (W_{\alpha_{11}} W_{-\alpha_{11}+ \delta})^\infty = \text {id}
  \,Ê\rangle $.} Its  formal structure is discussed in Appendix G. 

\begin{figure}[h]
\centering
   \includegraphics[width=14 cm]{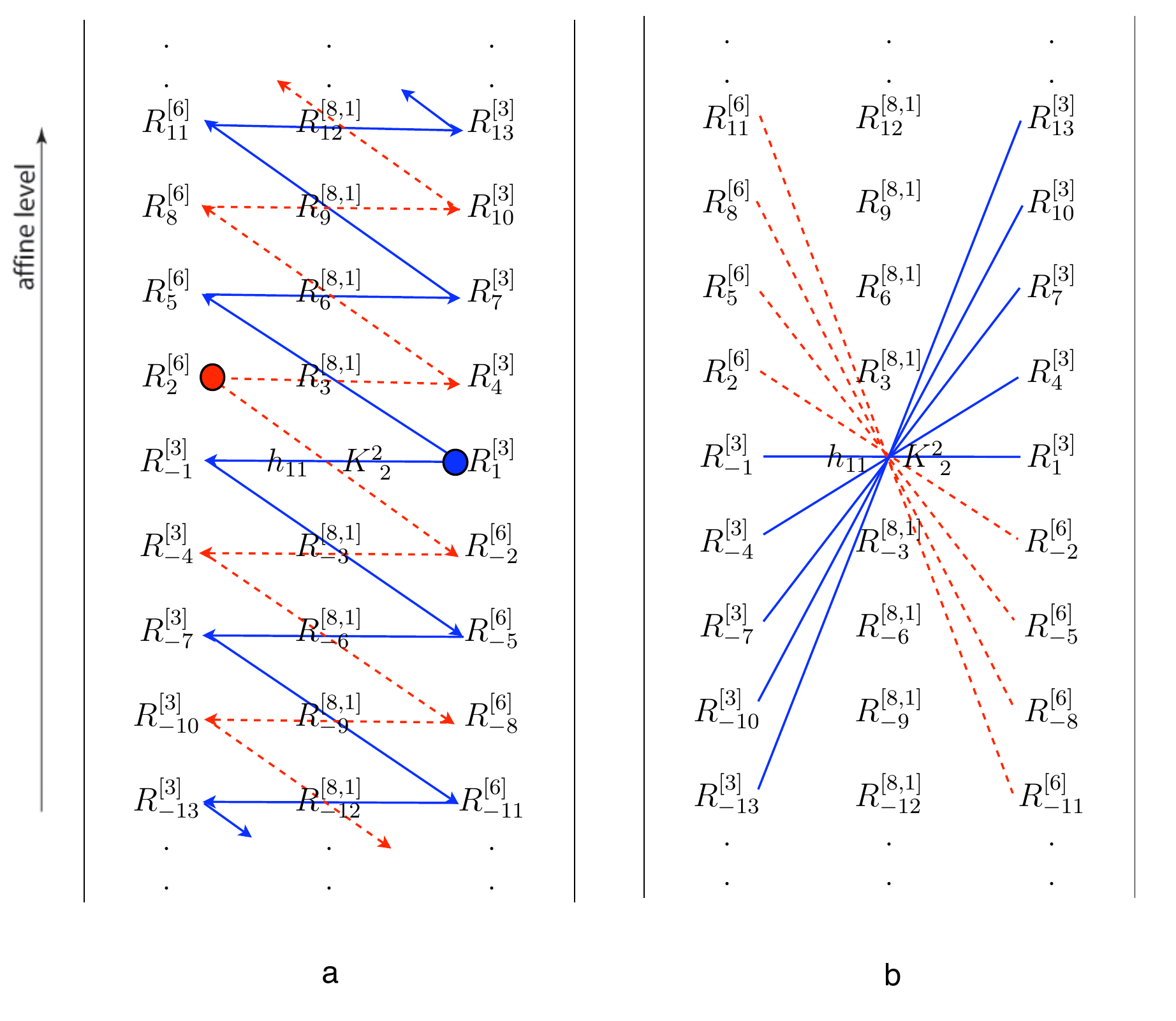}
 \caption {\label{ger1fig}\sl \small $A_1^+$ group for the M2-M5
 system. (a)  The Weyl 
 group: the M2 sequence is depicted by  solid lines and the M5
 sequence by dashed lines. Horizontal lines  represent Weyl reflexions
 $W_{\alpha_{11}}$ and diagonal ones $W_{-\alpha_{11}+ \delta}$. (b)
 The $SL(2)$ subgroups:   the solid lines label $SL(2)$ subgroups from
 the 3-tower Eq.(\ref{tower1})  and the dashed ones the $SL(2)$
 subgroups from the 6-tower Eq.(\ref{tower3}).  } 
\end{figure}

For the simply laced algebra considered here, with real roots normed
to square length 2,  the Weyl reflexion  $W_\alpha(\beta)$ in the
plane perpendicular to the real root $\alpha$ acting on the arbitrary
weight $\beta$ is given by 
\begin{equation}
\label{defWeyl}
W_\alpha(\beta)= \beta -\langle \beta , \alpha\rangle\,  \alpha
\end{equation}
Consider the Weyl reflexion $W_{-\alpha_{11} +\delta}$ acting on
$\alpha_{11} + n\,\delta$, which defines  $R^{[3]}_{1+3n}$ for $n\geq
0$ and $R^{[6]}_{1+3n}$ for $n<0$.  
One gets
\begin{equation}
\label{weyl1}
W_{-\alpha_{11} +\delta}(\alpha_{11}+n\,\delta) =
\alpha_{11}+n\,\delta -\langle \alpha_{11}+n\,\delta,  -\alpha_{11}
+\delta\rangle (-\alpha_{11} +\delta) = -\alpha_{11} +(n+2)\, \delta
\, , 
\end{equation}
where we have used Eq.(\ref{products}). The real root $ -\alpha_{11}
+(n+2)\, \delta$ defines the generator $R^{[6]}_{-1+3(n+2)}$ for $n\ge
0$ and $R^{[3]}_{-1+3(n+2)}$ for $n<0$.  
The Weyl reflexion has induced an $A_9$ level increase of four `units'
since $\d$ from (\ref{delta}) has three units of $A_9$ level and
$\alpha_{11}$ has one. 
Similarly, acting on the root $-\alpha_{11} +n\, \delta$, which
defines the generator $R^{[6]}_{-1+3n}$  for $n>0$  and
$R^{[3]}_{-1+3n}$ for $n\leq 0$, with the Weyl reflexion
$W_{\alpha_{11}}$  
one gets
\begin{equation}
\label{weyl2}
W_{\alpha_{11}}(-\alpha_{11}+n\,\delta)= -\alpha_{11}+n\,\delta
-\langle -\alpha_{11}+n\,\delta,  \alpha_{11} \rangle \alpha_{11} =
\alpha_{11} +n\, \delta \, . 
\end{equation}
This Weyl reflexion induces an $A_9$ level increase of two units. Thus acting
successively with the Weyl reflexions Eqs.(\ref{weyl1}) and
(\ref{weyl2}) on any root $\alpha_{11} + n\,\delta$ or in the reverse
order on the root $-\alpha_{11} +n\, \delta$ one induces a level
increase of six units, or equivalently of two affine levels. Of course
interchanging the initial and final roots and the order of the two
Weyl reflexions, one decreases the affine level by two units. Thus
starting from the roots $\alpha_{11}$ and $-\alpha_{11}+\delta$, we
obtain  the real roots defining all the generators of the
$R^{[3]}_{1+3n}$ and $R^{[6]}_{-1+3n}$ towers Eqs.(\ref{tower1}) and
(\ref{tower3}) and their negative counterparts. These form two
sequences depicted in Fig.2a. The `M2 sequence' originates
from the $\alpha_{11}$  root (and thus from the generator $R^{[3]}_1$)
and reads 
\begin{equation}
\label {M2seq}
\dots\stackrel{-\alpha_{11}+\delta}{\longleftarrow}R^{[3]}_{-7}\stackrel{\alpha_{11}}{\longleftarrow} 
R^{[6]}_{-5}\stackrel{-\alpha_{11}+\delta}{\longleftarrow}
R^{[3]}_{-1}\stackrel{\alpha_{11}}{\longleftarrow}\
\framebox{$R^{[3]}_1$}\stackrel{-\alpha_{11} +Ê\delta}{\longrightarrow}
R^{[6]}_5\stackrel{\alpha_{11} }{\longrightarrow}
R^{[3]}_7\stackrel{-\alpha_{11} +Ê\delta}{\longrightarrow}
R^{[6]}_{11}\stackrel{\alpha_{11}}{\longrightarrow}\dots  
\end{equation}
The `M5 sequence' originates from $-\alpha_{11}+\delta$ (and thus from
the generator $R^{[6]}_2$) and reads
\begin{equation}
\label {M5seq}
\dots\stackrel{\alpha_{11}}{\longleftarrow}R^{[6]}_{-8}\stackrel{-\alpha_{11}+\delta}{\longleftarrow}
R^{[3]}_{-4}\stackrel{\alpha_{11}}{\longleftarrow}
R^{[6]}_{-2}\stackrel{-\alpha_{11}+\delta}{\longleftarrow}\framebox{$R^{[6]}_2$}
\stackrel{\alpha_{11}
}{\longrightarrow} R^{[3]}_4\stackrel{-\alpha_{11}
  +Ê\delta}{\longrightarrow}
R^{[6]}_8\stackrel{\alpha_{11}}{\longrightarrow}
R^{[3]}_{10}\stackrel{-\alpha_{11} +Ê\delta}{\longrightarrow}\dots  
\end{equation}
Both sequences are represented in Fig.2a.

The Hodge duality relations Eqs.(\ref{dual1}) and (\ref{dual2}) will
play an essential role in the determinations of the $E_9$
BPS-branes. The Hodge dual generators  $R^{[3]}_1$ and $R^{[6]}_2$
have commutation relations  
\begin{equation}
\label{dualcom}
\left[R^{[3]}_1,R^{[6]}_2\right]= R^{[8,1]}_3 \, .
\end{equation}
The roots of the $E_9$ subalgebra do not contain $\alpha_2$ when expressed in terms of simple roots. Hence from Eq.(\ref{products}) any Weyl transformation from a $E_9$ real root leaves invariant (possibly up to a sign) the right hand side of Eq.(\ref{dualcom}). Therefore, the image of the basic pair
$R^{[3]}_1,R^{[6]}_2$ by any such $E_9$ Weyl transformation are pairs whose $A_9$ level sum is equal to three and we have:

\noindent
{\bf Theorem 1}
\it 
{The set of  Weyl transformations in $E_9$ mapping the $A_1^+$ group
  Eq.({\ref{geroch}) into itself either transforms the pair
    ($R^{[3]}_1,R^{[6]}_2$) into itself or into one of the pairs
    ($R^{[3]}_{1+3p} ,R^{[3]}_{-1-3(p-1)}$), ($R^{[6]}_{-1+3(p+1)}
    ,R^{[6]}_{1-3p}$) where $p$ is a positive integer.} \rm 

\noindent
This theorem applies  to the above Weyl transformations and is easily
checked from Fig.2a. 

The $A_1^+$ group  Eq.({\ref{geroch}) admits two infinite sets of
  $SL(2)$ subgroups 
\begin{eqnarray}
\label{sla}
\left[R^{[3]}_{1+ 3p},R^{[3]}_{-1-3p}\right]=h_{11}-p K^2{}_2 &,&
\left[h_{11}-p K^2{}_2 \, , R^{[3]}_{\pm(1+ 3p)}\right]=\pm 2R^{[3]}_{\pm (1+
  3p)}\quad (p\geq 0)\,,\\ 
\label{slb}
\left[R^{[6]}_{-1+ 3p},R^{[6]}_{1-3p}\right]=- h_{11}-p K^2{}_2 &,& 
\left[- h_{11}-p K^2{}_2\, , R^{[6]}_{\pm (-1+ 3p)}\right]
   =\pm 2R^{[6]}_{\pm (-1+ 3p)} \, (p >0)\,.\quad\quad
\end{eqnarray}
As all Weyl reflexions send opposite roots to opposite transforms, one has

\noindent
{\bf Theorem 2}
\it {The set of  Weyl transformations in $E_9$ mapping the $A_1^+$
  group Eq.({\ref{geroch}) into itself exchanges the $SL(2)$ subgroups
    between themselves.}\rm

\noindent
The subgroups Eq.(\ref {sla}) and (\ref {slb}) are depicted in Fig.2b. 

\subsubsection{The M2 sequence}

We take as representatives of the M2 sequence all the Weyl transforms
of the M2 representative Eq.(\ref {2M2}). The time coordinate is
9. Following  in Fig.2a the solid line towards positive step
generators, we encounter  Weyl transforms of the $SL(2)$ subgroup
generated by $(h_{11}\, , R^{[3]}_1\, ,R^{[3]}_{-1})$ represented by a
solid line in Fig.2b. Theorem 2 determines from  Eqs.(\ref{sla}) and
(\ref{slb}) the Weyl transform of the Cartan generators of
Eq.(\ref{2M2})  and we write 
\begin{eqnarray}
\label{borel3n}
{\cal V}_{1+6n}&=& \exp \left[\frac{1}{2} \ln H \, (h_{11}-2n K^2{}_2)\right]\, \exp\left [\frac{ 1}{H}\, R^{[3]}_{1+6n}\right]\qquad n\ge 0\\
\label{borel6n}
{\cal V}_{-1+6n}&= &\exp \left[\frac{1}{2} \ln H\, (-h_{11}-2n K^2{}_2)\right] \, \exp \left[\frac{1}{H}\, R^{[6]}_{-1+6n}\right] \qquad n>0\, .
\end{eqnarray}
We shall trade the tower fields $A^{[3]}_{1+6n}, A^{[6]}_{-1+6n}$
parametrising the generators in Eqs.(\ref{borel3n}) and
(\ref{borel6n}) in favour of the supergravity potential
$A_{9\,10\,11}$ and construct from them BPS solutions of 11D
supergravity. We have not indicated sign shifts induced by
the Weyl transformations in the tower fields from the sign of the
lowest level $n=0$ field which is taken to be $(+1/H)$. This is here
the only relevant sign,  as  discussed below. 

Let us consider explicitly the first two steps. These will introduce the two essential features of our construction: compensation and signature changes.

\bigskip
\noindent
{\bf $\bullet$ From level 1 to level 5: compensation}

Following in Fig.2a the solid line towards positive step generators, we first encounter the Weyl reflexion $W_{-\alpha_{11} +\delta}$ sending the level 1 generator $R^{[3]}_1$ to the level 5 generator $R^{[6]}_5$.  Eq.(\ref{borel6n}) for $n=1$ reads
\begin{equation}
\label{five}
{\cal V}_5= \exp\left [\frac{1}{2} \ln H\,(-h_{11}-2K^2{}_2)\right] \, \exp\left [\frac{1}{H}\, R^{[6]}_5\right]\, .
\end{equation}
As can be seen in Fig.2a,  the Weyl reflexion $W_{-\alpha_{11}
  +\delta}$ sending $R^{[3]}_1$ to $R^{[6]}_5$ sends its dual  $
R^{[6]}_2$ to $ R^{[6]}_{-2}$, in accordance with Theorem 1.  We call
$ R^{[6]}_{-2}$ the dual generator of $R^{[6]}_5$ and we get by acting
with  $W_{-\alpha_{11} +\delta}$ on  the dual representative for the
M2, Eq.(\ref {boreln1}), the `dual' representative of Eq.(\ref{five}), 
\begin{equation}
\label{prime5}
{\cal V}^\prime_5= \exp  \left[\frac{1}{2} \ln
  H\,(-h_{11}-2K^2{}_2)\right] \,\exp \left[-B\, R^{[6]}_{-2}\right]\,
. 
\end{equation}
The  $-$ sign in front of $B$ arises as follows. 
The   generators $-h_{11}-K^2{}_2\, ,\, R^{[6]}_2\, ,\,R^{[6]}_{-2}$
form an $SL(2)$ group, depicted by a dashed line in Fig.2b. We use the
representation\footnote{ $h_1$, $e_1$ and $f_1$ are the Chevalley
  generators of an  $SL(2)\subset E_9$.} 
\begin{equation}
\label{repsl}
h_1= \left[\begin{array}{cc}
1&0\\0&-1 \\
\end{array}\right]\,,\quad e_1=\left[\begin{array}{cc}
0&1\\ 0&0\\
\end{array}\right]\,,\quad f_1=\left[\begin{array}{cc}
0&0\\1&0\\
\end{array}\right]\, ,\quad K^2{}_2=\left[\begin{array}{cc}
1&0\\0&1\\
\end{array}\right]\, ,
\end{equation}
with  $ -h_{11}-K^2{}_2=h_1, R^{[6]}_2=e_1\, ,R^{[6]}_{-2}=f_1$, where
we have also included a representation for the central element $K^2{}_2$.
The Weyl reflexion $W_{-\alpha_{11} +\delta}$  is generated by the
group conjugation matrix $U_5$ of  $SL(2)$  \cite{Kac:book} 
\begin{equation}
\label{weyl5}
U_5=\exp R^{[6]}_{-2} \, \exp\,( - R^{[6]}_2 )\, \exp R^{[6]}_{-2}\, ,
\end{equation}
which can be represented by
\begin{equation}
\left[\begin{array}{cc}
1&0\\1&1\\
\end{array}\right]\left[\begin{array}{cc}
1&-1\\0&1\\
\end{array}\right]\left[\begin{array}{cc}
1&0\\ 1 &1\\
\end{array}\right] =\left[\begin{array}{cc}
0&-1\\1&0\\
\end{array}\right]\, ,
\end{equation}
and thus 
\begin{equation}
\label{trans5}
U_5 R^{[6]}_2 U^{-1}_5 =\left[\begin{array}{cc}
0&0\\-1&0\\
\end{array}\right] =-R^{[6]}_{-2}\, .
\end{equation}
One may verify that the conjugation matrix $U_5$ acting on the Cartan
generator of the representative of the M2 in dual form
Eq.(\ref{boreln1}) yields the same Cartan generator in the dual
representative ${\cal V}'_5$ in Eq.(\ref{prime5}) as in the direct
form ${\cal V}_5$, which was obtained from the level 1 representative
of the M2 Eq.(\ref{2M2}). 

We now write Eq.(\ref{prime5}) as an $SL(2)$ matrix times a factor
coming from the $K^2{}_2$ contribution. One has  
\begin{equation}
\label{matrix5}
{\cal V}^\prime_5= H^{-1/2}\left[\begin{array}{cc}
H^{1/2}&0\\0&H^{-1/2}\\
\end{array}\right]\left[\begin{array}{cc}
1&0\\ -B &1\\
\end{array}\right]\, ,
\end{equation}
The negative root in Eq.(\ref{prime5}) can be transferred to the
original Borel gauge by a compensating element of $K_{10}^{-}$. To
this effect we multiply on the left the matrix Eq.(\ref{matrix5}) by a
suitable element of the group  $SO(2)= SL(2)\cap K_{10}^{-}$. For a
well chosen $\theta$ we get 
\begin{eqnarray}
\overline{\cal V}^\prime_5&=&
H^{-1/2}\left[\begin{array}{cc}\cos\theta&\sin\theta\\-\sin\theta&\cos\theta\\ 
\end{array}\right]\left[\begin{array}{cc}
H^{1/2}&0\\0&H^{-1/2}\\
\end{array}\right]\left[\begin{array}{cc}
1&0\\ -B &1\\
\end{array}\right]\nonumber\\
&=& H^{-1/2}\left[\begin{array}{cc}
\displaystyle
\left(\frac{H}{H^2 + B^2}\right)^{-1/2}&0\\0& \displaystyle
\left(\frac{H}{H^2 + B^2}\right)^{1/2}\\
\end{array}\right]\left[\begin{array}{cc}
1& \displaystyle
\frac{-B}{H^2 + B^2}\\ 0 &1\\
\end{array}\right]\nonumber\\
\nonumber\\
\label{comp5}
&=&\exp \left[-\frac{1}{2}\ln (H^2 + B^2)\, K^2{}_2\right] \, \exp
\left[\frac{1}{2} \ln \frac{H}{H^2 + B^2}\, h_{11}\right]\, \exp
\left[\frac{-B}{H^2 + B^2}\, R^{[6]}_2\right]\, . 
\end{eqnarray}
Using the embedding relation Eq.(\ref{embedding}) we get from
Eq.(\ref{vielbein}) the metric\footnote{This solution of 11D
  supergravity has been derived previously in a different context
  \cite{Lozano-Tellechea:2000mc}.} encoded in the representative
Eq.(\ref{comp5})   
\begin{eqnarray}
{\rm Level }\ 5 &:& g_{11}=g_{22}=(H^2+B^2) \widetilde H^{1/3}\qquad
g_{33}=g_{44}=\dots =g_{88}= \widetilde H^{1/3}\nonumber\\ 
\label{metric5}
&&-g_{99}=g_{10\,10}=g_{11\,11}=\widetilde H^{-2/3}\, ,
\end{eqnarray}
where
\begin{equation}
\widetilde H =\frac{H}{H^2+B^2}\, .
\end{equation}
Using this metric and the duality equations Eq.(\ref{dual1}),
(\ref{dual2}), one obtains the supergravity 3-form potential $A_{9\,
  10\,11}$ dual to  $A_{3\,4\,5\,6\,7\,8}=-B/(H^2+B^2)$  
\begin{equation}
\label{potential5}
A_{9\, 10\,11}= \frac{1}{\widetilde H}\, ,
\end{equation}
and the dual representative of $\overline{\cal V}^\prime_5$, expressed
in terms of the potential Eq.(\ref{potential5}) is  
\begin{equation}
\label{fin5}
\overline{\cal V}_5=\exp\left [-\frac{1}{2}\ln (H^2 + B^2)\,
  K^2{}_2\right] \, \exp \left[\frac{1}{2} \ln \widetilde H\,
  h_{11}\right]\,  \exp\left [\frac{1}{\widetilde H}\,
  R^{[3]}_1\right] \, . 
\end{equation}
The dual pair $\widetilde H$ and $\widetilde B\equiv -B/(H^2+B^2)$ are
the conjugate harmonic functions defined by the analytic function
${\cal E}_2$ given by 
\begin{equation}
\label{analytic5}
{\cal E}_2= \frac{1}{{\cal E}_1}=\widetilde H + i\widetilde B\, .
\end{equation}
We shall verify later that the metric Eq.(\ref{metric5}) and the
potential Eq.(\ref{potential5}) solve the equations of motion of 11D
supergravity. 

\bigskip
\noindent
{\bf $\bullet$ From level 5 to level 7: signature change}

Pursuing further in Fig.2a the solid line towards positive step
generators, we encounter the Weyl reflexion $W_{\alpha_{11}}$  leading
from level 5 to the level 7 root $R^{[3]}_7$.  Eq.(\ref{borel3n})
reads 
\begin{equation}
\label{seven}
{\cal V}_7= \exp \left[\frac{1}{2} \ln H \, (h_{11}-2
  K^2{}_2)\right]\, \exp \left[\frac{1}{H}\, R^{[3]}_7\right]\, . 
\end{equation}
In the computation of the level 5 solution we have  
followed the sequence of duality transformations and the compensation
depicted on the second horizontal line  in  Fig.3. The sequence of
operations required to transform ${\cal V}_7$ to a representative
expressed in terms of the supergravity 3-form potential parametrizing
$R^{[3]}_1$ is depicted in the third line of Fig.3. As discussed below
in more details in the analysis of the full M2 sequence, all steps
appearing in the figure on the same column at  
levels 5 and 7 are related by the same Weyl transformation
$W_{\alpha_{11}}$. Hence one may short-circuit the first two dualities
and the first compensation and evaluate directly the Borel
representative pertaining to the last column of the level 5 line in
Fig.3. This amounts to take  
the Weyl transform by  $W_{\alpha_{11}}$
  of $\overline{\cal V}_5$ given in Eq.(\ref{fin5}). The generators
  $h_{11}\, , R^{[3]}_{-1}\, ,R^{[3]}_1$ generate as above an $SL(2)$
  group represented here by a solid line in Fig.2b. The Weyl
  conjugation matrix $U_7$ sending $R^{[3]}_1$ to $R^{[3]}_{-1}$ is 
  \begin{equation}
\label{weyl7}
U_7=\exp R^{[3]}_{-1}\, \exp\,( - R^{[3]}_1)\,  \exp R^{[3]}_{-1}\, ,
\end{equation}
which yields the result corresponding to Eq.(\ref{trans5}), namely
\begin{equation}
\label{trans7}
U_7\,R^{[3]}_1\, U^{-1}_7  =- R^{[3]}_{-1}\, .
\end{equation}
One gets
\begin{equation}
\label{in7}
\overline{\cal V}_7=\exp \left[-\frac{1}{2}\ln (H^2 + B^2)\,
  K^2{}_2\right] \, \exp \left[-\frac{1}{2} \ln \widetilde H\,
  h_{11}\right]\,  \exp\left[-\frac{1}{\widetilde H}\,
  R^{[3]}_{-1}\right]\, . 
\end{equation}
To convert the negative root into a positive one  we have to perform a
second compensation. Here a new phenomenon appears: the space-time
signature changes because  the temporal involution $\Omega$ does not
commute with the  above Weyl transformation. As explained in
Appendix~\ref{appw1}, the signature $(1,10, +)$ becomes $(2,9, -)$
with time coordinates 10 and 11 and negative kinetic energy for the
field strength. 
The signature change affects the compensation matrix. The intersection
of the $SL(2)$ group generated by $h_{11}\, , R^{[3]}_{-1}\,
,R^{[3]}_1$ with $K_{10}^{-}$ is not $SO(2)$ but $SO(1,1)$. Indeed the
transformed involution $\Omega^\prime$ resulting from the action of
the Weyl reflexion $\alpha_{11}$ yields $\Omega^\prime R^{[3]}_1= +
R^{[3]}_{-1}$, and the combination of step operators invariant under
$\Omega^\prime $ is the hermitian, hence {\em non-compact generator}
$R^{[3]}_1+R^{[3]}_{-1}$, implying   $SL(2)\cap K_{10}^{-}=SO(1,1)$.  

Recall that the field $1/\widetilde H$ in Eq.(\ref{in7}) is inherited
from Eq.(\ref{fin5}) which was obtained from Eq.(\ref{comp5}) using
Hodge duality. The latter is a differential equation and we have
hitherto chosen for simplicity the integration constant to be
zero. This choice would lead at level 7 to a singular compensating
matrix (see footnote~\ref{borelfn}) and we therefore will use instead
the field $ 1/\widetilde H-1$ 
(we could keep an arbitrary constant $\gamma\ne 0$ instead of $-1$ but that
would unnecessarily complicate notations). Using the matrices
Eq.(\ref{repsl}) with  $h_{11}=h_1\, , R^{[3]}_1=e_1\,
,R^{[3]}_{-1}=f_1$, we get for  a suitable choice of $\eta$ the
compensated representative 
\begin{eqnarray}
\overline{\overline{\cal V}}_7&=& (H^2+B^2)^{-1/2}\left[\begin{array}{cc}
\cosh\eta&\sinh\eta\\ \sinh\eta&\cosh\eta\\
\end{array}\right]\left[\begin{array}{cc}
\widetilde H^{-1/2}&0\\0&\widetilde H^{1/2}\\
\end{array}\right]\left[\begin{array}{cc}
1&0\\-\widetilde H^{-1}+1 &1\\
\end{array}\right]\nonumber\\
&=& (H^2+B^2)^{-1/2}\left[\begin{array}{cc}
(2-\widetilde H )^{1/2}&0\\0&(2-\widetilde H)^{-1/2}\\
\end{array}\right]\left[\begin{array}{cc}
1&1- (2-\widetilde H)^{-1}\\ 0 &1\\
\end{array}\right]\, ,\nonumber
\end{eqnarray}
or
\begin{equation}
\label{fin7}
\overline{\overline{\cal V}}_7=\exp \left[-\frac{1}{2}\ln (H^2+B^2)\,
  K^2{}_2 \right]\, \exp\left[ \frac{1}{2}\ln (2-\widetilde H )\,
  h_{11}\right]\, \exp \left[(1- \frac{1}{2-\widetilde H}) \,
  R^{[3]}_1\right]\, . 
\end{equation}
As shown below, this yields a solution of the supersymmetric exotic
partner of 11D supergravity with times in 10 and 11 and a negative
kinetic energy term. One has 
\begin{equation}
\label{potential7}
A_{9\,10\,11}=-\frac{1}{2-\widetilde H }\, ,
\end{equation}
where we dropped in the 3-form potential the irrelevant constant
$(-1)$. The metric encoded in the representative Eq.(\ref{fin7}) is,
from Eqs.(\ref{embedding}) and (\ref{vielbein}), 
\begin{eqnarray}
{\rm Level }\ 7 &:& g_{11}=g_{22}=(H^2+B^2) \widetilde{\widetilde H}^{1/3}\qquad g_{33}=g_{44}=\dots =g_{88}= \widetilde{\widetilde H}^{1/3}\nonumber\\
\label{metric7}
&&g_{99}=-g_{10\,10}=-g_{11\,11}=\widetilde{\widetilde H}^{-2/3}\, ,
\end{eqnarray}
where
\begin{equation}
\widetilde{\widetilde H} =2-\widetilde H\, .
\end{equation}
From the duality relations Eqs.(\ref{dual1}) and (\ref{dual2}), we see
that the field  $\widetilde{\widetilde B}$ dual to
$\widetilde{\widetilde H}$ is equal to $ -\widetilde B$. The dual pair
$\widetilde{\widetilde H}$ and $\widetilde{\widetilde B}$ are
conjugate harmonic functions associated to the analytic function 
\begin{equation}
\label{analytic7}
{\cal E}_3= 2-{\cal E}_2=\widetilde{\widetilde H} + i\widetilde
{\widetilde B}\, . 
\end{equation}

\bigskip
\noindent
{\bf $\bullet$ The complete M2 sequence}

The  M2 sequence is characterized by the generators $R^{[6]}_{-1+ 6n},
n>0$ and $R^{[3]}_{1+ 6n}, n\ge 0$. These are reached by following in
Fig.2a the solid line starting from $R^{[3]}_1$ towards the positive
roots. The representatives are given in Eqs.(\ref{borel3n}) and
Eq.(\ref{borel6n}).  
 
 \begin{figure}[h]
   \centering
   \includegraphics[width=17cm]{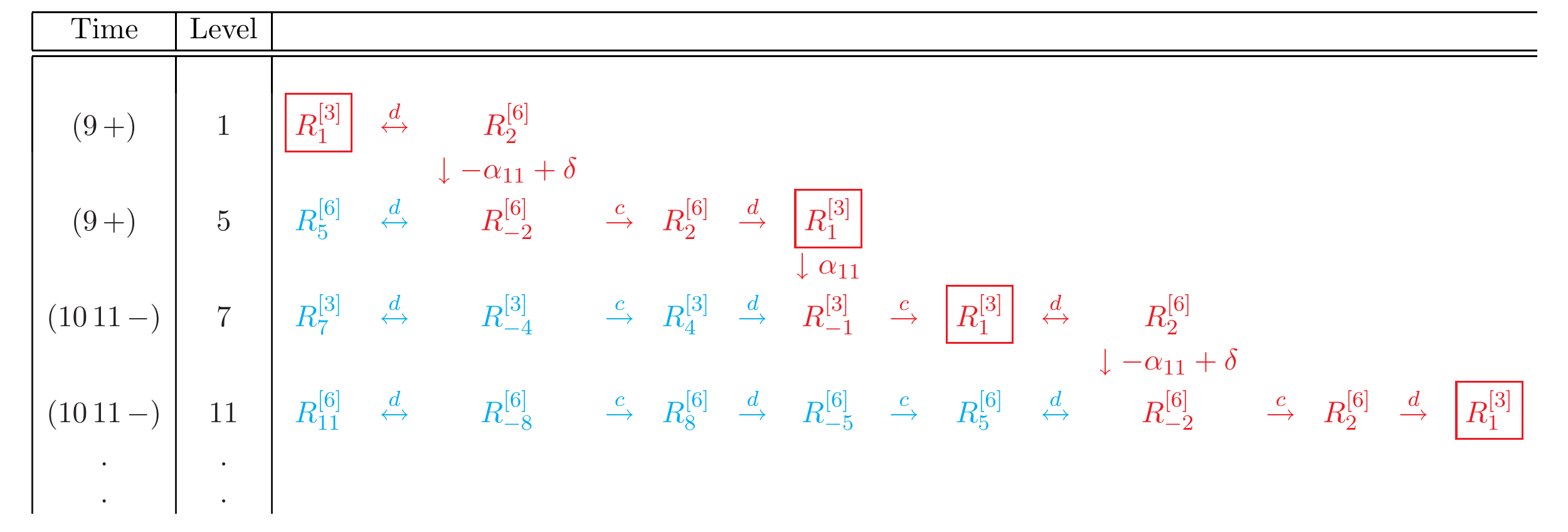}
 \caption {\sl \small The construction of BPS states for the M2
 sequence. The superscript $d$ labels a duality and the superscript
 $c$ labels a compensation. The horizontal rows are connected by Weyl
 reflexions as indicated.} 
 \label{second}
\end{figure}

A glance on Fig.3 shows that at any level of this sequence one may
 trade this representative either by performing the Weyl
 transformation $W_{\alpha_{11}}$ on the preceding level
 representative 
 expressed in terms of the generator $R^{[3]}_1$ (as was done to reach
 the level 7 from the level 5 representative), or  by  the Weyl
 transformation $W_{-\alpha_{11} +\delta}$ on its dual representative
 in terms of $R^{[6]}_2$ (as was done to reach the level 5 from the
 level 1 representative). In this way, one bypasses all the steps
 depicted on horizontal lines in Fig.3 which involve complicated
 duality relations and compensations  to solely perform a single
 compensation by a $SO(2)$ or $SO(1,1)$ matrix  and a known Hodge
 duality defined by Eqs.(\ref{dual1}) and (\ref{dual2}), as
 exemplified by the detailed analysis of the two first levels of the
 sequence.  The nature of the compensation needed  in the construction
 of the M2 sequence alternates at each step on a horizontal line of
 Fig.3 between $SO(2)$ and $SO(1,1)$ as shown in
 Appendix~\ref{appw1}. The representatives of the M2 sequence in terms
 of the supergravity fields are easily written in terms of complex
 potentials ${\cal E}_n$. These are obtained by operating on the
 analytic function ${\cal E}(z)_1= H(z,\bar z) + iB(z,\bar z), \, z=
 x^1+ix^2, \bar z= x^1-ix^2$, successively by  inversions  and
 translations according to 
 \begin{equation}
\label{Moebact}
{\cal E}(z)_{2n}= \frac{1}{{\cal E}(z)_{2n-1}}\,,\qquad{\cal
  E}(z)_{2n+1}= 2-{\cal E}(z)_{2n} \qquad n >0\, . 
\end{equation}
These formul\ae{} summarise the action of the $E_9$ Weyl group on BPS
states which are largely characterised by the harmonic functions
${\cal E}$. From Eq.(\ref{Moebact}) one sees that the action consists
of inversion and shift in a way very similar to the modular group
$SL(2,{\mathbb Z})$. In order to see that there is more than just the
action of an $SL(2,{\mathbb Z})$ one must consider the action of the
transformations on the full metric, including in particular the
conformal factor.
For the full solution one gets, defining ${\cal
  F}_{2n-1}={\cal E}_{1}{\cal E}_{3}\dots {\cal 
  E}_{2n-1} $ for $n >0$ (and ${\cal F}_{-1} \equiv 1$), the
straightforward generalisation of Eqs. (\ref{2M2}), (\ref{fin5}) and
(\ref{fin7}), 

\begin{eqnarray}
\label{seqM21}
{\cal V}_{1+6n}&=&\exp \left[-\frac{1}{2}\ln ({\cal F}_{2n-1}
  \bar{\cal F}_{2n-1})\, K^2{}_2\right] \, \exp\left[\frac{1}{2} \ln
  {\cal R}e\, {\cal E}_{2n+1}\, h_{11}\right]\,  \exp
\left[\frac{(-1)^{n}}{{\cal R}e\, {\cal E}_{2n+1}}\,
  R^{[3]}_1\right]\nonumber\\ 
n\geq 0\\
\nonumber\\
\label{seqM22}
{\cal V}_{-1+6n}&=&\exp \left[-\frac{1}{2}\ln( {\cal F}_{2n-1}
  \bar{\cal F}_{2n-1})\, K^2{}_2\right] \, \exp \left[\frac{1}{2} \ln
  {\cal R}e\, {\cal E}_{2n}\, h_{11}\right]\,  \exp
\left[\frac{(-1)^{n+1}}{{\cal R}e\, {\cal E}_{2n}}\,
  R^{[3]}_1\right]\nonumber\\ 
n>0
\end{eqnarray}
where ${\cal \bar{F}}$ denotes the complex conjugate of ${\cal F}$ and
the signatures in Eq.(\ref{seqM21})  are $(1,10,+)$ with time
in 9 for $n$ even and $(2,9, -)$ with times in 10,11 for $n$ odd. In
Eq.(\ref{seqM22}) the signatures are $(2,9, -)$ with times in 10,11
for $n$ even  and $(1,10,+)$  with time in 9 for $n$ odd. The detailed
analysis of the signatures for the M2 sequence is done in
Appendix~\ref{appw1} and the final results are summarised in Table 3.  
To interchange the role of even and odd $n$ in the above signatures,
one simply builds another M2 sequence starting from the  exotic M2 of
the theory $(2,9,-)$ whose metric is given in Eq.(\ref{M2metricE}). It
has two longitudinal times in 10 and 11 and one longitudinal spacelike
direction 9.

Eqs.(\ref{seqM21}) and (\ref{seqM22})  yield from Eqs.(\ref{vielbein})
and (\ref{embedding}) the metric and three-form potential 
\begin{eqnarray}
\label{M2odd}
ds^2&=& {\cal F}_{2n-1}\bar{\cal
 F}_{2n-1}H_{2n+1}^{1/3}[(dx^1)^2+(dx^1)^2]
 +H_{2n+1}^{1/3}[(dx^3)^2+\dots+(dx^8)^2]+\\ 
 &+& H_{2n+1}^{-2/3}[(-1)^{(n+1)}
 (dx^9)^2+(-1)^n(dx^{10})^2+(-1)^n(dx^{11})^2] \quad\qquad  (n \geq
 0)\nonumber \\A_{9\,10\,11}&=&\frac{(-1)^{n}}{H_{2n+1}} \nonumber\\ 
\nonumber\\
ds^2&=& {\cal F}_{2n-1}\bar{\cal
 F}_{2n-1}H_{2n}^{1/3}[(dx^1)^2+(dx^1)^2]
 +H_{2n}^{1/3}[(dx^3)^2+\dots+(dx^8)^2]+\label{M2even} \\ 
\nonumber
&+ &H_{2n}^{-2/3}[(-1)^n (dx^9)^2+(-1)^{n+1}(dx^{10})^2+(-1)^{(n+1)}
 (dx^{11})^2]\quad\qquad n>0 \nonumber\\ 
A_{9\,10\,11}&=&\frac{(-1)^{n+1}}{H_{2n}}\nonumber
\end{eqnarray}
with $H_p={\cal R}e\, {\cal E}_p$. We stress that an important effect
of the action of the affine Weyl group on the BPS solutions is the
change in the conformal factor which is expressed through ${\cal
  F}_{2n-1}\bar{\cal  F}_{2n-1}$. 

For each level on the M2-sequence, these equations satisfy the
equations of motion of 11D supergravity or of its exotic counterpart
outside the singularities of the functions $H_p$.  There, the factor
${\cal F}_{2n-1}\bar{\cal F}_{2n-1}$ can indeed be eliminated by a
change of coordinates and the functions $H_p$ are still harmonic
functions in the new coordinates. Eqs.(\ref{M2odd}) and (\ref{M2even})
have then the same dependence on $H_p$ as the M2 metric and 3-form
have on $H_1\equiv H$ and differ thus from the M2 solution only
through the choice of the harmonic function. They therefore solve the
Einstein equations. This is also discussed more abstractly in
section~\ref{analyticsec}. 
 
To obtain these results, we have chosen a particular path to reach
from level 1 the end of any horizontal line in Fig.3. Along this path,
all signs of the fields in the representatives were fixed by the
choice $+1/H$ at level 1 and by the Hodge duality relations
Eqs.(\ref{dual1}) and (\ref{dual2}).  Thus, we do not have to
explicitly take into account signs which might affect higher level
tower fields in Eqs.(\ref{borel3n}) and (\ref{borel6n}). 

The consistency of the procedure used in this Section to obtain
solutions related by U-dualities viewed as Weyl transformations rests
however on the arbitrariness of the path chosen to reach from level 1
the end of any horizontal line in Fig.3. Dualities for levels $l>2$
are in principle defined by the Weyl transformations.  Consistency is
thus equivalent to commuting  Weyl transformations with
compensations. Compensations and Weyl transformations do indeed
commute, as proven in Appendix E.  

\subsubsection{The M5 sequence}

We follow the same procedure as for the M2 sequence. We take as
representatives of the M5 sequence all the Weyl transforms of the M5
representative Eq.(\ref {2M5}) with fields Eq.(\ref{M5}) and time
coordinate 3. Following  in Fig.2a the dashed line towards positive
step generators, we encounter  Weyl transforms of the $SL(2)$ subgroup
generated by $(-h_{11}- K^2{}_2\, , R^{[6]}_2\, ,R^{[6]}_{-2})$
represented by a dashed line in Fig.2b.  Theorem 2 determines from
Eqs.(\ref{sla}) and (\ref{slb}) the Weyl transform of the Cartan
generators of Eq.(\ref{2M5})  and we write 
\begin{eqnarray}
\label{6Nborel}
{\cal V}_{2+6n}&=& \exp \left[\frac{1}{2} \ln H \, (-h_{11}-(
2n+1) K^2{}_2)\right]\, \exp \left[\frac{1}{H}\,
  R^{[6]}_{2+6n}\right]\qquad n\ge 0\\ 
\label{3Nborel}
{\cal V}_{-2+6n}&= &\exp\left[\frac{1}{2} \ln H\, (h_{11}-(2n-1)
  K^2{}_2)\right] \, \exp \left[\frac{1}{H}\, R^{[3]}_{-2+6n}\right]
\qquad n>0\, . 
\end{eqnarray}
We shall trade the tower fields $A^{[6]}_{2+6n}=A^{[3]}_{-2+6n}$  in
favour of the supergravity potential $A_{9\,10\,11}$ and construct
from them BPS solutions of 11D supergravity. 

For the M5 itself, we take the dual representative  $ {\cal
  V}^\prime_2$ expressed in terms of the 3-form potential, which is
given in Eq.(\ref{boreln2}). As previously the first two steps, levels
4 and 8, contain the essential ingredients of the whole sequence. 

\begin{figure}[h]
   \centering
   \includegraphics[width=17cm]{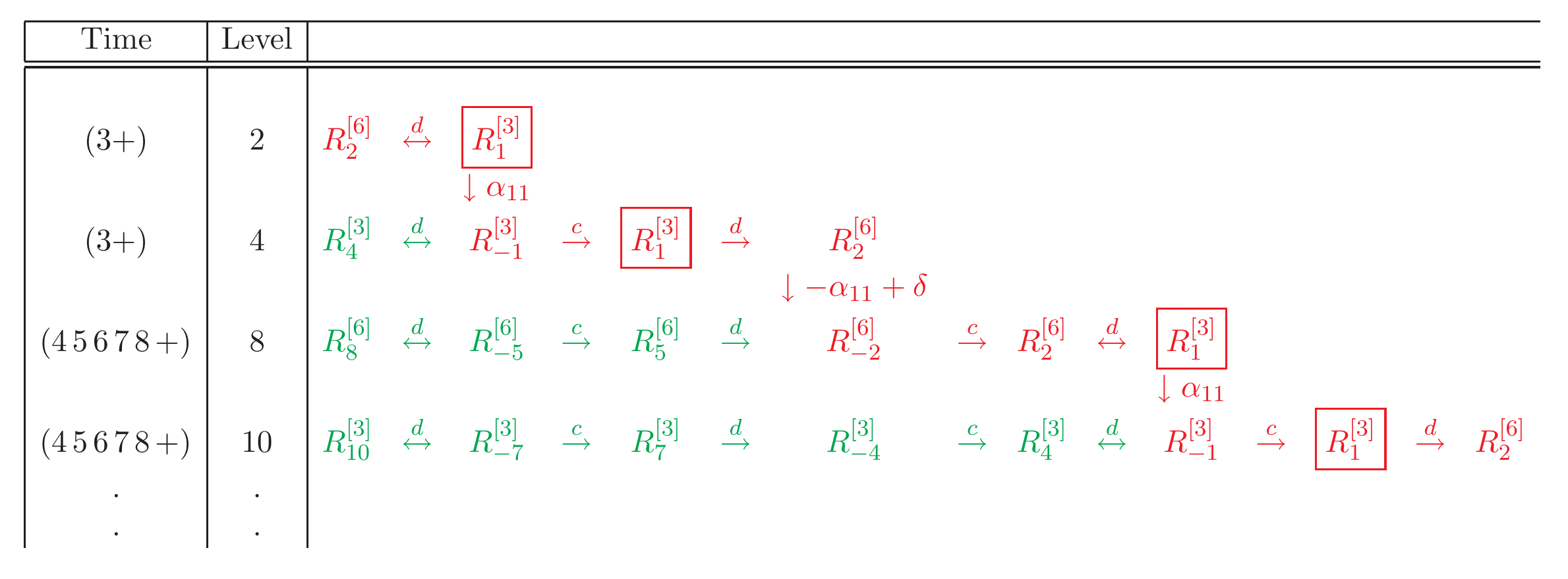}
 \caption {\sl \small The construction of BPS states for the M5
 sequence. The superscript $d$ labels a duality and the superscript
 $c$ labels a compensation.} 
\end{figure}

Following in Fig.2a the dashed line towards positive step generators,
we first encounter the Weyl reflexion $\alpha_{11} $ sending the level
2 generator $R^{[6]}_2$ to the level 4 generator $R^{[3]}_4$, or
equivalently, as exhibited in Fig.4, the dual generator $R^{[3]}_1$ to
the generator $R^{[3]}_{-1}$. Applying this Weyl reflexion to
Eq.(\ref{boreln2}) and performing an $SO(2)$ compensation we get the
representative 
\begin{equation}
\label{fin4}
\overline{\cal V}_4=\exp \left[-\frac{1}{2}\ln (H^2 + B^2)\,
  K^2{}_2\right] \, \exp \left[-\frac{1}{2} \ln \widetilde H\,
  (h_{11}+ K^2{}_2)\right]\,  \exp \left[\widetilde B\,
  R^{[3]}_1\right]\, , 
\end{equation} 
which yields $A_{9\, 10\,11}= \widetilde B$ and the
metric\footnote{This solution of 11 D supergravity has already been
  derived in a different context   \cite{Lozano-Tellechea:2000mc}.} 
\begin{eqnarray}
{\rm Level }\ 4 &:& g_{11}=g_{22}=(H^2+B^2) \widetilde H^{2/3}\qquad
-g_{33}=g_{44}\dots =g_{88}= \widetilde H^{-1/3}\nonumber\\ 
\label{metric4}
&&g_{99}=g_{10\,10}=g_{11\,11}=\widetilde H^{2/3}\, ,
\end{eqnarray}
with $\tilde{H}$ and $\tilde{B}$ as before.
The level 4 results are in  agreement with the interpretation of the
Weyl reflexion $W_{\alpha_{11}}$ as a double T-duality in the
directions 9 and 10 plus interchange of the two directions
\cite{Elitzur:1997zn, Obers:1998rn, Banks:1998vs, Englert:2003zs}. We
recover indeed  the level 4 metric and  3-form by applying  Buscher's
duality rules to the  M5 smeared in the directions $9,10$ and
$11$. This is shown in Appendix~\ref{appb}. 
The next step leads to level 8. As for the computation of the level 7
representative in the M2 sequence, we may skip the two first dualities
and the first compensation indicated in the third line of Fig.4. It
suffices to perform  the Weyl reflexion $W_{-\alpha_{11} +\delta}$ on
the dual representative of Eq.(\ref{fin4}) followed by a $SO(1,1)$
compensation and a Hodge duality. One gets 
\begin{equation}
\label{fin8}
\overline{\overline{\cal V}}_8=\exp\left[-\frac{1}{2}\ln (H^2+B^2)\,
  K^2{}_2\right]\, \exp\left[ -\frac{1}{2}\ln \widetilde{\widetilde
    H}\, (h_{11}+ K^2{}_2)\right]\, \exp \left[- \widetilde{\widetilde
    B} \, R^{[3]}_1\right]\, , 
\end{equation}
which yields $A_{9\, 10\,11}=-\widetilde{\widetilde B}$ and the metric
\begin{eqnarray}
{\rm Level }\ 8 &:& g_{11}=g_{22}=(H^2+B^2) \widetilde{\widetilde
  H}^{2/3}\qquad g_{33}=-g_{44}\dots =-g_{88}= \widetilde{\widetilde
  H}^{-1/3}\nonumber\\ 
\label{metric8}
&&g_{99}=g_{10\,10}=g_{11\,11}=\widetilde{\widetilde H} H^{2/3}\, .
\end{eqnarray}
As shown in Appendix~\ref{appw2} the signature is now $(5, 6,+)$ with
times in 4,5,6,7,8. 

The full M5 sequence is characterized by the roots $R^{[3]}_{-2+ 6n},
n>0$ and $R^{[6]}_{2+ 6n}, n\ge 0$. These are reached by following in
Fig.2a the dashed line starting at $R^{[6]}_2$ towards the positive
roots. The representative is defined by the Cartan generator given in
Eq.(\ref{sla}) or Eq.(\ref{slb}) and by the field $1/H$ multiplying
the positive root.  
As for the M2 sequence, the generalisation to all levels to the lowest
ones Eqs.(\ref{boreln2}), (\ref{fin4}) and (\ref{fin8}) is
straightforward. As indicated in Fig.4,  one obtains iteratively the
representatives in terms of the supergravity 3-form by solely
performing  a single compensation by a $SO(2)$ or $SO(1,1)$ matrix
and a known Hodge duality defined by Eqs.(\ref{dual1}) and
(\ref{dual2}).  One alternates after two steps between
representatives with a single time in 3 and  exotic ones with times in
4,5,6,7,8. The nature of the compensation changes at each step.  
One has
\begin{eqnarray}
{\cal V}_{2+6n}&=&\exp \left[-\frac{1}{2}\ln ({\cal F}_{2n-1}
  \bar{\cal F}_{2n-1})\, K^2{}_2\right] \, \exp \left[-\frac{1}{2} \ln
  {\cal R}e\, {\cal E}_{2n+1}\, (h_{11}+K^2{}_2)\right]\nonumber \\  
\label{seqM51}
&  \cdot &\exp \left[ (-1)^{n} {\cal I}m\, {\cal E}_{2n+1}\,
  R^{[3]}_1\right] \quad \quad n\geq 0\\ 
\nonumber\\
{\cal V}_{-2+6n}&=&\exp \left[-\frac{1}{2}\ln ({\cal F}_{2n-1}
  \bar{\cal F}_{2n-1})\, K^2{}_2\right] \, \exp\left[-\frac{1}{2} \ln
  {\cal R}e\, {\cal E}_{2n}\, (h_{11}+K^2{}_2)\right]\nonumber\\ 
\label{seqM52}
&\cdot& \exp\left[ (-1)^{n+1}{\cal I}m\, {\cal E}_{2n}\,
  R^{[3]}_1\right] \quad \quad 
n>0
\end{eqnarray}
where in Eq.(\ref{seqM51}) one has the signatures $(1,10,+)$ with time
in 3 for $n$ even and $(5,6, +)$ with times in 4,5,6,7,8 for $n$ odd,
and in Eq.(\ref{seqM52}) the signatures are  $(1,10, +)$  with time in
3 for $n$ odd and $(5,6,+)$ with times in 4,5,6,7,8 for $n$ even. As
previously it is always possible to interchange at each pair of levels
the two signatures by choosing an exotic M5 to initiate the
sequence. The detailed analysis of the signatures for the M5 sequence
and  of the compensations required is done in Appendix~\ref{appw2} and
summarised  in Table 5. 

These representative yield the metric and 3-form potential for all
states on the M5 sequence. We get from Eqs.(\ref{seqM51}) and
(\ref{seqM52}) 
\begin{eqnarray}
\label{M5odd}
ds^2&=& {\cal F}_{2n-1}\bar{\cal
 F}_{2n-1}H_{2n+1}^{2/3}[(dx^1)^2+(dx^1)^2]+H_{2n+1}^{2/3}[
 (dx^9)^2+(dx^{10})^2+(dx^{11})^2]\\&+&H_{2n+1}^{-1/3}[(-1)^{n+1}(dx^3)^2
 +(-1)^n(dx^4)^2\dots+(-1)^n(dx^8)^2] \qquad\qquad(n \geq
 0)\nonumber\\ 
 \nonumber A_{9\,10\,11}&=& (-1)^{n} B_{2n+1}\\
\nonumber\\
\label{M5even} 
ds^2&=& {\cal F}_{2n-1}\bar{\cal
  F}_{2n-1}H_{2n}^{2/3}[(dx^1)^2+(dx^1)^2]
+H_{2n}^{2/3}[(dx^9)^2+(dx^{10})^2+
  (dx^{11})^2]\\&+&H_{2n}^{-1/3}[(-1)^n(dx^3)^2+(-1)^{n+1}(dx^4)^2+\dots+(-1)^{n+1}(dx^8)^2]  \qquad (n>0)\nonumber \\ 
\nonumber
A_{9\,10\,11}&=& (-1)^{n+1} B_{2n}
\end{eqnarray}
with $B_p={\cal I}m\, {\cal E}_p$. 

For each level on the M5-sequence, these equations satisfy   the
equations of motion of 11D supergravity or of its exotic counterpart
outside the singularities of the harmonic functions $H_p$ and $B_p$.
There, the factor ${\cal F}_{2n-1}\bar{\cal F}_{2n-1}$ can indeed be
eliminated by a change of coordinates and the functions $H_p$ and
$B_p$ are still conjugate harmonic functions of the new
coordinates. Eqs.(\ref{M5odd}) and (\ref{M5even}) have then the same
dependence on $H_p$ and $B_p$ as the M5 metric and 3-form have on
$H_1\equiv H$ and $B_1\equiv B$ and differ thus from the M5 solution
only through the choice of the harmonic functions. They therefore
solve the Einstein equations.

\subsection{The gravity tower}

The affine $A_1^+$ group generated by $R^{[3]}_1\equiv R^{9\,10\,11}$ and
$R^{[6]}_2\equiv R^{3\,4\,5\,6\,7\,8} $ spans three towers of
generators. We found BPS solutions for each positive generator of the
3-tower Eq.(\ref{tower1}) and of the 6-tower Eq.(\ref{tower3}). All
these generators correspond to real roots while those in the third
tower Eq.(\ref{tower2}) generators correspond to null roots of
square length zero.  
Each generator of the third tower at level $3(n+1)\quad n\ge 0$
belongs to an irreducible representation of $A_8\subset E_9$ whose
lowest weight is the real root $\alpha_4 + 2\alpha_5
+3\alpha_6+4\alpha_7 +5\alpha_8 +3\alpha_9 +\alpha_{10} +3\alpha_{11}
+n\delta$. We now show that the  lowest weight generators belong to a
$A_1^+$ subgroup  of $E_9$ generated by $R^{4\,5\dots 10 \, 11\vert
  11}$ which sits at level 3 and by $K^3{}_{11}$, which is defined by
the level 0 real root
$\alpha_3+\alpha_4+\alpha_5+\alpha_6+\alpha_7+\alpha_8+\alpha_9+\alpha_{10}$. 

These two generators are related as follows
\begin{eqnarray}
\label{ger1}
K^3{}_{11} &\leftrightarrow&
\alpha_3+\alpha_4+\alpha_5+\alpha_6+\alpha_7+\alpha_8+\alpha_9+\alpha_{10}\equiv\lambda\,
,\\ 
\label{ger2}
R^{4\,5\dots 10 \, 11\vert 11} &\leftrightarrow& \alpha_4 + 2\alpha_5
+3\alpha_6+4\alpha_7 +5\alpha_8 +3\alpha_9 +\alpha_{10}
+3\alpha_{11}=-\lambda+\delta \, , 
\end{eqnarray}
where the last equality in Eq.(\ref{ger2}) is easily checked using
Eq.(\ref{delta}). They are  
Weyl transforms of the generators $R^{9\,10\,11}$ and
$R^{3\,4\,5\,6\,7\,8}$ of the the $A_1^+$ group defined in
Eq.(\ref{geroch}). To see this, first perform the Weyl transformation
interchanging 9 and 3. The $A_1^+$ generators  Eq.(\ref{geroch}) are
transformed to (all Weyl transforms of step generators are written up
to a sign) 
\begin{eqnarray}
R^{9\,10\,11}&\to &R^{3\,10\,11}\\
R^{3\,4\,5\,6\,7\,8}&\to & R^{4\,5\,6\,7\,8\,9}\, ,
\end{eqnarray}
defined by the roots $\alpha_3 +
\alpha_4+\alpha_5+\alpha_6+\alpha_7+\alpha_8 +\alpha_{11}$ and
$\alpha_4+2\alpha_5+3\alpha_6+4\alpha_7+5\alpha_8+4\alpha_9
+2\alpha_{10} +2\alpha_{11}$. 
Then perform the Weyl reflexion  $W_{\alpha_{11}}$ to get the generators
\begin{eqnarray}
R^{3\,10\,11} &\to &  K^3{}_9\\
R^{4\,5\,6\,7\,8\,9} &\to &  R^{4\,5\,6\,7\,8\,9\,10\,11\vert 9}\, ,
\end{eqnarray}
defined by the roots $\alpha_3 +
\alpha_4+\alpha_5+\alpha_6+\alpha_7+\alpha_8$ and  
$\alpha_4+2\alpha_5+3\alpha_6+4\alpha_7+5\alpha_8+4\alpha_9
+2\alpha_{10} +3\alpha_{11}$. Finally perform the Weyl
transformation exchanging 9 and 11 to get 
\begin{eqnarray}
K^3{}_9 &\to &  K^3{}_{11}\\
R^{4\,5\,6\,7\,8\,9\,10\,11\vert \,9} &\to &
R^{4\,5\,6\,7\,8\,9\,10\,11\vert \,11}\, , 
\end{eqnarray}
whose defining roots are $\lambda$ and $-\lambda +\delta$. 
The transformed Cartan generators are $K^3{}_3 -K^{11}{}_{11}$ and
$-K^2{}_2 +K^{11}{}_{11}-K^3{}_3$. Under these transformations, the
M2-brane generator is mapped onto the  Kaluza-Klein wave generator in
the direction 11. The M5-brane generator is mapped to the dual
Kaluza-Klein monopole generator $R^{4\,5\,6\,7\,8\,9\,10\,11\vert
  \,11}$. These generate the `gravity $A_1^+$ group' conjugate in $E_9$
to the `brane $A_1^+$ group' Eq.(\ref{geroch}). 

We now find the BPS solutions of 11D pure gravity (which are of course
solution of 11D supergravity) associated to each positive real root of
the gravity $A_1^+$ group. One could redo  the analysis of the M2-M5
system starting from the representatives of the KK-wave and
KK6-monopole given in Eqs.(\ref{bgr1}) and  (\ref{bgr2}) and the
duality relations Eq.(\ref{Pdual2}).  It is however simpler to take
advantage of the Weyl mapping of the two $A_1^+$ subgroups of $E_9$ 
\begin{eqnarray}
\label{bragramap}
R^{9\,10\,11} &\leftrightarrow &  K^3{}_{11} \\
\label{bragra1}
R^{4\,5\,6\,7\,8\,9} &\leftrightarrow &  R^{4\,5\,6\,7\,8\,9\,10\,11\vert 11}\\
 \alpha_{11} \leftrightarrow   \lambda\quad &, & \quad\delta\leftrightarrow\delta\, .
  \label{bragra2}
\end{eqnarray}

The generators $R^{[3]}_{1+3n}$ of the 3-tower Eq.(\ref{tower1}) are
mapped to generators of level $3n$. We label these generators
$R^{[0]}_{3n}$ ($R^{[0]}_{0}\equiv K^3{}_{11}$). The generators
$R^{[6]}_{-1+3n}$ of the 6-tower Eq.(\ref{tower3}) are also mapped to
generators of level $3n \, (n>0)$. We label these generators $\bar
R^{[8,1]}_{3n}$ ($\bar R^{[8,1]}_{3}\equiv
R^{4\,5\,6\,7\,8\,9\,10\,11\vert 11}$).  In the mapping the signature
changes as shown in Appendix~\ref{appsg}. In particular, the KK-wave
$R^{[0]}_{0}$ yields a single time in 3 and  the KK6-monopole $\bar
R^{[8,1]}_{3}$ becomes exotic with two times 9 and 10. This mapping of
the M2-M5 sequences of Fig.2a to the gravity sequences is illustrated
in Fig.5. To the M2 sequence corresponds a wave sequence starting with
the KK-wave and to the M5 sequence a monopole sequence staring with
the (exotic) KK6-monopole. Note that there is a duplication in each
sequence of states with the same level $3n$ for $n>0$. We shall show
that this duplication is spurious in the sense that the two states are
related by a switch of coordinates.  

\begin{figure}[h]
\centering
   \includegraphics[width=15cm]{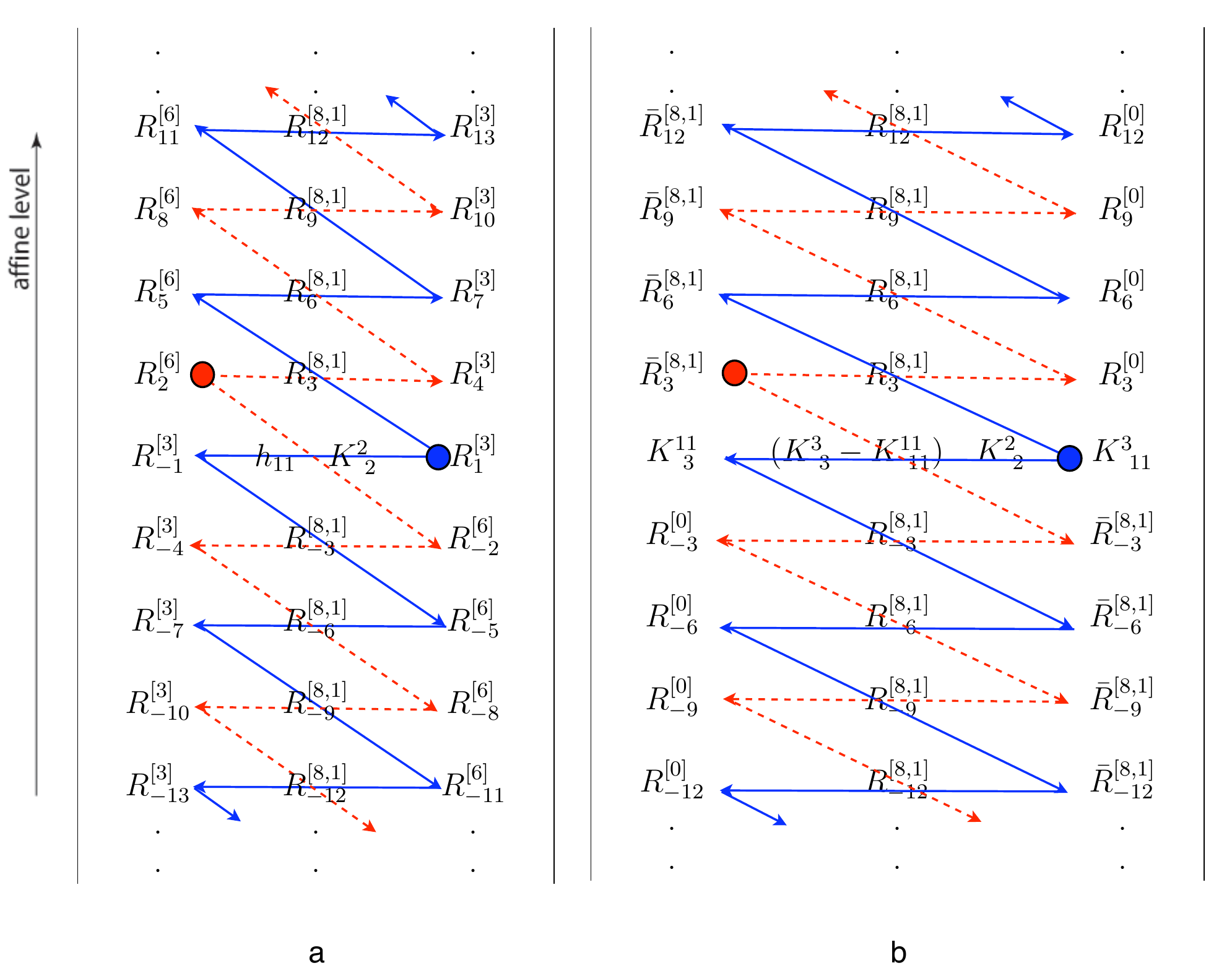}
 \caption {\sl \small Mapping of the brane $A_1^+$ group (Fig.5a) to the
 gravity $A_1^+$ group (Fig.5b). M2 and wave sequences are depicted by
 solid lines, M5 and monopole sequences by dashed lines. In Fig.5a
 (Fig.5b) horizontal lines represent Weyl reflexions by
 $W_{\alpha_{11}}$ ($W_\lambda$), diagonal lines by
 $W_{-\alpha_{11}+\delta}$ ($W_{-\lambda+\delta}$).} 
\end{figure}

From the correspondence we immediately get from the representatives of
the M2 sequence (\ref{borel3n}), (\ref{borel6n}), and of the M5
sequences (\ref{6Nborel}) and (\ref{3Nborel}), the representatives of
the KK-wave sequence  Eqs.(\ref{KW1}), (\ref{KW2}) and of the
KK-monopole sequence Eqs.(\ref{KM1}), (\ref{KM2}) in terms of the
$R^{[0]}_{3n}$ and $\bar R^{[8,1]}_{3n'}$ generators 
\begin{eqnarray}
\label{KW1}
{\cal V}_{6n}&= &\exp \left[\frac{1}{2} \ln H \,
  (K^3{}_3-K^{11}{}_{11} -2n K^2{}_2)\right]\, \exp
\left[\frac{1}{H}\, R^{[0]}_{6n}\right]\qquad n\ge0 \\ 
\label{KW2}
{\cal V}_{6n'}&=& \exp \left[\frac{1}{2} \ln H\,
  (-K^3{}_3+K^{11}{}_{11} -2n' K^2{}_2)\right] \, \exp
\left[\frac{1}{H}\, \bar R^{[8,1]}_{6n'}\right]\qquad n'>0\\ 
\label{KM1}
{\cal V}_{3+6n'}&=& \exp \left[\frac{1}{2} \ln H \,
  (-K^3{}_3+K^{11}{}_{11} -(2n'+1) K^2{}_2)\right] \, \exp
\left[\frac{1}{H}\, \bar R^{[8,1]}_{6n'+3}\right]\ n'\ge0\\ 
\label{KM2}
{\cal V}_{-3+6n}&=& \exp \left[\frac{1}{2} \ln H\,
  (K^3{}_3-K^{11}{}_{11} -(2n-1) K^2{}_2)\right]\, \exp
\left[\frac{1}{H}\, R^{[0]}_{6n-3}\right] \quad n>0\ \, . 
\end{eqnarray}
In these equations we distinguish the representatives of the [0]-tower
depicted in the right column of Fig.5b from those of the [8,1]-tower
depicted in the left column by labelling the former by $n$ and the
latter by $n'$. 

To get the representatives for the wave sequence in terms of the
gravitational potential $A_{3}^{(11)}$ given by Eq.(\ref{potential0}),
we apply the mapping Eqs.(\ref{bragramap}), (\ref{bragra1}) and
(\ref{bragra2}) to the representative of the M2 sequence in terms of
the supergravity 3-form potential\footnote{We have added an
  integration constant $-1$ to the field $A_3^{(11)}$ as in the
  discussion below Eq.(\ref{potential}).}, Eqs.(\ref{seqM21}) and
(\ref{seqM22}) 
\begin{eqnarray}
\label{seqG01}
{\cal V}_{6n}&=&\exp \left[-\frac{1}{2}\ln ({\cal F}_{2n-1} \bar{\cal
 F}_{2n-1})\, K^2_{\ 2}\right] \, \exp \left[\frac{1}{2} \ln {\cal
 R}e\, {\cal E}_{2n +1}\, (K^3{}_3-K^{11}{}_{11})\right] \nonumber \\ 
 && \exp \left[(-1)^{n} \bigg(\frac{1}{{\cal R}e\, {\cal
 E}_{2n+1}}-1\bigg)\, K^{3}_{\ 11}\right] \qquad n\geq 0\\ 
\nonumber\\
\label{seqG02}
{\cal V}_{6n^\prime}&=&\exp \left[-\frac{1}{2}\ln ({\cal F}_{2n'-1}
  \bar{\cal F}_{2n'-1})\, K^2_{\ 2}\right] \, \exp [\frac{1}{2} \ln
  {\cal R}e\, {\cal E}_{2n^\prime}\,(K^3{}_3-K^{11}{}_{11})] 
\nonumber\\  
&& \exp \left[ (-1)^{n^\prime+1} \bigg( \frac{1}{{\cal R}e\, {\cal
  E}_{2n^\prime}} -1 \bigg)\, K^{3}_{\ 11}\right] \qquad n^\prime>0 
\end{eqnarray}
where in Eq.(\ref{seqG01}) one has the signatures $(1,10,+)$ with time
in 3 for $n$ even and in 11 for $n$ odd, and in Eq.(\ref{seqG02}) the
signatures are  $(1,10,+)$ with time in 3 for $n^{\prime}$ odd and in
11 for $n^\prime$ even (see Appendix~\ref{appsg}). 

It is proven in Appendix~\ref{apprg} that the KK-wave sequence
contains a redundancy of the solutions for $n>0$, namely  
Eq.(\ref{seqG01}) and Eq.(\ref{seqG02})  lead to identical metric up
to interchange of the time coordinates 3 and 11. The full wave
sequence for $n>0$ has metric: 
\begin{eqnarray}
\label{KKW}
ds^2_{6 n^\prime}&=& {\cal F}_{2n'-1}\bar{\cal F}_{2n'-1} \Big
 [(dx^1)^2+(dx^2)^2 \Big ] + (-1)^{n^\prime}
 H^{-1}_{2n^\prime}(dx^3)^2+\Big [(dx^4)^2\dots+(dx^{10})^2 \Big ]
 \nonumber\\ 
 &+& (-1)^{n^\prime +1} H_{2n^\prime} \Big [ dx^{11} -  \Big(
 (-1)^{n^\prime +1}   H^{-1}_{2n^\prime}+ (-1)^{n^\prime} \Big)dx^3
 \Big]^2  \, , 
 \end{eqnarray}
where $H_p={\cal R}e\, {\cal E}_p$. For $n=0$ it is given by
Eq.(\ref{G0metric1}). All metrics in the KK-wave are solutions of 11D
supergravity.  The factor ${\cal F}_{2n'-1}\bar{\cal F}_{2n'-1}$ can
again be eliminated by a (singular) coordinate change, preserving the
harmonic character of $H_p$.  

From the representatives of the M5 sequence in terms of the
supergravity 3-form potential, Eqs.(\ref{seqM51}) and (\ref{seqM52}),
we get the representatives for the monopole sequence in terms of the
gravitational potential $A_{3}^{~(11)}$ 
\begin{eqnarray}
\label{seqG31}
{\cal V}_{3+6n'}&=&\exp \left[-\frac{1}{2}\ln ({\cal F}_{2n'-1}
  \bar{\cal F}_{2n'-1})\, K^2_{\ 2}\right] \, \exp \left[-\frac{1}{2}
  \ln {\cal R}e\, {\cal E}_{2n'+1}\,
  (K^3{}_3-K^{11}{}_{11}+K^2_2)\right]\nonumber\\ 
&&  \exp \left[ (-1)^{n'} {\cal I}m\, {\cal E}_{2n'+1}\,  K^{3}_{\
  11}\right]\qquad n'\geq0\\ 
\nonumber\\
\label{seqG32}
{\cal V}_{-3+6n}&=&\exp \left[-\frac{1}{2}\ln ( {\cal F}_{2n-1}
  \bar{\cal F}_{2n-1})\, K^2_{\ 2}\right] \, \exp \left[-\frac{1}{2}
  \ln {\cal R}e\, {\cal E}_{2n}\,
  (K^3{}_3-K^{11}{}_{11}+K^2_2)\right]\nonumber\\  && \exp
\left[(-1)^{n+1 } {\cal I}m\, {\cal E}_{2n}\,  K^{3}_{\ 11}\right] 
\qquad n>0
\end{eqnarray}
where in Eq.(\ref{seqG31}) one has  the signatures  $(2,9, -)$  with
time in 9,10 for $n'$ even and $(5,6,+)$ with time in 4,5,6,7,8 for
$n'$ odd, and in Eq.(\ref{seqG32}) the signatures are    $(2,9, -)$
with time in 9,10 for $n$ odd and $(5,6,+)$ with time in 4,5,6,7,8 for
$n$ even (see Appendix~\ref{appsg}). 

In analogy with the KK-wave sequence,  the metric in
Eqs.(\ref{seqG31}) and (\ref{seqG32}) are equivalent up to a
redefinition of the time coordinates (see Appendix~\ref{apprg}).
There is thus only one gravity tower, the left and the right tower of
Fig.5b are equivalent, each of them contains the full wave and
monopole sequences. 

The full monopole sequence has the metric: 
\begin{eqnarray}
ds^2_{3+ 6n'} &=&\!  \!  {\cal F}_{2n'-1}\bar{\cal F}_{2n'-1}
 H_{2n'+1}\Big [(dx^1)^2+(dx^2)^2 \Big ] + H_{2n'+1}(dx^3)^2+(-1)^{n'}
 \Big [(dx^4)^2\dots+(dx^{8})^2 \Big ]  \nonumber \\ \label{mono} 
 &+&\!  \! \!(-1)^{n'+1}  \Big [(dx^9)^2+(dx^{10})^2 \Big ]   +
 H^{-1}_{2n'+1} \Big [ dx^{11} -  \Big( (-1)^{n'}   B_{2n'+1}\Big)dx^3
 \Big]^2 \, ,  
 \end{eqnarray}
 where $B_p = {\cal I}m\, {\cal E}_p$.\\
 Again the metric of the monopole sequence solve the Einstein equations.
 
 The generators $R_{1+3p}^{[3]}$, $R_{2+3p}^{[6]}$, $\bar
 R_{3+3p}^{[8,1]}$, $p\ge 0$, and $K^3{}_{11}$ span the M2, M5 and
 gravity towers for positive real roots and define distinct BPS
 solutions. All positive real roots of $E_9$ can be reached from these
 by permuting coordinate indices in $A_8$ or equivalently by
 performing Weyl transformations $W_{\alpha_i}$ from the  gravity line
 depicted in Fig.1 with nodes 1 and 2 deleted. In this way we reach
 all $E_9$ positive real roots and the related BPS solutions. In what
 follows we shall keep the above notation for all  towers of positive
 real roots differing by $A_8$ indices, and specify the coordinates
 when needed.

\subsection{Analytic structure of BPS solutions and the Ernst
    potential} 
\label{analyticsec}

We have obtained an infinite U-duality multiplet of $E_9$ BPS solutions of
11D supergravity depending on two non-compact space variables. This was
achieved by analysing various $A_1^+ \equiv A_1^{(1)}$
subalgebras of $E_9$, which allow us to reach all positive roots within such a
subalgebra from sequences of Weyl reflexions starting from basic BPS
solutions reviewed in Section 2. A striking feature of the method is that
each solution is determined by a pair of conjugate harmonic functions $H_p$
and $B_p$ which can be combined into an analytic function ${\cal E}_p
=H_p+iB_p$, where $p$ characterises the level of the solution. This feature
emerges from the action of the affine $A_1^+$ subgroup on the
representatives and is clearly not restricted to supergravity. In this
section, we establish the link with another
$A_1^+$ subgroup of $E_9$, namely the Geroch group of general relativity. As
is well known  \cite{Geroch:1970nt, Geroch:1972yt, Breitenlohner:1986um},
the latter acts on stationary axisymmetric (or colliding plane wave)
solutions in four space-time dimensions (which can be embedded consistently
into 11D supergravity) via `non-closing dualities'
generating infinite towers of higher order dual potentials. Here
we explain the action of the Geroch group on BPS solutions, for which
the so-called {\em Ernst potential} (see (\ref{Ernst}) below)
is an analytic function, and hence is entirely analogous to the function
${\cal E}$ encountered above. As we will see this action
(so far not exhibited in the literature to the best of our knowledge)
`interpolates' between free field dualities and the full non-linear
action of the Geroch group on non-analytic Ernst potentials --- exactly
as for the M2-M5 sequence discussed in section~\ref{m2m5sec}. To keep
the discussion simple we will restrict attention to four-dimensional
Einstein gravity with two commuting Killing vectors, that is,
depending only on two (spacelike) coordinates.

Before we specialise to the case of BPS solutions we present the more
general formalism. The general line element in this case is of the
form 
\be\label{linelem}
ds^2 = \D^{-1} e^{2\s} (dx^2+dy^2) + (-\r^2\D^{-1} +\D \mmp^2) dt^2
+2\D \mmp dt\,dz +\D dz^2.
\ee
Here, $\p_t$ and $\p_z$ are Killing vectors, hence the metric coefficients
depend only on the space coordinates $(x,y)\equiv (x^1,x^2)$.
Furthermore, we have adopted a conformal frame for the $(x,y)$ 
components of the metric, with conformal factor $e^{2\s}$. $\mmp$ is 
the called the Matzner--Misner potential and related to the Ehlers
potential $\ep$ through the duality relation
\be\label{dualrel}
\eps_{ij}\p_j \ep = \r^{-1}\D^2 \p_i \mmp,
\ee
where $i,j=1,2$. There is no need to raise or lower indices, as the
metric in $(x,y)$ space is the flat Euclidean metric, with
$\eps_{12}=\eps^{12}=1$. Therefore the inverse duality relation 
is $\eps_{ij}\p_j \mmp = - \r\D^{-2}\p_i \ep$.

The vacuum Einstein equations for the line element Eq.(\ref{linelem})
in terms of the Matzner--Misner potential $\mmp$ read
\be\label{eommm}
\D\p_i (\r\p_i\D) &=& \r \left(\p_i\D \p_i\D -\r^{-2}\D^4\p_i
  \mmp\p_i \mmp\right)\nn\\
\r\D^{-1}\p_i (\r \p_i \mmp) &=& 2 \r \p_i \left(\frac\r\D\right) \p_i
\mmp \, , 
\ee
Rewritten in terms of the Ehlers potential $\ep$ these give, using
Eq.(\ref{dualrel}),
\be\label{eome}
\D \p_i(\r\p_i \D) &=& \r \left(\p_i\D\p_i\D - \p_i\ep \p_i\ep \right)\nn\\
\D\p_i(\r \p_i \ep) &=& 2 \r \p_i\D \p_i\ep \, ,
\ee
where the two sets of equations Eqs.(\ref{eommm}) and (\ref{eome}) are related
by the so-called Kramer--Neugebauer transformation $\ep\leftrightarrow \mmp,
\D \leftrightarrow \r/\D$. In addition, there are equations for $\r$ 
and the conformal factor $\s$. These are two (compatible) first order
equations for the conformal factor  
\be\label{constraint}
\r^{-1}\p_{(x} \r \p_{y)}\s &=& \frac14 (\D^{-1}\p_x\D)(\D^{-1}\p_y\D)
  +\frac14 (\D^{-1}\p_x \ep)(\D^{-1}\p_y \ep),\nn\\
\r^{-1}\p_x \r \p_x\s - \r^{-1}\p_y \r \p_y\s &=&
       + \frac14 (\D^{-1}\p_x\D)^2 + \frac14 (\D^{-1}\p_x \ep)^2  \nn\\
       && -\frac14 (\D^{-1}\p_y\D)^2 -  \frac14 (\D^{-1}\p_y \ep)^2 \, ,
\ee
while $\r$ satisfies the two-dimensional Laplace equation without source
\be
\p_i\p_i\r =0\, .
\ee
A second order equation for $\s$ can be deduced by varying $\r$, or
alternatively from the constraints and the dynamical equations for the 
metric Eq.(\ref{eommm}) [or Eq.(\ref{eome})]; it reads
\be\label{conf}
\p_i\p_i \s = -\frac14\r\D^{-2} \big( \p_i\D\p_i\D + \p_i \ep \p_i \ep \big),
\ee
If $\r$ is different from a constant (as is the case generally with
axisymmetric stationary or colliding plane wave solutions), we can 
integrate the first order Eqs.(\ref{constraint}), which determine
the conformal factor up to one integration constant; the second order
equation Eq.(\ref{conf}) is then automatically satisfied as a consequence 
of the other equations of motion. On the other hand, as we 
will see below, the BPS solutions are characterized by $\r = \rm constant$, 
for which the l.h.s. of Eq.(\ref{constraint}) vanishes identically (whence 
the r.h.s. must also vanish identically). In this case, we are left with
the second order Eq.(\ref{conf}), and the conformal factor is only
determined modulo a harmonic function in $(x,y)$. 

The equations of motion Eq.(\ref{eome}) can be rewritten conveniently
in terms of the {\em complex Ernst potential} [cf .Eq.(\ref{analytic})]
\be\label{Ernst}
\cE = \D + i \ep,
\ee
satisfying the Ernst equation
\be\label{ernsteq}
\D \p_i (\r \p_i \cE) = \r \p_i\cE\p_i\cE.
\ee
As we will see below this equation is trivially satisfied for BPS
solutions in the sense that both sides vanish identically.

\subsubsection{BPS solutions}

In our analysis of the 11-dimensional gravity tower, the lowest level
BPS solution is the KK-wave   Eq.(\ref{G0metric1}). It stems from the
generator $K^3_{~11}$ with time in 3. In 4D gravity with $x^1,x^2$ as
non compact space variables, the corresponding wave solution is
associated to the Chevalley generator $K^3_{~4}$ depicted  in Fig.6 by
the node 3. Taking the timelike direction to be 3, we get
[$(x,y)\equiv (x^1,x^2)$, $(t,z) \equiv (x^3,x^4)$] 
\be
\label{kksol2d}
ds^2 =  dx^2 + dy^2 + (\D-2) dt^2 - 2 (1-\D) dt\,dz + \D dz^2.
\ee
Here, $\D=\D(x,y)$ is a {\em harmonic function} in $x,y$, which, for the
brane with a  source at $x=y=0$ we choose to be $\D=\frac12 
\ln (x^2+y^2) = \ln |\z|$ in terms of the complex coordinate
\be
\z = x + i y.
\ee 
Comparing Eq.(\ref{kksol2d}) with Eq.(\ref{linelem}),  
we see that for this  BPS solution the general fields $\mmp,\s$
and $\r$ are expressed 
in terms of $\D$ as\footnote{The constant $b$ for $\mmp$ should be chosen
as $1$ in order to obtain an asymptotically flat solution in more
than four space-time dimensions. From the point of view of the
two-dimensional reduction,
however, it does not matter and can be chosen arbitrarily. Note also,
that constant shifts of $\mmp$ are part of the Matzner-Misner $SL(2)$,
see below.}
\be\label{kksol2da}
 e^{2\s}=\D,\quad\quad \mmp= b - \D^{-1},\quad\quad
\r=1.
\ee
Using the duality relation Eq.(\ref{dualrel}) one obtains the Ehlers
potential $\ep$ up to an integration constant. Indeed, as already
mentioned above, with Eq.(\ref{kksol2da}), 
the duality relations Eq.(\ref{dualrel}) just become the  Cauchy--Riemann 
equations for the Ernst potential (\ref{Ernst}), to wit
\be
\p_x \ep = -\p_y \D \quad , \quad \p_y \ep = \p_x \D \, ,
\ee
or, in short notation, $\eps_{ij}\p_j\D = - \p_i \ep$. Conversely,
for Eq.(\ref{dualrel}) to reduce to the Cauchy--Riemann relations, we must
have $\mmp= 1/\D + {\rm constant}$ and $\r$ constant. {\it Therefore, the
Cauchy--Riemann equations for the Ernst potential are equivalent to
the BPS (`no force')  
condition and may thus be taken as the defining equations for BPS 
solutions}. In a supersymmetric context these (first order) equations
would be equivalent to the Killing spinor conditions defining the
BPS solution.

For $\D =  \frac12 \ln (x^2 + y^2)$, we immediately obtain
\be
\ep = \arctan\left(\frac{y}{x}\right) + {\rm constant} = \arg(\z) + {\rm constant} \, ,
\ee
whence the Ernst potential is simply
\be
\cE(\z) = \ln |\z| + i \arg(\z) + {\rm constant} = \ln \z + {\rm constant} \, ,
\ee
and so is an {\em analytic} function of $\z$. It is then easy
to see that the equations of motion and the constraint equations are 
satisfied for {\em any} analytic Ernst potential $\cE$ if $\r$ is
constant (and in particular, with $\r=1$). Namely both the Ernst equation 
Eq.(\ref{ernsteq}) as well as Eq.(\ref{constraint}) reduce to the identity 
$0=0$ for all such solutions. Because Eq.(\ref{constraint}) is void, the 
conformal factor must then be determined from the second order equation 
Eq.(\ref{conf}). For holomorphic $\cE$, Eq.(\ref{conf}) can be
rewritten as
\be\label{conf1}
\p_\z \p_{\bar{\z}} \s = -\frac12\r\,
\frac{\p_\z \cE \p_{\bar{\z}} \bar{\cE}}{(\cE + \bar{\cE})^2}
\ee
and only in this case the solution to this equation can be given in
closed form. It reads 
\be
\s(\z,\bar{\z}) = \frac12\ln \, (\cE + \bar{\cE})
\ee
The ambiguity involving harmonic functions left by Eq.(\ref{conf})
is related to the covariance of the equations of motion under 
{\it conformal analytic coordinate transformations} of the complex 
coordinate $\z=x+iy$, which leave the 2-metric in diagonal form, viz.
\be
\z \longrightarrow \z'=f(\z) \;\; .
\ee
As is well known, the conformal factor transforms as
\be
\s(\z,\bar\z) \longrightarrow \s (\z,\bar\z) 
    + \frac12 \ln \big| f(\z)\big|^2 \, ,
\ee
under such transformations, where the second term on the r.h.s. is
indeed harmonic. We already used this fact  when we removed the
conformal factors $ {\cal F}_{2n-1}\bar{\cal F}_{2n-1}$  to prove that
the metric in Eqs.(\ref{M2odd}), (\ref{M2even}), (\ref{M5odd}),
(\ref{M5even}),  (\ref{KKW}) and (\ref{mono}) solve the Einstein
equations outside the singularities of the harmonic functions. 

\subsubsection{Action of Geroch group}

The Geroch group for $(3+1)$-dimensional gravity in
the stationary axi-symmetric case discussed here is affine 
$\widehat{SL(2)}$ with central extension ($\equiv A_1^+$); this is the same
structure 
we encountered for the M2 and M5 towers. It is depicted in Fig.6 by
the Dynkin diagram formed by the nodes 3 and 4. Extending the diagram
with node 2 to the overextended $A_1^{++}$, we may as previously
identify the central charge with a Cartan generator of $A_1^{++}$
($-K^2_{~2}$ in $E_{10}\equiv E_8^{++}$). Adding the node 1 leads to
the very extended $A_1^{+++}$ which is the pure gravity counterpart
(in $ D=4$) of $E_{11}$. 

\begin{figure}[h]
\centering
 \includegraphics [width=4cm]{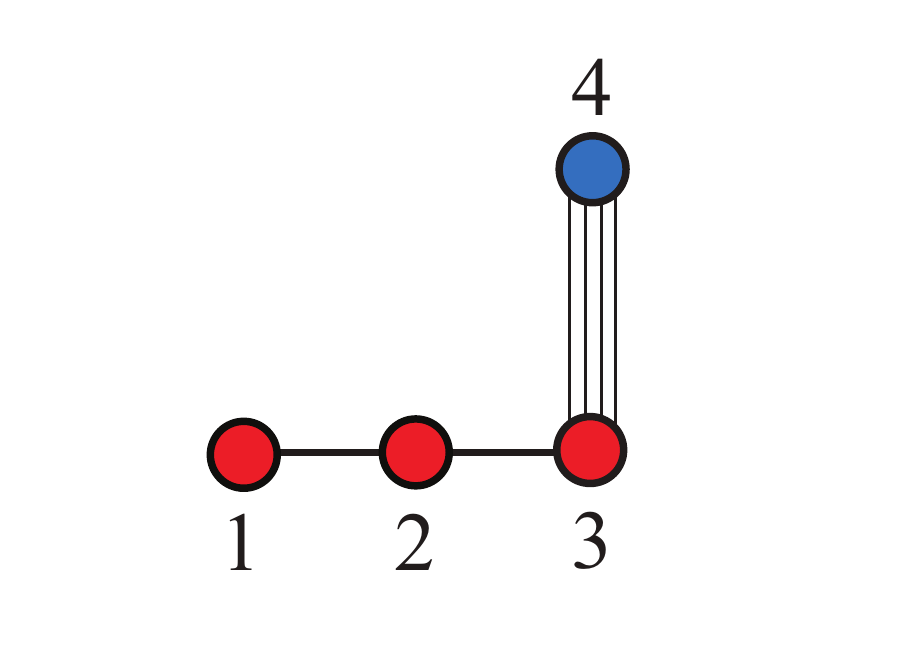}
\caption{\sl \small Dynkin diagram of $ A_1^+$ and
  its `horizontal' extensions $A_1^{++}$ and $A_1^{+++}$. The
  Matzer--Misner $SL(2)$ is represented by the node 3  and the Ehlers
  $SL(2)$ by the node 4.}  
\end{figure}

The two distinguished $SL(2)$ subgroups of $A_1^+$
corresponding to the Matzner--Misner and the Ehlers cosets appear
in different real forms. The timelike Killing vector
$\p_t$ turns the Matzner--Misner coset into $SL(2)/SO(1,1)$ (with
non-compact denominator group), whereas the Ehlers coset is
$SL(2)/SO(2)$ (with compact denominator group).
With 3 or 4 as time direction,  the temporal involution Eq.(\ref{map})
leaves indeed invariant  the Lorentz generator $\Omega\, (
K^3{}_4+K^4{}_3)=K^3{}_4+K^4{}_3$ in the Matzer--Misner $ SL(2)$, but
preserves the rotation generator 
 $\Omega \,(R^{4 \vert 4}-R_{4 \vert 4})=R^{4 \vert 4}-R_{4 \vert 4}$
in the Ehlers  $ SL(2)$. Here $R^{4 \vert 4}$  is  the simple
positive step operator corresponding  in four dimensions to the
11-dimensional generator   $R^{4\,5\,6\,7\,8\,9\,10\,11\vert \,11}$ of
Section 3.3.

For the underlying split real algebra $A_1^+$ we use a simple
Chevalley--Serre basis consisting of $e_i,f_i, h_i$ ($i=3,4$) and
derivation $d$ (see also Appendix~\ref{affapp}). The central element 
is $c=h_3+h_4$. The index `4' refers to the Ehlers
$SL(2)$ and the index `3' refers to the Matzner--Misner
$SL(2)$, as depicted in Fig.6.

\medskip
\noindent
{\bf $\bullet$ Ehlers group}

The Ehlers $SL(2)$ acts by M\"obius transformations on the
(analytic) Ernst
potential. Using the standard notation for M\"obius
transformation generators\footnote{Note that these are M\"obius
  transformations on $\cE$ and {\em not} on the complex coordinate $\z$.
  The system admits an additional
  {\em conformal symmetry} acting on the complex coordinate $\z$
  \cite{Julia:1996nu}, which shows again that {\em any} analytic Ernst
  potential solves the equations of motion.}
  \be
L_{-1} = - \frac{\p}{\p\cE},\quad\quad L_0 = -\cE \frac{\p}{\p\cE},
\quad\quad L_{1} = - \cE^2 \frac{\p}{\p\cE} \, ,
\ee
the Ehlers generators are\footnote{The factors $i$ are understood
from studying the invariant bilinear form for $(L_{-1},L_0,L_1)$
which is non-standard from the Kac--Moody point of view.}
\be
e_4 = i L_{-1},\quad\quad h_4 = 2 L_0,\quad\quad f_4 = i L_1 \, ,
\ee
with the resulting transformation of the real and imaginary
components of $\cE$
\begin{align}
f_4 \D &= 2 \D \ep,&\quad\quad f_4 \ep = & \ep^2-\D^2,&\nn\\
h_4 \D &= -2 \D,&\quad\quad  h_4 \ep = & -2 \ep,&\nn\\
e_4 \D &= 0,&\quad\quad e_4 \ep =& -1.&
\end{align}

\noindent
{\bf $\bullet$ Matzner--Misner group}

The infinitesimal action of the Matzner--Misner group on general
$SL(2)/SO(1,1)$ coset fields $\D$ and $\mmp $ is (for $\r=1$)
\begin{align}\label{mmtrm}
f_3 \D &= -2 \D \mmp,&\quad\quad f_3 \mmp = & \mmp^2+\D^{-2},&\nn\\
h_3 \D &= 2 \D,&\quad\quad  h_3 \mmp = & -2 \mmp,&\nn\\
e_3 \D &= 0,&\quad\quad e_3 \mmp =& -1.&
\end{align}

In order to compute its action on the Ernst potential $\cE$, we have to
exploit the duality relation Eq.(\ref{dualrel}) between the potentials
$\mmp$ and $\ep$. It is straightforward
to work out the action of $e_2$ and $h_2$, with the result
\be
h_3 \ep = 2 \ep,\quad\quad e_3 \ep =0,
\ee
where some constants have been fixed from the commutation
relations. Finally, the action of $f_2$ on $\ep$ follows from
\be
\eps_{ij}\p^j (f_3 \ep) =f_3 (\D^2 \p_i \mmp) =
   - 2 \D^2 \mmp \p_i\mmp + \D^2 \p_i (\D^{-2}).
\ee
Using  $\mmp = b - \D^{-1}$ and the Cauchy--Riemann equation, this
yields
\be
\eps_{ij}\p^j (f_3 \ep) = -2b \p_i \D = -2 b \eps_{ij}\p^j \ep.
\ee
Therefore we find
\be\label{f2B1}
f_3 \ep = -2b \ep,
\ee
setting an integration constant equal to zero.
We note that in order to satisfy $[e_3,f_3]=h_3$ on
$\ep$ we need to have a non-trivial action of $e_3$ on the
integration constant $b$, namely $e_3b =-1$ which is consistent
with the general shift property of the Matzner--Misner group
Eq.(\ref{mmtrm}) and $\mmp=b-\D^{-1}$ for this solution. In a sense,
one can view the Matzner--Misner group as acting via M\"obius transformations
on the variable $b$. However, closing this action with the Ehlers
M\"obius transformations on $\cE$ leads one to introduce new
constants (notably in $f_4 \mmp$) which transform non-trivially
under the remaining generators. For
completeness we note the transformation rules
\be
e_4 \mmp = 0,\quad\quad h_4 \mmp = 2 \mmp,
\ee
which are true generally and
\be
f_4 \mmp = 2 \ep \D^{-1} +\l \, , 
\ee
where $\l$ is an example of a new constant. This last relation is
true {\em on the solution} $\mmp=b-\D^{-1}$; generally the result would
be some non-local expression.

Combining Eq.(\ref{mmtrm}) and Eq.(\ref{f2B1}),
the action of $f_2$ on the full Ernst potential is
\be
\label{finact}
f_3 \cE =  -2b\cE - 2 \, .
\ee
This shows that the action of $f_3$ on $\cE$ does not yield a new
transformation, but simply a linear combination of previous ones
(to wit, $e_4$ and $h_4$). Hence, under the action of the Matzner--Misner
$SL(2)$, a BPS solution will remain a BPS solution.\footnote{The
  constant parameters $b,\l,\ldots$ do not influence the analyticity of
  the solution, although they are essential for the action of the
  Geroch group.} The formula  Eq.(\ref{finact}) agrees with our
findings in Eq.(\ref{Moebact}).

This almost `trivial' action of the Geroch group on the BPS solutions --
which essentially acts only via M\"obius transformations on the Ernst 
potential $\cE$ -- confirms our previous finding for M2 and M5 branes
(\ref{Moebact}), but is in marked contrast to its action on non-BPS
stationary axisymmetric 
solutions \cite{Breitenlohner:1986um}. There $\cE(x,y)$ is {\em not}
analytic, and $\r(x,y)$ is a non-constant function, often identified
with a radial coordinate (so-called Weyl canonical coordinates).
When starting from the vacuum solution to obtain say, the Schwarzschild
or Kerr solution, the $(x,y)$ dependence of the Ernst potential is
precisely the one induced by the $(x,y)$ dependence of the spectral
parameter whose coordinate dependence, in turn, hinges on the coordinate
dependence of $\r$. Since we have $\r=1$ for BPS solutions, this
mechanism does not work, confirming our conclusion that the action
of the Geroch group cannot turn an analytic Ernst potential into
a non-analytic one, hence leaves the class of BPS solutions stable.
 
The results of this section can be summarised by saying that the Weyl
group of $A_1^+$ acts via shifts and inversions on the complex Ernst
potential and at the same time transforms the conformal factor but
leaves invariant the set of analytic Ernst potentials.

\setcounter{equation}{0}

\section{Dual formulation of the $E_9$ multiplet}

\subsection{Effective actions}

We showed in Section 2.2 that the basic magnetic BPS solutions of 11D
supergravity (M5 and KK6-monopole) smeared in all directions but one
are expressible in terms of the dual potentials $A_{3\,4\,
  5\,6\,7\,8}$ and  $A_{4\,5\,6\,7\,8\,9\,10\,11\vert 11}$
parametrising the Borel generators $R^{[6]}_2$ and $ \bar
R^{[8,1]}_3$. In higher non-compact transverse space dimensions these
potentials are related by Hodge duality to the supergravity fields
$A_{9\,10\,11}$ and $A_{i}^{~(11)}$. The dual potentials take on the
solutions, up to an integration constant, the same value $1/H$ as do
the fields $A_{9\,10\,11}$  and $A_3^{~(11)}$ for the basic electric BPS
solutions, namely the M2-branes at level 1 and the KK-waves at level
0. $H$ is, in any number of non-compact transverse spacelike
directions, a harmonic function with $\delta$-function singularities
at the location of the sources.  

In Section 3 we  constructed  BPS solutions of 11D supergravity in two
transverse spacelike directions for all $E_9$ positive real roots. We
shall label such description of the BPS states in terms of the
supergravity metric and 3-form the `direct' description.  
Each solution was obtained by relating through dualities and
compensations the `generalised dual potential' $1/H$ parametrising an
$E_9\subset E_{10}$ positive root in the Borel representative of
$E_{10}$ to the supergravity metric and 3-form. 

In this Section, we will present a space-time description of the BPS
states directly in terms of the generalised dual potentials. We 
label it  the `dual' description. We shall show that the dual
description of the BPS solutions can be derived from gauge fixed
effective actions 
\begin{equation}
\label{gendual}
{\cal S}^{(11)}_{\{q\}} =\frac{1}{16\pi G_{11}}\,\int d^{11}x
\sqrt{\vert g\vert}\left(R^{(11)}- \epsilon{1\over 2   
} F_{i\,\{q\}} F^{i\,\{q\}}\right)\, ,
\end{equation}
where $i$ runs over the two non-compact dimensions $1,2$ and
$\epsilon$ is $+1$ if the action involves a single time coordinate (or
an odd number of time coordinates) and $-1$ if the number of time
coordinates is even. $F_{i\,\{q\}}=\partial_i A_{\{q\}}$ where $\{q\}$
stands for the tensor indices of the $A_p^{[N]}$ potential
multiplying $R_p^{[N]}$ in the Borel representative. Here $p$ is the
level and $[N]$  labels a tower $[3]$, $[6]$, $[0]$ or $[8,1]$ for any
set of $A_8\subset E_9$ tensor indices.  The set of indices is fixed
by the  $A_1^+$ group selected by a Weyl transformed in $A_8$ of the
$A_1^+$ subgroup of $E_{10}$ chosen in  Eq.(\ref{geroch}).

We first consider the M2-M5 system of Section 3.2. Explicitly
$A_{1+3n}^{[3]}= A_{9\,10\,11\, ,\, [3\,4\,\dots \,10\,11].n}$ for the
3-tower depicted in the right column of Fig.2a  and $A_{2+3n}^{[6]}=
A_{3\,4\,5\,6\,7\,8\, ,\, [3\,4\,\dots \,10\,11].n}$  for the 6-tower
depicted in the left column. Here the symbol $n$  is the number of times the
antisymmetric set of indices  $[3\,4\,\dots \,10\,11]$ must be taken.
That this is the correct tensor structure follows from the structure
of the `gradient representations'
in~\cite{Damour:2002cu,Nicolai:2003fw}.

For all BPS states in the M2-M5 system, we take 
\begin{equation}
\label{Hfield}
A_p^{[N]}=1/H\, .
\end{equation}
The metric associated to $A_p^{[N]}$ is encoded in the Borel
representatives of the M2 sequence Eqs.(\ref{borel3n}),
(\ref{borel6n}) and the M5 sequence (\ref{6Nborel}),
(\ref{3Nborel}). We combine Eqs.(\ref{borel3n}) and (\ref{3Nborel}) to
form the 3-tower Eq.(\ref{tower1}) and Eqs.(\ref{borel6n}) and
(\ref{6Nborel}) to form the 6-tower Eq.(\ref{tower3}) . We have 
\begin{eqnarray}
\label{bor3}
{\cal V}_{1+3n}&=& \exp \left[\frac{1}{2} \ln H \, (h_{11}-n
  K^2{}_2)\right]\, \exp \left[\frac{1}{H}\,
  R^{[3]}_{1+3n}\right]\qquad n\ge 0 \\ 
\label{bor6}
{\cal V}_{2+3n}&= &\exp \left[\frac{1}{2} \ln H\, (-h_{11}-(n+1)
  K^2{}_2)\right] \, \exp \left[\frac{1}{H}\, R^{[6]}_{2+3n}\right]
\qquad n\ge 0 \, , 
\end{eqnarray}
which yield for the 3-tower Eq.(\ref{bor3}) the metric
Eq.(\ref{metric3}) and for the 6-tower Eq.(\ref{bor6}) the metric
Eq.(\ref{metric6})  
\begin{eqnarray}
 \label{metric3}
&&\vert g_{11}\vert=\vert g_{22}\vert=H^{1/3 +n}\nn\\ 
&&\vert g_{33}\vert =\vert g_{44}\vert =\dots =\vert g_{88}\vert
 =H^{1/3}\qquad \vert g_{99}\vert =\vert g_{10\,10}\vert =\vert
 g_{11\,11}\vert = H^{-2/3}\, ,\\ 
\nonumber \\
\label{metric6}
&&\vert g_{11}\vert=\vert g_{22}\vert=H^{2/3 +n}\nn\\
 &&\vert g_{33}\vert =\vert g_{44}\vert =\dots =\vert g_{88}\vert
 =H^{-1/3}\qquad \vert g_{99}\vert =\vert g_{10\,10}\vert =\vert
 g_{11\,11}\vert = H^{2/3}\, . 
\end{eqnarray}
For the time components of the metric one multiplies the absolute
values of the metric components by a minus sign. The time components
for the 3- and 6-towers are specified in Section 3.2 for the Weyl
orbits initiated by the M2 with time in 9 and by the M5 with time in
3. Note that we could as well take the  Weyl orbit initiated by the M2
with times in 9, 10 and by the M5 with times in 4, 5, 6, 7,
8. Alternatively we could mix the two orbits to avoid for instance at
all level exotic solutions and have always time in 9 for the M2
sequence and in 3 for the M5 sequence depicted in Fig.2a. Note that in
all cases, climbing the 3-tower or the 6-tower by steps of one unit of
$n$ amounts to alternate between BPS states on the M2 sequence and the
M5 sequence. We shall comment on this feature in Section 6. 

We now verify that the matter term in Eq.(\ref{gendual}) solves the
Einstein equations with  $A_p^{[N]}=1/H$ and  the metric given in
Eqs.(\ref{metric3}) and (\ref{metric6}). For the 3-tower the matter
Lagrangian reads 
\begin{eqnarray}
\label{matter3}
{\cal L} &=&-\epsilon\frac{1}{2} \sqrt{\vert g\vert }  F_{i\,\{q\}}
F^{i\,\{q\}}\nn\\&=& 
-\epsilon \frac{1}{2}\sum_{i=1}^2  \sqrt{\vert g\vert}  g^{ii} g^{99}
g^{10\,10} g^{11\,11} [g^{33}g^{44}\dots g^{11\,11}]^n
\left(\partial_i A_{9\,10\,11\, ,\, [3\,4\,\dots
    \,10\,11].n}^{[3]}\right)^2\, , 
\end{eqnarray}
while for the 6-towers one gets
\begin{eqnarray}
\label{matter6}
{\cal L}
=-\epsilon\frac{1}{2}\sum_{i=1}^2  \sqrt{\vert g\vert }  g^{ii} g^{33}
g^{44} g^{55}g^{66}g^{77}g^{88} [g^{33}g^{44}\dots g^{11\,11}]^n
\left(\partial_i A_{3\,4\,5\,6\,7\,8\, ,\, [3\,4\,\dots
    \,10\,11].n}^{[6]}\right)^2\, . 
\end{eqnarray}
One computes from  Eqs.(\ref{matter3}) and (\ref{matter6}) the
energy-momentum tensors for the 3-tower 
\begin{eqnarray}
\label{tensor3}
&&T^1_1=-T^2_2= -\frac{1}{4} H^{-7/3 -n} \,\left[(\partial_1 H)^2
  -(\partial_2 H)^2\right]\,,\qquad T^1_2=T^2_1 = -\frac{1}{2} H^{-7/3
  -n} \,[\partial_1 H\partial_2 H]\nn\\&&T^3_3=T^4_4=\dots
T^8_8=-\frac{2n-1}{4} H^{-7/3 -n}\, [(\partial_1H)^2
  +(\partial_2H)^2]\\ 
&& T^9_9=T^{10}_{10}=T^{11}_{11}=-\frac{2n+1}{4} H^{-7/3 -n} \,
\left[(\partial_1H)^2 +(\partial_2H)^2\right]\nn\, , 
\end{eqnarray}
and for the 6-tower
\begin{eqnarray}
\label{tensor6}
&&T^1_1=-T^2_2=- \frac{1}{4} H^{-8/3 -n} \,\left[(\partial_1 H)^2
  -(\partial_2 H)^2\right]\,,\qquad T^1_2=T^2_1 = -\frac{1}{2} H^{-8/3 -n}
\,[\partial_1 H\partial_2 H]\nn\\
&&T^3_3=T^4_4=\dots T^8_8=-\frac{2n+1}{4} H^{-8/3 -n}\, 
  \left[(\partial_1H)^2  +(\partial_2H)^2\right]\\ 
&& T^9_9=T^{10}_{10}=T^{11}_{11}=-\frac{2n-1}{4} H^{-8/3 -n} \, 
  \left[(\partial_1H)^2 +(\partial_2H)^2\right]\nn\, .
\end{eqnarray}
The fact that $T^\mu_\nu$ in Eqs.(\ref{tensor3}) and (\ref{tensor6})
does not depend on $\epsilon$ results from the cancellation between
negative signs arising from the kinetic energy term in the action
Eq.(\ref{gendual})  and from the concomitant even numbers  of time
metric components. From the metric Eqs.(\ref{metric3}) and
(\ref{metric6}) one easily verifies that the Einstein equations 
\begin{equation}
\label{einstein}
R^\mu_\nu-\frac{1}{2}\delta^\mu_\nu R=T^\mu_\nu\, ,
\end{equation}
with $T^\mu_\nu$ given by Eqs.(\ref{tensor3}) and (\ref{tensor6}), are
satisfied. This result holds for  $\epsilon=\pm1$ because the left
hand side of Eq.(\ref{einstein}) turns out to be independent of
$\epsilon$. 

Using the mapping Eqs.(\ref{bragramap}), (\ref{bragra1}),
(\ref{bragra2}) of the brane $A_1^+$ group Eq.(\ref{geroch}) to the
gravity towers $A_1^+$ group, one maps the 3-tower to the 0-tower and
the 6-tower to the [8,1]-tower depicted respectively on the right and
left columns of Fig.5b.  
We combine Eqs.(\ref{KW2}) and (\ref{KM1}) to form the [8,1]-tower. One has
\begin{equation}
{\cal V}_{3+3n}= \exp \left[\frac{1}{2} \ln H \,
  (-K^3{}_3+K^{11}{}_{11} -(n+1) K^2{}_2)\right] \, \exp
\left[\frac{1}{H}\, \bar R^{[8,1]}_{3n+3}\right] \qquad n \ge 0\, . 
\end{equation}
The corresponding dual metric are
\begin{eqnarray}
 \label{metric0}
&&\vert g_{11}\vert=\vert g_{22}\vert=H^{n+1}\nn\\ 
&&\vert g_{33}\vert =H\qquad \vert g_{11\,11}\vert = H^{-1}\\
&&\vert g_{aa}\vert =1\qquad a\neq 1,2,3,11
\nonumber \, .
\end{eqnarray}
As expected, up to the interchange of the coordinates 3 and 11, the
same result holds for the redundant 0-tower (except for the level 0 of
$E_{10}$, which is the KK-wave). The verification of the Einstein
equations derived from the actions Eq.(\ref{gendual})  duplicates that
of the M2-M5 system. Note that in this generalized dual formulation,
the metric of the gravity tower are all diagonal. 

The above results for the 3- and 6-towers and for the gravity tower
have been established for a chosen set of tensor indices, namely the
set determined by the choice of the $A_1^+$ subgroup of $E_{10}$ for
which $R^{[3]}_1$ is identified with $R^{9\, 10 \, 11}$. The validity
of the effective action Eq.(\ref{gendual}) for all $E_9\subset E_{10}$
fields associated to its positive real roots follows from permuting
the $A_8$ tensor indices, that is from performing   Weyl
transformations of the gravity line.

\subsection{Charges and masses}
\label{chargesec}

The charge content of the $E_9$ BPS solutions is easier to analyse in
the dual description because the dual potential is not mixed with
fields arising from the compensations. {\em Outside the sources}\/,
the equation of motion for the dual field $A_{\{q\}}$ is from
Eq.(\ref{gendual}) 
\begin{equation}
\label{genfield}
\sum_{i=1}^2\partial_i \left(\sqrt{\vert g \vert}g^{\{q\}}\partial_i
A_{\{q\}}\right)=0 \, , 
\end{equation}
where 
\begin{eqnarray}
\label{lap3}
&& g^{\{q\}}=g^{ii}g^{99} g^{10\,10} g^{11\,11} [g^{33}g^{44}\dots
  g^{11\,11}]^n\qquad\qquad~~~\hbox {3-tower : level  (1+3$n$)}\\ 
\label{lap6}
&& g^{\{q\}}=g^{ii}g^{33} g^{44} g^{55}g^{66}g^{77}g^{88}
      [g^{33}g^{44}\dots g^{11\,11}]^n \qquad ~\,\hbox{ 6-tower :
	level  (2+3$n$)} \\ 
\label{lapKK}
&& g^{\{q\}}=g^{ii} g^{44} g^{55}\dots g^{10\,10}
(g^{11\,11})^2[g^{33}g^{44}\dots g^{11\,11}]^n \quad \hbox{[8,1]-tower
  : level (3+3$n$)}\, , 
\end{eqnarray}
with $n\ge 0$. The appearance of the $n$-fold blocks of antisymmetric
metric factors $g^{33}g^{44}\cdots g^{11\,11}$ is again due to the
embedding of $E_9$ in $E_{10}$, see~\cite{Kleinschmidt:2006dy}.  
From Eqs.(\ref{metric3}), (\ref{metric6}),
(\ref{metric0}), and from the embedding relation in $E_{11}$
Eq.(\ref{embedding}), we get  for all towers and hence, by permutation
of tensor indices in $A_8$, for all $E_9$ BPS states $\sqrt{\vert
  g\vert }g^{\{q\}} =\pm H^2$. As, up to an integration constant, one
has always $A_{\{q\}}=1/H$, the field equation Eq.(\ref{genfield})
reduces to 
\begin{equation}
\label{allH}
\sum_{i=1}^2\partial_i \partial_i H=0 \, .
\end{equation}
Here, as for the KK-monopole  discussed in Section 2.2,
Eq.(\ref{allH}) is valid outside the sources and the latter are
determined by fixing the singularities of the function $H$. Labelling
the positions of the smeared M2 by $x^1_k,x^2_k$ and their charges by
$q_k$, one takes  
\begin{equation}
\label{source2}
H(x^1,x^2)=\sum_k \frac{q_k}{2\pi}\ln  \sqrt{(x^1-x^1_k)^2+(x^2-x^2_k)^2}\, ,
\end{equation}
and,  in analogy with Eq.(\ref{h}), the extension of Eq.(\ref{allH})
including the sources  reads\footnote{We  fix the M5 charges by the
  Weyl transformation relating the M2 to the M5, which for convenience
  was not explicitly used in our general derivation of the $E_9$ BPS
  solutions in Section 3.} {\em for all $E_9$ BPS solutions} 
\begin{equation}
\label{fullH}
\sum_{i=1}^2\partial_i \partial_i H=\sum_k \frac{q_k}{2\pi}\delta
({\bf x - x_k})\, . 
\end{equation}
Thus we obtain, for all BPS states, the same charge value as for the
M2, as expected from  U-dualities viewed as $E_9$ Weyl reflexions.  
Our identification of $q_k$ with a charge is however not the
conventional one as long as our solutions with 2 non-compact space
dimensions have not been identified with static solutions in 2+1
space-time dimensions. This raises the question whether we are allowed
to decompactify the time. This will be examined in the following
Sections 4.2.1 and 4.2.2. 
We wish to stress that decompactification of time or space dimensions
is {\em not} the same as `unsmearing'. The latter term refers to the
undoing of the smearing process by which the dependence of the
harmonic functions characterising our BPS solutions is reduced by one
or more variables through compactification. Thus unsmearing implies
decompactification of space dimensions but the converse is not
necessarily true. When it is true we call the decompactified
dimensions `transverse'.  For the basic BPS solution smeared to two
space dimensions, unsmearing is of course possible up to the space
dimensions of the defining solution given in Section 2.2 (8 for the
M2, 5 for the M5, 9 for the KK-wave and 3 for the KK-monopole). {\em
  For all higher level BPS states in 2 non-compact space dimensions,
  unsmearing is impossible.} It is indeed straightforward to show that
the Einstein equation Eq.(\ref{einstein}) is not satisfied if the
harmonic function $H$ entering the right hand side of the equation is
extended to three dimensions. 

One verifies that in the dual formulation all $E_9$ BPS states can be
smeared to one space dimension, the charge being still defined by
Eq.(\ref{fullH}) with $i$ equal to 1. These solutions are also
solutions of the  $\sigma$-model  $S^{brane}$ Eq.(\ref{full}). 

A criterion for decompactification of longitudinal spacelike
directions  and of timelike directions will be obtained from the
requirement that the tensions should be finite. These quantities will
be evaluated in the string context from string dualities and uplifting
to eleven dimensions.  For each BPS state characterised by a dual
potential $A_p^{[N]}$ 
we define an action $\cal A$ given in Planck units by the product of
all spatial and temporal compactification radii, each of them at a
power equal to the number of times the corresponding index occurs in
$A_p^{[N]}$. 
One gets from Eqs.(\ref{lap3}), (\ref{lap6}) and (\ref{lapKK}) the
action ${\cal A}_l$ of the level $l$ solution\footnote{Similar actions
  were considered also in~\cite{Brown:2004jb}.} 

\begin{eqnarray}
\label{act3}
&&{\cal A}_{1+3n}=\frac{1}{l_p^{9n+3}}\, R_9 R_{10} R_{11} [R_3
  R_4\dots R_{11}]^n\qquad\qquad~~~\hbox {3-tower : level  (1+3$n$)}\\ 
\label{act6}
&& {\cal A}_{2+3n}=\frac{1}{l_p^{9n+6}}\,R_3 R_4R_5R_6R_7R_8 [R_3 R_4\dots R_{11}]^n\qquad ~\,\hbox{ 6-tower : level  (2+3$n$)} \\
\label{actKK}
&& {\cal A}_{3+3n}=\frac{1}{l_p^{9(n+1)}}\,R_4R_5\dots R_{10}(R_{11})^2 [R_3 R_4\dots R_{11}]^n\quad \hbox{ [8,1]-tower : level  (3+3$n$)} \, ,
\end{eqnarray}
where $l_p$ is the 11-dimensional Planck constant ($l_p^9=8\pi
G_{11}$). We identify for non-exotic states $\cal A$ to $MR_t$ where
$M$ is the mass of the source and $R_t$ the compactification time
radius. We derive in Appendix \ref{appm7}  the actions $\cal A$,
Eqs.(\ref{act3}), (\ref{act6}) and (\ref{actKK}), from the
interpretation in the context of string theory of the Weyl reflexions
used to construct the BPS solutions, both for exotic and non-exotic
states. Requiring finiteness of the action density implies that
$A_{\{q\}}$ be linear in the radii for those directions, spatial or
temporal, which can be decompactified. For non-exotic states this is
equivalent to requirement of finite tension.

It is immediately checked that for the basic branes $n=0$  one obtains
the correct mass formula for the M2 (time in 9) Eq.(\ref{massl1}), the
M5 (time in 3) Eq.(\ref{mm5}) and the KK-monopole (time in 4)
Eq.(\ref{mkk}). The KK-monopole mass is in agreement with the
calculation of the ADM mass of the unsmeared KK6-monopole
\cite{Bombelli:1986sb}. 
Our criterion confirms that time and all longitudinal space dimensions
can be taken to be non-compact except for the Taub-Nut direction 11 of
the KK6-monopole which occurs quadratically in Eq.(\ref{actKK}) and
hence cannot be decompactified. This fact is in agreement with the
fact that the  KK monopole solution in 3 transverse space dimensions,
characterised by an harmonic function $H=1+q/r$ where $r^2 \equiv
(x^1)^2+(x^2)^2+(x^9)^2$,  has, in order to avoid a conical
singularity, its radius $R_{11} \propto q$ \cite{Hawking:1976jb,
  Gibbons:1979xm, Sen:1997js} and hence finite. 

We now examine further the nature of the BPS solutions for level
higher than 3, that is outside the realm of the basic BPS solutions of
Section 2.2.

\subsubsection{From level 4 to level 6}

The actions ${\cal A}_l$ defined in Eqs.(\ref{act3}), (\ref{act6}) and
(\ref{actKK}) are in agreement with the computation of the masses for
level 4 Eq.(\ref{ml4}) with time in 3, level 5 Eq.(\ref{massl5}) with
time in 9, level 6 Eq.(\ref{massl6}) with time in 3 obtained in
Appendix C in the string context. Thus time occur linearly in $\cal A$
and can be non-compact. Examining the dependence of $\cal A$ in the
spatial radii, we see that 
the spacelike directions that can be decompactified are $(4,5,6,7,8)$
for $l=4$, $(10,11)$ for $l=5$ and none for $l=6$.  

U-duality requires that the dimensionally reduced metric in 2+1
dimensions  be identical for  these solutions and in addition be
equivalent to the (2+1)-dimensional metric for the basic BPS solutions
of Section 2.2.\footnote {\label{bpsfn}All BPS solutions for $l\le 6$ should then
  form a multiplet  of $E_8$ which is the  symmetry of 11D
  supergravity reduced to (2+1) dimensions.}  We now show that this
requirement is fulfilled, both in the direct and in the dual
formalism. 

To perform the dimensional reduction we write in general the
11-dimensional metric in the following form 
\begin{equation}
 ds^{2}= g_{\mu \nu} dx^\mu \, dx^\nu + \sum_r h_{rr}(dx^r)^2 \, ,
 \end{equation}
 where $\mu, \nu=1,2,a$ and $r\neq 1,2,a$, labelling $a$ as the time
 coordinate. To find the canonical Einstein action in $d= 3$
 dimensions, the components of the reduced metric have to be Weyl
 rescaled  
\begin{eqnarray}\label{reduc}
 \widetilde{g}_{\mu \nu} = g_{\mu \nu}\,  h^{\frac{1}{d-2}} =g_{\mu
 \nu}\,  h\, , 
 \end{eqnarray}
 where $h= \det\, h_{rs}$.
 
We first consider the level 4 state. In the direct formulation the
level 4 metric is given by 
Eq.(\ref{metric4}) and for the dimensional reduction to $3$ dimensions
we get using Eq.(\ref{reduc}) with $h= \widetilde{H}^{1/3}$ and time
in 3 
\begin{eqnarray}
ds^2_{l=4,3D}=  \widetilde{g}_{\mu \nu} dx^\mu \, dx^\nu= H [(dx^1)^2+
  (dx^2)^2]- (dx^3)^2\, . 
 \end{eqnarray}
The same metric is obtained in the reduction of the dual metric
Eq.(\ref{metric3}) with $n=1$, where now  $h= {H}^{-1/3}$. Similarly
the level 5 solution with time in 9, given in the direct formulation
by Eq.(\ref{metric5}) and in the dual one by Eq.(\ref{metric6}) with
$n=1$ with respectively $h= \widetilde{H}^{2/3}$ and $h= {H}^{-2/3}$,
yields from Eq.(\ref{reduc}) 
\begin{equation}
ds^2_{l=5,3D}=  \widetilde{g}_{\mu \nu} dx^\mu \, dx^\nu= H [(dx^1)^2+
  (dx^2)^2]- (dx^9)^2\, . 
 \end{equation}
The level 6 solution with the timelike direction 3 is given in the
direct formulation by Eq.(\ref{KKW}) with $n^\prime=1$, namely 
\begin{eqnarray}
\label{level6sol}
ds^2_{l=6}&=& (H^2+B^2)  [(dx^1)^2+(dx^2)^2  ] - \widetilde
 H^{-1}(dx^3)^2+[(dx^4)^2\dots+(dx^{10})^2  ] \nonumber\\ 
 &+& \widetilde H  [ dx^{11} -  (  \widetilde  H^{-1}-1 )dx^3 ]^2  \, .
 \end{eqnarray}
  and in the dual formulation by Eq.(\ref{metric0}) with $n=1$, that is
\begin{equation}
\label{level6soldu}
ds^2_{l=6}= H^2 [(dx^1)^2+(dx^2)^2  ] - H(dx^3)^2+
[(dx^4)^2\dots+(dx^{10})^2  ]+H^{-1} ( dx^{11} )^2  \, , 
 \end{equation}
Reducing the level $6$ metric Eqs.(\ref{level6sol}) and
(\ref{level6soldu}) to $3$ dimensions, we again find 
\begin{equation}
ds^2_{l=6,3D}=  \widetilde{g}_{\mu \nu} dx^\mu \, dx^\nu= H [(dx^1)^2+
  (dx^2)^2]- (dx^3)^2 \,. 
 \end{equation}
One easily checks that the same 3-dimensional metric  (with suitable
time coordinate)  are recovered for all basic BPS solutions recalled
in Section 2. 

We have thus verified that all the BPS solutions of 11D supergravity,
for levels $l \leq 6$  are equivalent in 2+1 dimensions.  These
solutions constitute an $E_8\subset E_9$ multiplet of branes (see
footnote~\ref{bpsfn}).  The  $E_8$ multiplet is the same as the one studied
some time ago algebraically as a consequence of M-theory compactified
on $T^8$ for which the masses of the different BPS states of the
multiplet has been  
derived (see \cite{Elitzur:1997zn} and in particular Table 11), and
their  space-time interpretation  was  obtained in reference
\cite{Lozano-Tellechea:2000mc}. We recover here these results from the
Weyl group  of $E_9$ endowed with the temporal involution,  and from
the interpretation of these Weyl transformation in the context of
string theory. This $E_9$ containing a timelike direction  is the
correct setting to describe all  the BPS solutions with two unsmeared
spacelike directions  in a group theoretical language. We summarise
below for all the levels $0< l\le 6$ its mass content\footnote{For the
  level 3 KK-monopole potential we have put the time in 4 as in
  Section 2 instead of 9, 10 in the general metric Eq.(\ref{mono}) to
  avoid here exotic states.} in the form ${\cal A}_l =M R_t$, where
${\cal A}_l$ is the level $l$ action Eqs.(\ref{act3}), (\ref{act6})
and (\ref{actKK}), $M$  the mass and $R_t$ the time radius (which can
be taken to $\infty$),  to exhibit their striking relation with the
$E_9$ dual potentials. 

\begin{center}
\begin{tabular}{|c|cccccc|}
\hline
$l$&1&2&3&4&5&6\\
\hline
\hline
time &9&3&4&3&9&3\\
\hline
$E_9$ field &$A_{9\,10\,11}$&$A_{3\dots8}$&$A_{4\dots 10\,11,11}$&$A_{3\dots11, 9\,10\,11}$&$A_{3\dots11, 3\dots 8}$&$A_{3\dots11,4\dots11,  11}$\\
\hline
&&&&&&\\
${\cal A}_l \!=\!MR_t$  &$\frac{R_9R_{10}R_{11}}{l_p^3}$&$\frac{R_3 \dots R_8}{l_p^6}$&$\frac{R_4 \dots R_{11}.R_{11}}{l_p^9}$&$\frac{R_3 \dots R_{11}.R_9R_{10}R_{11}}{l_p^{12}}$&$\frac{R_3 \dots R_{11}.R_3\dots R_8}{l_p^{15}}$&$\frac{R_3 \dots R_{11}.R_4\dots R_{11}. R_{11}}{l_p^{18}}$\\ 
&&&&&&\\
\hline
\end{tabular}
\end{center}

 These BPS states  already reach levels beyond the classical levels $l
 \leq 3$ for which a dictionary between $E_{10}$ fields  and
 space-time fields depending on one coordinate
 exists~\cite{Damour:2002cu}. In the next section we discuss the
 solutions for $l>6$.

\subsubsection{The higher level solutions}

The action formulae Eqs.(\ref{act3}), (\ref{act6}) and (\ref{actKK})
are all in agreement with the evaluation of $\cal A$ in the string
context, as seen from Section C.5. No time or longitudinal space
radius occurs linearly in $\cal A$. Thus time is compact and the only
non-compact space radii are the transverse two dimensions.  

All these states can  only be reached from basic non-exotic solutions
through a timelike T-duality. 

U-duality does no more imply that the dimensionally reduced metric in
(2+1) dimensions should be identical to that of the basic ones and one
indeed verifies  that they are distinct.  However metric and the
induced dilaton field are expected to be identical when reduced to 2
dimensions. This is indeed the case as we now show.  

Reducing all the solutions down to three dimensions\footnote{The
  spacelike or timelike nature of $x^r$ is irrelevant for this
  argument.} 1, 2  and $r$, with $r\in 3 \dots 11$   , we
get\footnote{In the dual formalism this formula holds with
  $\widetilde{g}_{11}$ depending only on $H$.} 
\begin{equation}
\label{metriclg}
ds^2_{l,3D}=  \widetilde{g}_{11}(H,B)  [(dx^1)^2+
  (dx^2)^2]+(dx^r)^2.
 \end{equation}
Because for all the solutions we have unity in front of $(dx^r)^2$,
reducing on $x^r$ down to two dimensions and performing a Weyl
rescaling all the metric reduce to a flat two-dimensional space with
zero dilaton field.

\setcounter{equation}{0}

\section{Transcending 11D supergravity}

We have seen in Section 4 that all $E_9$ BPS solutions (including the
basic ones discussed in Section 2.2) can be smeared to one space
dimension in the dual formalism. There is in fact no such description
in the direct formalism, except for the M2 and the KK-wave. More
generally, the dual formalism in one non-compact space dimension is
equivalent to the  $\sigma$-model Eq.(\ref{full})  restricted to a
single  $l> 0$ field. In that case indeed  the matter term is the same
in both formalisms as no covariant derivative arises in the
$\sigma$-model \cite{Damour:2002cu, Englert:2004ph} and the Einstein
term  \cite{Banks:1998vs} coincides with its  level zero. 

We now consider BPS solutions obtained from positive real roots of
$E_{10}$ not present in $E_9$. These solutions can be obtained by
performing Weyl transformations on 
$E_9$ BPS solutions smeared to one dimension. They  depend on one
non-compact space variable and may have no counterpart in 11D
supergravity. We illustrate the construction of such solutions by one
example. 

\begin{table}[h]
\begin{center}
\begin{tabular}{|cc|cccccccccc|c|}
\hline
$IIA$&$M$-theory&1&2&3&4&5&6&7&8&9&10&11\\
\hline
D6 & KK6& \,&$\times$&$\times$&$\times$&$\times$&$\times$&$\times$&$\times$&\, & \,& $\times$ \\
D8 & M9& \,&$\times$&$\times$&$\times$&$\times$&$\times$&$\times$&$\times$&$\times$&$\times$&$\times$ \\
\hline
\end{tabular}
\caption{\sl \small The D8 brane obtained from the D6 brane by
  performing the Weyl  
reflexion $W_{\alpha_{11}}$, i.e. a double T-duality plus exchange of
the directions $x^9$ and $x^{10}$.} 
\end{center}
\end{table}

We shall obtain the M9 (namely the `uplifting'\footnote{There is a
  no-go theorem stating that massive 11-dimensional supergravity does
  not exist \cite{Bautier:1997yp}. The concept of uplifting the D8
  seems thus puzzling. However  a definition of  `massive 11D
  supergravity' has been proposed \cite{Bergshoeff:1997ak} for a
  background with an isometry generated by a spatial Killing
  vector. In that theory the M9 solution does exist
  \cite{Bergshoeff:1998bs}. Existence of the M9 is also suggested by
  the study of the central charges of the M-theory superalgebra
  \cite{Townsend:1997wg, Hull:1997kt}.}  of the D8 brane of massive
Type IIA supergravity)\footnote{The metric of the D8 brane has
  previously been discussed in the context of $E_{11}$ in
  \cite{West:2004st}.} by performing a Weyl transformation on the
KK6-monopole smeared in all directions but one described in Section
2.2.3. We start with a D6 
along the spatial directions $2,3,\dots,7$ and we choose 8 as time
  coordinate.  The D6 is smeared in the directions 9 and 10 and thus
  depends only on the non-compact variable $x^1$. 
The uplifting to M-theory of the D6 yields a KK6-monopole with
  Taub-NUT direction 11 (see Table 1). 
To obtain the D8 and its uplifting M9, we  perform the Weyl reflexion
  $W_{\alpha_{11}}$ which may be viewed as a double T-duality in the
  directions 9 and 10 plus exchange of the two radii  
\cite{Elitzur:1997zn, Obers:1998rn, Banks:1998vs, Englert:2003zs}.

The KK6-monopole, in the longitudinal directions $2,\dots ,7$ with
timelike direction 8 and Taub-NUT direction 11,  smeared in  9 and 10,
is described in the  $\sigma$-model by a  
level 3 generator\footnote{This level 3 step operator contains the
  index 2 and belongs to a  $E_9$ conjugate in $E_{10}$ to the $E_9$
  we used in the previous sections.}  $R^{2345678\,11\vert11}$. The
solution is given in Eq.(\ref {G3}) up to a permutation of
indices. Defining $p^a = -h^{~a}_a$ one has 

\begin{eqnarray}
\label{modkkmono}&&p^1=p^9=p^{10} = {1\over 2} \ln H(x^1) \nonumber \\
&&p^{11}=-{1 \over 2} \ln 
H(x^1)\nonumber\\ && p^i=0 \qquad i=2 \dots 8 \nonumber\\
& &A_{2345678\,11\vert 11}={1 \over H(x^1)} .
\end{eqnarray}

The level 3 root $\alpha^{(3)}$  corresponding to
$R^{2345678\,11\vert11}$  is
$\alpha^{(3)}=\alpha_2+2\alpha_3+3\alpha_4+4\alpha_5+5\alpha_6+6\alpha_7+7\alpha_8+4\alpha_9+\alpha_{10}+3\alpha_{11}$.
Performing the Weyl 
reflexion $W_{\alpha_{11}}$, we get
$W_{\alpha_{11}}(\alpha^{(3)})\equiv \alpha^{(4)}
=\alpha^{(3)}+\alpha_{11}$. 
The root $\alpha^{(4)}$  is the lowest weight of the $A_9$ irreducible
representation of level 4 \cite{Nicolai:2003fw} whose 
Dynkin labels are $(2,0,0,0,0,0,0,0,0)$. This $A_9$ representation in
the decomposition of the adjoint representation of $E_{10}$  is not in
$E_9$. The action of    $W_{\alpha_{11}}$ on the Cartan fields is
\cite{Englert:2003zs} 
\begin{eqnarray}
&& p^{\prime a}= p^a+ {1 \over 3} (p^9+p^{10}+p^{11}) \qquad a=2 \dots
  8\noindent \\ 
&&p^{\prime a}= p^a- {2 \over 3} (p^9+p^{10}+p^{11}) \qquad a=9,10,11.
\end{eqnarray}
Using the embedding  relation Eq.(\ref{embedding}), the Weyl transform
of Eqs.(\ref{modkkmono})  yields the solution 
\begin{eqnarray}
\label{mett1}
&&p^{\prime 1}={4\over 6} \ln H(x^1)\\
\label{mett2}
&& p^{\prime a}= {1\over 6} \ln H(x^1) \qquad a=2 \dots 10 \noindent \\
\label{mett3}
&&p^{\prime 11}= - {5\over 6} \ln H(x^1). \\ \label{complicated}
&&A_{23456789 \,10\,11\vert 11\vert 11}={1 \over H(x^1)}\, .
\end{eqnarray}
The level 4 field Eq.(\ref{complicated}) contains the antisymmetric
 set of indices $2,3\dots 11$. These are not apparent in its $A_9$
 Dynkin labels given above but are needed in the 11-dimensional metric
 stemming from the embedding $E_{10}\subset E_{11}$ encoded in 
 Eq.(\ref{embedding}). The $A_{10}\subset E_{11} $ Dynkin labels of
 this  field in this embedding are indeed
 $(2,0,0,0,0,0,0,0,0,1)$. From Eq.(\ref{vielbein}) one gets the
 11-dimensional metric  
\begin{equation}
\label{M9}
ds^{2}_{M9}= H^{4/3} (dx^1)^2+H^{1/3} [(dx^2)^2+\dots
  -(dx^8)^2+(dx^9)^2+(dx^{10})^2] 
+H^{-5/3} (dx^{11})^2\, .
\end{equation}
One verifies the validity of the dual formalism equation
Eq.(\ref{genfield}) for the field Eq.(\ref{complicated}), namely  
\begin{equation}
\label{tenfield}
\frac{d}{dx^1} \left(\sqrt{\vert g \vert}g^{22}g^{33}\dots g^{10\,10}[
  g^{11\,11}]^3\frac{d}{dx^1}A_{23456789 \,10\,11\vert 11\vert
  11}\right)= \frac{d^2}{d(x^1)^2}H(x^1)=0 \, . 
\end{equation}
The metric Eq.(\ref{M9}) describes the   M9  \cite{Bergshoeff:1998bs},
which reduced to 10 dimensions gives the D8~\cite{Bergshoeff:1996ui}
of  massive type IIA. The computation of the M9-mass in the string
context Eq.(\ref {1110}) agrees with our general action formula 
\begin{equation}
{\cal A}_{M9}= R_8 M_{M9}={R_2R_3R_4R_5R_6R_7R_8R_9  R_{10} R^3_{11}
  \over l_p^{12}}\, , 
\end{equation}
indicating that $R_{11}$ is compact.

\setcounter{equation}{0}
\section {Summary and comments}

We have constructed an infinite $E_9$ multiplet of  BPS solutions of
11D supergravity and of its exotic counterparts depending on  two
non-compact variables. These solutions  are related by U-dualities
realised as  Weyl transformations of the $E_9$ subalgebra of $E_{11}$
in the regular embedding $E_9\subset E_{10}\subset E_{11}$. Each BPS
solution stems from an $E_9$ potential $A^{[N]}_p$ multiplying the
generator  $R^{[N]}_p$ in the Borel representative of the coset space
$E_{10}/K^-_{10}$ where $K^-_{10}$ is invariant under a temporal
involution.  $A^{[N]}_p$ is related to the supergravity 3-form and
metric through dualities and compensations. This $E_9$ multiplet of
states split into three classes according to the  level $l$.  For
$0\le l \le 3$ we recover the basic BPS solutions, namely, the
KK-wave, the M2, the M5 and the KK-monopole. For $4\le l \le 6$, the
solutions have 8 longitudinal space dimensions. We argue that for
higher levels, all 9 longitudinal directions, including time, are
compact.  Each  BPS solution can be mapped to a solution of a dual
effective action of gravity coupled to matter expressed in terms of
the $E_9$ potential $A^{[N]}_p$. In the dual formulation the BPS
solutions can be smeared to one non-compact space dimension and
coincides then with solutions of the $E_{10}$ $\sigma$-model build
upon $E_{10}/K^-_{10}$. The $\sigma$-model yields in addition an
infinite set of BPS space-time solutions corresponding to all real
roots of $E_{10}$ which are not roots of $E_9$. These appear to
transcend 11D supergravity, as exemplified by the lowest level $(l=4)$
solution which is identified to the M9.

The relation between the  11D supergravity 3-form and metric, and the
$E_9$ potentials  $A^{[N]}_p$  has significance beyond the realm of
BPS solutions. To see this we first recall that the  $E_9$ potentials
can be organised in towers defined by decomposing $E_9\subset E_{10}$
into  $A_1^+$ subgroups with central charge $-K^2{}_2\in E_{10}$.  
Each of these $A_1^+$ subgroup contains  two `brane' towers $[N]=[3],[6]$
or one `gravity tower' $[N]=[8,1]$ of real roots (the two gravity towers
$[N]=[8,1]$ and [0] are redundant except for the lowest level representing
the KK-wave). We first examine the brane towers. 
 
The recurrences of the 3-tower $A^{[3]}_{1+3n}$ alternate in nature at
each step: they switch from states on the M2 sequence to those on the
M5 sequence. This feature is illustrated in  Fig.2a where the 3-tower
recurrences are depicted on the right column. On the other hand, each
recurrence of the 3-tower is related by {\em duality-compensation
  pairs} to the supergravity 3-form potential which we denote by
$(A^{[3]}_1)_q$,  as seen in the horizontal lines of both Fig.3 and
Fig.4. Here we designated by the integer $q$ the number of
duality-compensation pairs needed to reach the field $(A^{[3]}_1)_q$
from $A^{[3]}_{1+3n}$. In the realm of BPS states studied in this
paper each field $A^{[3]}_{1+3n}$ defines a different BPS solution of
11D supergravity defined by $(A^{[3]}_1)_q$ and the related
metric. Comparing Fig.2a with Fig.3 and 4 we see that  $q$ is equal to
$n$. This relation expresses the fact that {\em the number of steps
  needed to climb the 3-tower up to the field 
$A^{[3]}_{1+3n}$ is equal to the number of duality-compensation pairs
  needed to reach from $A^{[3]}_{1+3n}$ the 11D supergravity 3-form
  defining the corresponding BPS solution}. However the relation $q=n$
does not rely on the BPS character of the solution and hence has
general significance, which can be pictured as follows.  

Were all the compensations matrices put equal to unity  no new
solutions could be generated by Weyl transformations from any solution
of 11D supergravity defined by its  3-form $A^{[3]}_1$ (or from its
Hodge dual) and metric. Indeed, because of the Weyl equivalence of all
dualities depicted in Fig.3 and Fig.4, one would simply get  
\begin{eqnarray}
\label{comp01}
A^{[3]}_{1+3n} &\stackrel{\rm I}{=}& A^{[3]}_1\qquad n \,{\rm even}\\
A^{[3]}_{1+3n} &\stackrel{\rm I}{=}& A^{[6]}_2\qquad n \,{\rm odd}\, ,
\end{eqnarray}
where $A_2^{[6]}$ is taken to be the Hodge dual of $A_1^{[3]}$  and
the superscript I means that all compensations have been formally
equated to unity. A similar analysis of the 6-tower depicted in the
left column of Fig.2a would yield 
\begin{eqnarray}
\label{comp02}
A^{[6]}_{2+3n} &\stackrel{\rm I}{=}& A^{[6]}_2\qquad n \,{\rm even}\\
A^{[6]}_{2+3n} &\stackrel{\rm I}{=}& A^{[3]}_1\qquad n \,{\rm odd}\, .
\end{eqnarray}
The same phenomenon would appear in the gravity tower, where it is
somewhat hidden in the redundant 0-tower of Fig.5b. 

The non-trivial content of the $E_9$ tower potentials is entirely
due to compensations. These prevent `duals of duals' to be equivalent
to unity and one may view the $E_9$ towers as defining `non-closing
dualities', familiar from the standard Geroch group. They translate
through the compensation process the genuine non-linear structure of
gravity.

\section*{Acknowledgments}
We are greatly indebted to Philippe Spindel  for many interesting and
fruitful  discussions and we are grateful to Marc Henneaux for his support and encouragement.   
Laurent Houart  thanks Riccardo Argurio for many
illuminating conversations. Fran\c cois Englert thanks Sasha Belavin
for useful comments. 
Fran\c cois Englert and Laurent Houart are grateful to the
Max-Planck-Institut f\"ur Gravitationsphysik at Potsdam, and Axel Kleinschmidt and
Hermann Nicolai to the
Universit\'e Libre de Bruxelles, for their  warm
hospitality. 

This work was supported in part by IISN-Belgium (convention 
4.4511.06 and convention 4.4505.86) by the European Commission FP6 RTN
programme MRTN-CT-2004-005104, and by 
the Belgian Federal Science Policy Office through the 
Interuniversity Attraction Pole P VI/11.

\newpage
\appendix
\setcounter{equation}{0}

\section{Signature changes and compensations} 
\label{appw}
 Expressing a Weyl transformation $W$  as a conjugation by a group
element
$U_W$ of $E_{11}$ ($E_{10}$), one defines the involution
$\Omega^\prime$ operating on 
the conjugate elements by
\begin{equation}
\label{newinvolve}
\Omega^\prime (T^\prime)=U_W\,\Omega(\underbrace{U^{-1}_W T^\prime U_W}_{T}) \, U^{-1}_W\, ,
\end{equation}
where $T$ and $T^\prime$ are any conjugate pair of generators in
$E_{11}$ ($E_{10}$). The subgroup invariant under $\Omega$ is
conjugate to the subgroup invariant under $\Omega^\prime$. However,
Weyl reflexions in general do not commute with the temporal
involution~\cite{Keurentjes:2004bv,Keurentjes:2004xx}. 

\subsection{The gravity line}
First consider the gravity line of $E_{11}$ endowed with the temporal involution Eq.(\ref{map}). The Weyl reflexion $W_{\alpha_1}$ in the hyperplane perpendicular to
$\alpha_1$ changes the time index in the  $A_{10}$ tensors from 1 to 2.
Indeed  applying Eq.(\ref{newinvolve}) to
the Weyl reflexion $W_{\alpha_1}$  generates from 
$\Omega_1\equiv\Omega$ a new involution $\Omega_2\equiv\Omega^\prime$ such that
\begin{eqnarray}
\label{permute}
&&U_1\, \Omega_1 K^2_{\ 1} \, U^{-1}_1= \rho K^2_{\ 1}= \rho\Omega_2
 K^1_{\ 2}\nonumber\, ,\\
&&U_1\, \Omega_1 K^1_{\ 3} \, U^{-1}_1= \sigma K^3_{\ 2}=
\sigma\Omega_2
 K^2_{\ 3} \, ,\\
&&U_1\, \Omega_1 K^i_{\ i +1} \, U^{-1}_1= -\tau K^{i+1}_{\ \, i}=
\tau\Omega_2  K^i_{\ i +1}\quad i >2\, .\nonumber
\end{eqnarray}
Here $\rho,\sigma,\tau$ are plus or minus signs which may arise
as step operators are representations of the Weyl group 
up to signs. Eq.(\ref{permute}) illustrate the general result
that such signs always cancel in the determination of
$\Omega^\prime$ because they are identical in the Weyl transform of
corresponding positive and negative roots, as
their commutator is in the Cartan subalgebra which  forms a true
representation of the Weyl group. The content of Eq.(\ref{permute}) is
represented in Table 2. 
The signs below the generators of the gravity
line indicate the sign in front of the
 negative step operator obtained by the involution: a
minus sign is in agreement with the conventional Chevalley involution
and indicates that the indices in
$K^m_{\ m +1}$ are both either space or time indices while a plus sign
indicates that one index must be time and the other  space.
\begin{table}[h]
\begin{center}
\begin{tabular}{|c|ccccc|c|}
\hline
gravity line&$K^1_{\ 2}$&$K^2_{\ 3}$&$K^3_{\ 4}$&$\cdots$&$K^{D-1}_{\
D}$&time coordinate\\
\hline\hline
$\Omega_1$&$+$&$-$&$-$&$-$&$-$&1\\
\hline$\,\Omega_2$&$+$&$+$&$-$&$-$&$-$&2\\
\hline
\end{tabular}
\caption{\sl \small Involution switches from $\Omega_1$  to
$\Omega_2$ in $E_{11}$ due to the Weyl reflexion $W_{\alpha_1}$}
\end{center}
\end{table}

\noindent
Table 2 shows that
the  time coordinates in
$E_{11}$ must now be identified either with 2, or with all indices
$\neq 2$. We choose the first description, which leaves
unaffected coordinates attached to planes invariant under the Weyl
transformation. More generally, performing  Weyl reflexions from roots of the
gravity line, we can identify the time index to any $A_{10}$ tensor index.

The Weyl transformations on the gravity line of $E_{11}$ (or $E_{10}$)
simply changes the time coordinate but do not modify the global
signature $(1,10)$ (or $(1,9)$).  This need not be the case for Weyl
transformations from roots pertaining to higher levels. We shall
determine  the different signatures for the M2 and M5 sequences. 
In order to do that, we have to study the effect of the Weyl
reflexions $W_{\alpha_{11}}$ and $W_{-\alpha_{11}+\delta}$ on  the
involutions characterising  the M2 and the M5 we started with. 
We consider separately the two sequences. We will also see that the
nature of the compensation 
transformations ($SO(2)$ or $SO(1,1)$) is determined and follows from
this analysis. Finally we shall consider the signatures induced on the
gravity tower by the mapping Eqs.(\ref{bragramap}), (\ref{bragra1})
and (\ref{bragra2}). 

\subsection{The brane towers}
\subsubsection{Signatures of the M2 sequence}
\label{appw1}
We start with the conventional M2 described by the solution
Eq.(\ref{2M2}) of 11-dimensional supergravity 
with the signature (1,10, +). Here the first entry denotes the number
of timelike directions 
 (in our case the single direction 9), the second denotes the number
of spacelike directions and the third 
gives the sign of the kinetic energy term in the action ($+$ being the
usual one). We want to determine the space-time 
signature for all the solutions of the M2 sequence. The M2 of level 1
is characterised by 
the involution  $\Omega_{l_1}\equiv \Omega_9$ fixing 9 as a time coordinate.

To determine the signature at  level 5, we  perform the Weyl
transformation $W_{-\alpha_{11}+\delta}$ to find  
the corresponding  new involution $\Omega_{l_5}$.
The generators of the gravity line affected by this reflexion are $
K^2_{\ 3}$ and $K^8_{\ 9}$. From  
Eq.(\ref{newinvolve}) we have (from now on, we drop irrelevant signs
$\rho,\sigma,\tau$ appearing in Eq.(\ref{permute})) 
\begin{eqnarray}
\label{ww1a}
\Omega_{l_5}
 K^2{}_ 3 &=&U_{W_{-\alpha_{11}+\delta}}\, \Omega_{l_1} R^{2\, 4\,5 \,6 \,7 \,8 } \, U^{-1}_{W_{-\alpha_{11}+\delta}}\nn\\
 &=&- K^3{}_ 2 \\
\Omega_{l_5}
 K^8{}_ 9&=& U_{W_{-\alpha_{11}+\delta}}\, \Omega_{l_1} R_{3\, 4\,5 \,6 \,7 \,9 } \, U^{-1}_{W_{-\alpha_{11}+\delta}}
 \label{ww1ab}\nn\\
&=& +K^9{}_8\, .
\end{eqnarray}
The signature of the level 5 solution is thus unchanged, the only time coordinate is still 9.
The action of the involution $\Omega_{l_5}$ on $R^{9\, 10\, 11}$  follows from  $W_{-\alpha_{11}+\delta}(\alpha_{11})= 2\delta  - \alpha_{11}$. We get 
\begin{equation}
\label{ww1b}
\Omega_{l_5}
 R^{\,  9 \, 10\, 11}= U_{W_{-\alpha_{11}+\delta}}\, \Omega_{l_1} R^{[6]}_5\, U^{-1}_{W_{-\alpha_{11}+\delta}}= + R_{\,  9 \, 10\, 11}\, .
 \end{equation}
As 9 is a timelike coordinate, this yields the usual sign for this generator and one sticks to the signature $(1,10,+)$  as it should. We have 
$\Omega_{l_5}\, (R^{[6]}_2-R^{[6]}_{-2})=R^{[6]}_2-R^{[6]}_{-2}$ and the compensation  for the $SL(2)$ of level 5 (see Eqs.(\ref{matrix5})-(\ref{comp5})) lies in      its $SO(2)$ subgroup.

We now  perform the Weyl reflexion $ W_{\alpha_{11}}$   to reach level 7. We have

\begin{eqnarray}
\Omega_{l_7} K^8{}_ 9 &=& U_{W_{\alpha_{11}}}\, \Omega_{l_5} R^{\,8 \,10 \,11 } \, U^{-1}_{W_{\alpha_{11}}} \nn \\ \label{ww2a}
&=& - K^9{}_8\\
 \label{ww2b}
\Omega_{l_7}
R^{\, 9\, 10\, 11} &=& U_{W_{\alpha_{11}}}\, \Omega_{l_5} R_{\, 9\, 10\, 11} \, U^{-1}_{W_{\alpha_{11}}}\nn\\
&=& +R_{\, 9\, 10\, 11}\, .
\end{eqnarray}
From Eq.(\ref{ww2a}), we deduce immediately that the time coordinates are now 10 and 11 and from
Eq.(\ref{ww2b}) we deduce that the sign of the kinetic terms is the `wrong' one, namely it corresponds to the (2,9,-) theory as it should \cite{Hull:1998vg, Hull:1998ym}. We have $\Omega_{l_7}\, (
R^{\, 9\, 10\, 11} +R_{\, 9\, 10\, 11})=R^{\, 9\, 10\, 11} +R_{\, 9\, 10\, 11}$ and  the compensation for the $SL(2)$ of level 7 solution Eq.(\ref{fin7}) lies in its $SO(1,1)$   subgroup.

\begin{table}[h]
\begin{center}
\begin{tabular}{|c|c|c|c|}
\hline
levels ($n>0$)&
times & $(t,s,\pm)$ & compensation \\
\hline\hline
1&9&$(1,10,+) $&$--$\\
\hline 
5&9&$(1,10,+) $&$SO(2)$\\
\hline
 1+6n , n odd&10,11&$(2,9,-) $&$SO(1,1)$\\
 -1+6(n+1), n odd &10,11&$(2,9,-) $&$SO(2)$\\
\hline 
1+6n , n even&9&$(1,10,+) $&$SO(1,1)$\\
 -1+6(n+1), n even &9&$(1,10,+) $&$SO(2)$\\
\hline
\end{tabular}
\caption{\sl \small Involution switches from $\Omega_{l_1}$ to
$\Omega_{l_{\pm 1+6n}}$ in the M2 sequence due to the application of
  the successive Weyl reflexions  $W_{-\alpha_{11}+\delta}$ and
  $W_{\alpha_{11}}$} 
\end{center}
\end{table}

\noindent
We can now repeat the analysis to all levels of the M2 sequence. We use $ W_{-\alpha{11}+\delta}$  to go from level $1+6n$ to level $-1+6(n+1)$. Replacing in Eqs.(\ref{ww1a}), (\ref{ww1ab}) and  (\ref{ww1b})  
$\Omega_{l_1}$ by $\Omega_{l_{1+6n}}$ and $\Omega_{l_5}$ by $\Omega_{l_{-1+6(n+1)}}$, we see that the signature of the theory is unchanged. Furthermore,  analysing the action of   $\Omega_{l_{-1+6(n+1)}}$  on $R^{[6]}_2$, we conclude that the compensation for level  $-1+6(n+1)$ is always an $SO(2)$ one.
We use $W_{\alpha_{11}}$  to go from level $-1+6(n+1)$ to level $1+6(n+1)$. Replacing in Eqs.(\ref{ww2a}) and  (\ref{ww2b}) $\Omega_{l_5}$ by $\Omega_{l_{-1+6(n+1)}}$ and $\Omega_{l_7}$ by $\Omega_{l_{1+6(n+1)}}$, we see that theories $(1,10,+)$ and $(2,9,-)$ are interchanged. The action of $\Omega_{l_{1+6(n+1)}}$ on  $R^{[3]}_1$ shows that the compensation at level  $1+6(n+1)$ lies always in $SO(1,1)$. The results are summarised in Table 3.

\subsubsection{Signatures of the M5 sequence}
\label{appw2}
We start with the non-exotic M5 described by Eq.(\ref{2M5}) solution of 11D supergravity
with the signature (1,10, +) and time in 3. We want to determine the space-time
signature of all the solutions of the M5 sequence depicted in Fig.2. The M5 of level 2 is characterised by
the involution  $\Omega_{l_2} \equiv \Omega_3$ fixing 3 as a time coordinate.

To determine the signature of  level 4, we  perform the Weyl reflexion $ W_{\alpha_{11}}$  to find 
the  new involution $\Omega_{l_4}$.
The generator of the gravity line affected by this reflexion is $K^8{}_9$. From 
Eq.(\ref{newinvolve}) we have
\begin{eqnarray}
\label{ww52a}
&&\Omega_{l_4}
 K^8_{\ 9}=U_{W_{\alpha_{11}}}\, \Omega_{l_2} R^{\,8 \,10 \,11 } \, U^{-1}_{W_{\alpha_{11}}}= - K^9_{\ 8} \\ \label{ww52b}
&&\Omega_{l_4}
R^{\, 9\, 10\, 11}= U_{W_{\alpha_{11}}}\, \Omega_{l_2} R_{\, 9\, 10\, 11} \, U^{-1}_{W_{\alpha_{11}}}= -R_{\, 9\, 10\, 11}\, ,
\end{eqnarray}

From Eq.(\ref{ww52a}), we deduce immediately that there is no change of signature and thus 3 remains the only timelike direction. From
Eq.(\ref{ww52b}) the sign of the kinetic terms is unchanged and  the phase is still
(1,10,+) as it should. We have $\Omega_{l_4} (R^{\, 9\, 10\, 11} -R_{\, 9\, 10\, 11})=R^{\, 9\, 10\, 11} -R_{\, 9\, 10\, 11}$. Hence the coset characterising the level 4 solution is
$SL(2)/SO(2)$ and the compensation at level 4 (see Eq.(\ref{fin4})) lies in $SO(2)$.

\begin{table}[h]
\begin{center}
\begin{tabular}{|c|ccccccccc|c|c|}
\hline
level &$K^2_{\ 3}$&$K^3_{\ 4}$&$K^4_{\ 5}$&$K^5_{\ 6}$&$K^6_{\
7}$&$K^7_{\ 8}$&$K^8_{\ 9}$&$K^9_{\ 10}$&$K^{10}_{\ 11}$&
times & $(t,s,\pm)$ \\
\hline\hline
4&$+$&$+$&$-$&$-$&$-$&$-$&$-$&$-$&$-$&3&$(1,10,+) $\\
\hline 8& $-$&$+$&$-$&$-$&$-$&$-$&$+$&$-$&$-$&4,5,6,7,8&$(5,6,+)
$\\
\hline
\end{tabular}
\caption{\sl \small Involutions at level 4 and 8.}
\end{center}
\end{table}

\vskip -.5cm
To determine the signature of  level 8, we  perform the Weyl reflexion $W_{-\alpha{11}+\delta}$ to find 
the  new involution $\Omega_{l_8}$.
We have
\begin{eqnarray}
\label{ww51a}
&&\Omega_{l_8}
 K^2{}_3=U_{W_{-\alpha_{11}+\delta}}\, \Omega_{l_4} R^{2\, 4\,5 \,6 \,7 \,8 } \, U^{-1}_{W_{-\alpha_{11}+\delta}}= - K^3{} _2 \nonumber \\
&&\Omega_{l_8}
 K^8{}_9 = U_{W_{-\alpha_{11}+\delta}}\, \Omega_{l_4} R_{3\, 4\,5 \,6 \,7 \,9 } \, U^{-1}_{W_{-\alpha_{11}+\delta}}= +K^9{}_8\, .
\end{eqnarray}
The flip of sign in  $K^2_{\  3}$ and $K^8_{\  9}$ 
illustrated in Table 4 shows that the resulting theory comprises the 5 time coordinates 4,5,6,7,8.
The involution $\Omega_{l_8}$ acts on $R^{9\, 10\, 11}$ according to
 \begin{equation}
\label{ww51b}
\Omega_{l_8}
 R^{\,  9 \, 10\, 11}=U_{W_{-\alpha_{11}+\delta}}\, \Omega_{l_4} R^{[6]}_5\, U^{-1}_{W_{-\alpha_{11}+\delta}}= - R_{\,  9 \, 10\, 11}\, .
 \end{equation}
The directions 9,10,11 being spacelike, the action of the involution on $R^{\,  9 \, 10\, 11}$ yields the `right' kinetic energy terms. The new theory is $(5,6,+)$  as it should \cite{Hull:1998vg, Hull:1998ym}.
We have  $\Omega_{l_8}(R^{[6]}_2+R^{[6]}_{-2})=R^{[6]}_2+R^{[6]}_{-2}$. Hence the compensation
at level 8 (see Eq.(\ref{fin8})) lies in $SO(1,1)$.

\begin{table}[h]
\begin{center}
\begin{tabular}{|c|c|c|c|}
\hline
levels ($n>0$)&
times & $(t,s,\pm)$ & compensation \\
\hline\hline
2&3&$(1,10,+) $&$--$\\
\hline 
4&3&$(1,10,+) $&$SO(2)$\\
\hline
 2+6n , n odd&4,5,6,7,8&$(5,6,+) $&$SO(1,1)$\\
 -2+6(n+1), n odd &4,5,6,7,8&$(5,6,+) $&$SO(2)$\\
\hline 
2+6n , n even&3&$(1,10,+) $&$SO(1,1)$\\
 -2+6(n+1), n even &3&$(1,10,+) $&$SO(2)$\\
\hline
\end{tabular}
\caption{\sl \small Involution switches from $\Omega_{l_2}$ to
$\Omega_{l_{\pm 2+6n}}$ in the M5 sequence due to the application of the successive Weyl reflexions  $W_{\alpha_{11}}$ and $W_{-\alpha_{11}+\delta}$. }
\end{center}
\end{table}

\vskip -.5cm
We can repeat the analysis to find the signature of all  the levels of the M5 sequence.
Again we find only the two signatures found at the lower levels alternating every two steps while, as for the M2 sequence, the nature of the compensation alternates at each step.
The results are summarised in Table 5.

\subsection{The gravity towers}
\label{appsg}
The gravity towers were obtained by performing Weyl transformations on the brane towers. In section 3.3. we have showed that the M2 sequence is mapped to the wave sequence and the M5 sequence to the monopole sequence. We shall take advantage of this Weyl mapping to find the signatures of the gravity towers.

The brane towers comprise $4$ different signatures:
\begin{itemize}
\item $(1,10,+)$ with time in $9$ and $(2,9,-)$ with time in $10$ and $11$ for the M2 sequence,
\item  $(1,10,+)$ with time in $3$ and $(5,6,+)$ with time in $4,5,6,7,8$  for the M5 sequence.
\end{itemize}

We perform the Weyl mapping on the $4$ signatures in three steps.  The first Weyl transformation $W_{(1)}$ interchanges 9 and 3 on the gravity line. The second Weyl transformation is the Weyl reflexion $W_{(2)}\equiv W_{\alpha_{11}}$  and the last Weyl transformation $W_{(3)}$ interchanges 9 and 11 on the gravity line. 

The first and last Weyl transformations $W_{(1)}$ and $W_{(3)}$ permute tensor indices and do not alter the global signature. Only  $W_{(2)}$ can change the global signature $(t,s,\pm)$. There are $2$ simple roots affected by $W_{(2)}\equiv W_{\alpha_{11}}$: $\alpha_8$ and $\alpha_{11}$ defining respectively the generators $K^8_{\ 9}$ and $R^{9\, 10\, 11}$. 
From Eq.(\ref{newinvolve}) we get, dropping irrelevant signs, 
\begin{eqnarray}\label{k89}
\Omega' \, K^8_{\ 9}= s_1\, K^9_{\ 8}= U_{W_{(2)}} \, \underbrace {\Omega \,  R^{8\, 10\, 11}}_{s_1\, R_{8\, 10\, 11}} \, U_{W_{(2)}}^{-1} \\
\label{r91011}
\Omega' \, R^{9\, 10\, 11}= s_2\, R_{9\, 10\, 11}= U_{W_{(2)}} \, \underbrace {\Omega \,  R_{9\, 10\, 11}}_{s_2\, R_{9\, 10\, 11}} \, U_{W_{(2)}}^{-1}\, .
\end{eqnarray}
where $s_1$ and $s_2$ are signs. The possible change of signatures by the Weyl transformation $W_{(2)}$ will be deduced from the signs $s_1$ and $s_2$.

\subsubsection{Signatures of the wave sequence}

\noindent
{\bf $\bullet$ Mapping of the signature $(1,10,+)$ with time in $9$}

The global signature $(1,10,+)$ is not modified by $W_{(1)}$ but the time coordinate is no longer in $9$ but in $3$. From Eqs.(\ref{k89}) and (\ref{r91011}), with $s_1=-1 $ and $s_2=-1$, we deduce that the signature is unchanged by the second Weyl reflexion $W_{(2)}$. The last transformation $W_{(3)}$ does not change the signature either. 

The signature $(1,10,+)$ with time in $9$ is thus mapped by the three successive Weyl transformations to the signature $(1,10,+)$ with time in $3$.

\medskip
\noindent
{\bf $\bullet$ Mapping of the signature $(2,9,-)$ with time in $10$ and $11$}

The first Weyl transformation $W_{(1)}$ does not modify the signature. We then perform the Weyl transformation $W_{(2)}$. From Eq.(\ref{k89}) with $s_1=+1$ we find that the time coordinate becomes $9$ and from Eq.(\ref{r91011}) with $s_2=+1$, we deduce the sign of the kinetic term. This sign is the `usual' one and the signature becomes $(1,10,+)$ with time coordinate $9$. The last Weyl reflexion $W_{(3)}$ does not change the global signature but puts the time coordinate  in $11$.

The signature $(2,9,-)$ with times in $10$ and $11$ is thus mapped by the three successive Weyl transformations to the signature $(1,10,+)$ with time in $11$.

\medskip
\noindent{\bf $\bullet$ Signatures of the wave sequence}

\begin{table}[h]
\begin{center}
\begin{tabular}{|c|c|c|}
\hline
levels ($n, n' >0$)&
times & $(t,s,\pm)$  \\
\hline\hline
0&3&$(1,10,+) $\\
\hline 
6n , n odd&11&$(1,10,+) $\\
6n, n even&3&$(1,10,+)  $\\
\hline 
6n' , n' odd&3&$(1,10,+) $\\
6n', n' even&11&$(1,10,+)  $\\
\hline
\end{tabular}
\caption{\sl \small Signatures of the wave sequence}
\end{center}
\end{table}

The coset representatives of the M2 sequence $\mathcal{V}_{1+6n}$ Eq.(\ref{seqM21}) and $\mathcal{V}_{-1+6n}$ Eq.(\ref{seqM22}) are mapped respectively to the coset representatives of the wave sequence $\mathcal{V}_{6n}$ Eq.(\ref{seqG01}) and $\mathcal{V}_{6n'}$ Eq.(\ref{seqG02}). All the signatures of the wave sequence are summarised in Table 6.

\subsubsection{Signatures of the monopole sequence}

\noindent
{\bf $\bullet$ Mapping of the signature $(1,10,+)$ with time in $3$}

The global signature $(1,10,+)$ is not modified by $W_{(1)}$ but the time coordinate is no longer in $3$ but in $9$. We then perform the Weyl transformation $W_{(2)}$. From Eq.(\ref{k89}) with $s_1=-1$, we find that the time coordinates become $10$ and $11$ and from Eq.(\ref{r91011}) with $s_2=+1$ we deduce the sign of the kinetic term. This sign is the `wrong' one and the signature becomes $(2,9,-)$. The last Weyl reflexion $W_{(3)}$ does not change the global signature $(2,9,-)$ but puts the time coordinates  in $9$ and $10$. 

The signature $(1,10,+)$ with time in $3$ is mapped by the three successive Weyl transformations to the signature $(2,9,-)$ with times in $9$ and  $10$.

\medskip
\noindent
{\bf $\bullet$ Mapping of the signature $(5,6,+)$ with times in $4, 5, 6, 7, 8$}

The first Weyl transformation $W_{(1)}$ does not modify the signature. From Eqs.(\ref{k89}) and (\ref{r91011}) with $s_1=+1 $ and $s_2=-1$ we see that the signature is invariant  under the second Weyl reflexion $W_{(2)}$.  The last Weyl reflexion $W_{(3)}$ also leaves the signature unchanged. 

The signature $(5,6,+)$ with time in $4, 5, 6, 7, 8$ is left invariant by the Weyl mapping.

\begin{table}[h]
\begin{center}
\begin{tabular}{|c|c|c|}
\hline
levels ($n, n' >0$)&
times & $(t,s,\pm)$  \\
\hline\hline
3&9,10&$(2,9,-) $\\
\hline 
3+ 6n' , n' odd&4,5,6,7,8&$(5,6,+) $\\
3+6n', n' even&9,10&$(2,9,-)  $\\
\hline 
-3+6n , n odd&9,10&$(2,9,-) $\\
-3+6n, n even&4,5,6,7,8&$(5,6,+)  $\\
\hline
\end{tabular}
\caption{\sl \small Signatures of the monopole sequence}
\end{center}
\end{table}

\noindent
{\bf $\bullet$ Signatures of the monopole sequence}

The coset representatives of the M5 sequence $\mathcal{V}_{2+6n}$ Eq.(\ref{seqM51}) and $\mathcal{V}_{-2+6n}$ Eq.(\ref{seqM52}) are mapped respectively to the coset representatives of the monopole sequence $\mathcal{V}_{3+ 6n'}$ Eq.(\ref{seqG31}) and $\mathcal{V}_{-3+ 6n}$ Eq.(\ref{seqG32}). All the signatures of the monopole sequence are summarised in Table 7.

 \setcounter{equation}{0}

 \section{ Coset representatives of the gravity tower}
 \label{appcg}

We want to rewrite
\begin{equation}
\label{wava}
{\cal V}_1= \exp [h_3{}^3(K^3{}_3-K^{11}{}_{11})+\,h_3{}^{11}K^3{}_{11} ]\, ,
\end{equation}
in terms of a product of two exponentials, the first one containing only the Cartan generators, namely
we want to determine $A_{3}^{~(11)}$ in 
\begin{equation}
\label{wavb}
{\cal V}_2= \exp [h_3{}^3(K^3{}_3-K^{11}{}_{11})]\exp [A_{3}^{~(11)} \,K^3{}_{11}]\, .\end{equation}
In terms of the $SL(2)$ matrices defined in Eq.(\ref{repsl}), $  K^3{}_3-K^{11}{}_{11}=h_1$ and $  K^3{}_{11}=e_1$.  One has
\begin{equation}
\label{nu1}
{\cal V}_1=\sum_{n=0}^{\infty} {1\over n!} \left[\begin{array}{cc}h_3{}^3&h_3{}^{11}\\0&-h_3{}^3\\\end{array}\right]^n =\left[\begin{array}{cc}e_{11}{}^{11}&-e_3{}^{11}\\0&e_3{}^3\\\end{array}\right]
\end{equation}
\begin{equation}
\label{vielkk}
e_3{}^3=e^{-h_3{}^3} \qquad e_{11}{}^{11}=e^{h_3{}^3}\qquad e_3{}^{11}=- {h_3{}^{11} \over 2  h_3{}^3} \,  ( e^{h_3{}^3}-  e^{-h_3{}^3})\, ,
\end{equation}
where the $e_{\mu} {}^{n}$ are the vielbein Eq.(\ref{vielbein}) characterising the KK-solution (see also Appendix B of \cite{Englert:2003py}). On the other hand, from Eq.(\ref{vielkk}), one gets
\begin{equation}
\label{nu2}
{\cal V}_2=\left[\begin{array}{cc}e_{11}{}^{11}&e_{11}{}^{11} \, A_{3}^{~(11)}\\0&e_3{}^3\\\end{array}\right]\, .
\end{equation}
Equating ${\cal V}_1$  Eq.(\ref{nu1}) and ${\cal V}_2$ Eq.(\ref{nu2}), we have
\begin{equation}
\label{kka}
A_{3}^{~(11)}=-e_3{}^{11} (e_{11}{}^{11})^{-1}\, .
\end{equation}
and the metric corresponding to the representative Eq.(\ref{wava}) with time in 3  is thus
\begin{equation} ds^2=(dx^1)^2+(dx^2)^2+(dx^4)^2+\dots +(dx^{10})^2-(e_3{}^3)^2 (dx^3)^2
+ (e_{11}{}^{11})^2\left[ dx^{11} -A_{3}^{~(11)} dx^3 \right]^2\, .
  \label{ametric}
\end{equation}

\setcounter{equation}{0}

\section{Masses and U-duality}
\label{appm}
We first review the KK6-monopole mass and then derive all the masses or actions ${\cal A}_l$ of the $E_9$ multiplet in the M-theory context from the Weyl reflexions interpreted as T-dualities.
Their spacelike or timelike nature  is discussed.

\subsection{The level $3$ solution}
\label{appm3}

The KK6-monopole mass can be derived from the M5 mass using the relation between 11-dimensional supergravity and type IIA theory and T-duality. We recall that the relations between the 11-dimensional parameters $R_{11}$ and the Planck  length  $l_p$ and the string parameters $g$ and $l_s$ are (we neglect all numerical factors):
\begin{eqnarray}
\label{1110}
l_p= g^{1/3}\, l_s \qquad ,\qquad R_{11}= g \, l_s \,.
\end{eqnarray}
On the other hand if one compactifies a direction of type IIA theory  on a circle of radius R, the T-duality along this direction acts as:
\begin{eqnarray}
\label{tdualp}
R \rightarrow \frac{l_s^2}{R}\qquad , \qquad g \rightarrow \frac{g l_s}{R} .
\end{eqnarray}
We start with a M5 along the directions $4 \dots 8$ and 3 is the longitudinal timelike direction. The mass of this elementary M5 is given by
\begin{equation}
\label{mm5}
M_{l=2}={R_4 \dots R_8 \over l_p^6}.
\end{equation}
We smear this M5 along the directions 10 and 11. Reducing along the eleventh direction, using Eq.(\ref{1110}), one obtains a NS5 in type IIA then performing a T-duality along the direction 10, using Eq.(\ref{tdualp}) one get the KK5-monopole of type IIA with  Taub-NUT direction 10 
\begin{equation}
\label{mkk5}
M_{kk5}={R_4 \dots R_8  R^2_{10} \over g_s^2 \, l_s^8}\, .
\end{equation}
Uplifting back to eleven dimension, using Eq.(\ref{1110}), one obtains a KK6 monopole with longitudinal directions $4 \dots 8, 11$ and with Taub-NUT direction 10 .
Using Eqs.(\ref{1110}), (\ref{tdualp}), one find the mass of this KK6-monopole\footnote{The KK6-monopole discussed here has timelike direction is 3 and the Taub-NUT direction is 10. The level 3 KK6-monopole of the gravity tower in the mapping from the 6-tower in Sections 3 and 4 have timelike directions 9 and 10 and Taub-NUT direction 11. Note that in Section 2.2.3 the non-exotic KK6-monopole has its single timelike direction in 4 and Taub-NUT direction in 11. }
\begin{equation}
\label{mkk}
M_{l=3}={R_4 \dots R_8 R_{11} R^2_{10} \over l_p^9}.
\end{equation}

\subsection{The level $4$ solution}
\label{appm4}
To go to level 4 from the M5 considered in Eq.(\ref{mm5}) we perform as in Section 3.2.3 the Weyl reflexion $W_{\alpha_{11}}$ interpreted here as a double T-duality plus exchange in the direction 9 and 10. We thus smear the KK5-monopole with mass given in Eq.(\ref{mkk5}) in the direction 9 and perform a further T-duality in this direction. From 
Eqs.(\ref{1110}), (\ref{tdualp}), we find the mass of the level 4 solution \begin{equation}
\label{ml4}
M_{l=4}={R_4 \dots R_8 R^2_9 R^2_{10}R^2_{11}  \over l_p^{12}}.
\end{equation}

\subsection{The level $5$ solution}
\label{appm5}

As in Section 3.2.2, to   reach the level $5$ solution we perform,
the Weyl reflexion $W_{-\alpha_{11}+\delta} \equiv W_{-\alpha_{11}+\delta}$ sending the level $1$ generator $R_{1}^{[3]}$ to the level $5$ generator $R_{5}^{[6]}$. 
  
We  first decompose the Weyl reflexion $W_{-\alpha_{11}+\delta}$  in terms of simple Weyl reflexions
which have an interpretation in terms of permutations of coordinates and double T-duality plus exchange of the directions
9 and 10. This decomposition will permit us to compute the mass of the level 5 solution and also to check that the timelike direction 9 is unaffected by $W_{-\alpha_{11}+\delta}$.  We write
\begin{eqnarray} 
W_{-\alpha_{11}+\delta}= s_i \, s_j \ldots s_k\, ,
\end{eqnarray}
where $s_i \equiv s_{\alpha _i}$ is the simple Weyl reflexion corresponding to the simple root $\alpha_i$. To perform this decomposition, we can use the following lemma (whose proof is straightforward computing both sides of the equality) :

\noindent
{\bf Lemma}
\it 
{If a real root $\gamma$ can be written as the sum of two real roots $\gamma = \gamma_1 + \gamma_2$ such that $ <\gamma_1, \gamma_2>= -1$, then the Weyl reflexion $s_{\gamma }$ can be decomposed as $s_{\gamma } = s_{\gamma_1 }\, s_{\gamma_2 }\, s_{\gamma_1 }$  .} \rm

\noindent

One may write the root $- \alpha _{11}+ \delta$ as the sum $\gamma_1 + \gamma_2 $ where $\gamma_1$ is the root associated to $R^{345}$ and $\gamma_2$ is the root associated to $R^{6\, 7\, 8}$. Using the lemma, we get
\begin{eqnarray}
W_{-\alpha_{11}+\delta} = s_{\gamma_1+ \gamma_2 } = s_{\gamma_1 }\, s_{\gamma_2 }\, s_{\gamma_1 }\, ,
\end{eqnarray}
with
\begin{eqnarray}
s_{\gamma_1}&=& \underbrace{T_{39}\, T_{4\, 10}\, T_{5\, 11}}_{\mathcal{R}_1}\,  s_{\alpha_{11}} T_{5\, 11}\, T_{4\, 10}\, T_{39} \\
s_{\gamma_2}&=&\underbrace{ T_{69}\, T_{7\, 10}\, T_{8\, 11}}_{\mathcal{R}_2}\,  s_{\alpha_{11}} T_{8\, 11}\, T_{7\, 10}\, T_{69}\, ,
\end{eqnarray}
where $T_{ij}$ is the  Weyl reflexion of the gravity line permuting
the $i$ and $j$ indices namely it permutes the compactified radii $R_i
\leftrightarrow R_j$ and $ s_{\alpha_{11}}$ is the simple Weyl
reflexion with respect to $\alpha_{11}$ interpreted in type IIA as a
double T-duality in the directions 9 and 10 followed by an exchange of
the two radii.  We can directly check that there are no timelike
T-dualities when we perform the Weyl reflexion $W_{\alpha_{11}}$. The
reason is that the $T_{69}$ and $T_{39}$ replace always the time
coordinate 9 by 3 or 6 before any double T-duality.

We can now compute the mass of the level $l=5$ solutions by applying the Weyl reflexion $W_{-\alpha_{11}+\delta}$ to the expression Eq.(\ref{massl1}) for the  mass of the M2 with longitudinal spacelike directions 10, 11 and smeared in all spacelike directions but two 1 and 2 . The M2 mass is given by
\begin{eqnarray}
\label{massl1}
M_{M2}=  \frac{R_{10}\, R_{11}}{l_p^{3}}\, .
\end{eqnarray}
 To perform the sequence of simple  Weyl reflexions in the decomposition of $W_{-\alpha_{11}+\delta}$, we insist on two points :
\begin{itemize}
\item the permutations of the radii $R_i \leftrightarrow R_j$ are always performed in $11$ dimensions ;
\item the T-dualities are performed in $10$ dimensions.
\end{itemize}
Then, when reaching any $s_{\alpha_{11}}$ in a sequences of Weyl reflexions, we reduce the 11-dimensional theory to type IIA to perform the double T-duality plus exchange of radii,  and then do an uplifting to $11$ dimensions. 
Performing successive permutations  of radii, reduction on type IIA, T-duality, exchange of radii, uplifting to $11$ dimensions, and so on, we find the mass of the level $5$ solution 
\begin{eqnarray}
\label{massl5}
M_{M2}=  \frac{R_{10}\, R_{11}}{l_p^{3}}\quad \stackrel {W_{-\alpha_{11}+\delta}} {\longrightarrow } \quad M_{l=5}=  \frac{(R_3\, R_4 \ldots R_8)^{2}R_{10}\, R_{11}}{l_p^{15}} \,.
\end{eqnarray}

\subsection{The level 6 solution}
\label{appm6}

To obtain the mass of the level 6 solution we start from the expression for the level 5 mass  Eq.(\ref{massl5}) and we use the brane to gravity tower map Eqs.(\ref{bragramap})-(\ref{bragra2}).  In the string language this amounts to
perform  3 steps: first, exchange the radii $R_9$ and $R_3$ in 11 dimensions, second,  reduce to type IIA and, using Eqs.(\ref{1110})-(\ref{tdualp}), perform a double T-duality in the directions 9 and 10 plus exchange of the radii $R_9$ and $R_{11}$, finally uplift back to 11 dimensions and exchange the radii $R_9$ and $R_{11}$. The result is\footnote{Note that as the previous cases no timelike T-duality has been performed.}
 \begin{eqnarray}
\label{massl6}
M_{l=6}=   \frac{(R_4 \ldots R_{10})^2 \, R_{11}^3}{l_p^{18}} \,.
\end{eqnarray}

\subsection{Beyond level 6}
\label{appm7}

To cope with time-like T-dualities and exotic states involving more than one timelike direction, we first compactify time to a radius $R_t$ for the non-exotic states.  
We define the action ${\cal A}= M R_t$, where $M$ is the mass. Applying the relations Eqs.(\ref{1110}) and (\ref{tdualp}) to ${\cal A}_l= M_l R_t$ for the U-dualities performed in Sections C.1-C.4 leaves all computations of masses unchanged. As $\cal A$ treats space and time symmetrically it is natural to assume that U-duality can be extended to timelike T-dualities by applying the relations Eqs.(\ref{1110}) and (\ref{tdualp}) directly to $\cal A$ for both space and time radii.

We can now compute  $A_l$ for any level $l$ from  the $l \leq 6$ non-exotic solutions: we use the double T-duality plus inversion of radii encoded in the Weyl transformations $W_{\alpha_{11}}$ and $W_{-\alpha_{11}+\delta}$ used in Section 3 to reach {\em all} $E_9$ BPS states. It is easy to show by recurrence that one recovers in this way for all levels the results Eqs.(\ref{act3}), (\ref{act6}) and (\ref{actKK}). 

Indeed, consider first the brane towers depicted in Fig.2a.  Perform the Weyl transformation $W_{\alpha_{11}}$ and assume that  the formula Eq.(\ref{act6}) is true for level $2+3m$. Using Eqs.(\ref{1110}) and (\ref{tdualp}) one obtains  Eq.(\ref{act3}) for $n=m+1$. Perform the Weyl transformation $W_{-\alpha_{11}+\delta}$ and assume that the formula Eq.(\ref{act3}) is valid for level  $1+3m$. Using the decomposition of $W_{-\alpha_{11}+\delta}$ described in Section \ref{appm5} one finds  Eq.(\ref{act6}) for $n=m+1$. The validity of assumption at the levels $l< 6$ yields Eqs.(\ref{act3}) and (\ref{act6}).

For the [8,1]-gravity tower, assume that the formula  Eq.(\ref{act6}) is valid for all level $2+3m$ and use the gravity tower map Eqs.(\ref{bragramap})-(\ref{bragra2})   translated in string language as in Section \ref{appm6}. One finds Eq.(\ref{actKK}) for $n=m$, QED.

\setcounter{equation}{0}

\section{Level 4 by Buscher's duality }
\label{appb}
We review the Buscher formulation \cite{Buscher:1987sk,
  Buscher:1987qj}  of T-duality in 10-dimensional superstring theories
for backgrounds admitting one Killing vector\footnote{Here we are
  interested  in NS-NS backgrounds  as it was discussed originally
  \cite{Buscher:1987sk, Buscher:1987qj}, thus the formula apply not
  only for type II but also to the bosonic string in 26 dimensions. We
  will not need here the generalisation to R-R backgrounds
  \cite{Bergshoeff:1995as}.} . 
In Buscher's  construction one starts with a manifold $\cal M$ with
  metric $g_{ij}$  in the string frame, dilaton background $\phi$ and
  NS-NS background potential  $b_{ij}$. If the background is invariant
  under $x^{10}$ translations, it becomes under T-duality in the
  direction 10 $(a,b=1 \dots 9)$ 
\begin{eqnarray}
\label{busch}
&& \widetilde{g}_{10 \, 10} = 1/g_{10\, 10}, \qquad \widetilde{g}_{10 \, a}=b_{10\, a}/g_{10\, 10} \nonumber\\
&& \widetilde{g}_{ab} = g_{ab}-(g_{10\, a}g_{10 \, b}-b_{10\, a}b_{10\, b})/g_{10 \, 10} \nonumber\\
&& \widetilde{b}_{10 \, a}=g_{10\, a}/g_{10\, 10}  \\
&& \widetilde{b}_{ab} = b_{ab}-(g_{10\, a}b_{10 \, b}-b_{10\, a}g_{10\, b})/g_{10 \, 10} \nonumber\\
&&\widetilde{\phi}= \phi-{1\over 2} \ln g_{10\, 10} \, .\nonumber
\end{eqnarray}

We apply these transformations to a M5 brane with longitudinal spacelike directions $4 \dots 8$ and longitudinal timelike direction 3 to generate the level 4 BPS solution. We smear the M5 in the directions 9, 10, 11.  We perform a double T-duality in the directions 9 and 10. 
Upon dimensional reduction along $x^{11}$ to type IIA, this M5 yields a NS5 brane smeared in the directions 9,10 with non-compact transverse directions are
1 and 2. The  smeared NS5  is given in the string frame by 
\begin{eqnarray}
\label{sns5}
&& ds^2_{NS5}=  -(dx^3)^2 +(dx^4)^2+\dots  +(dx^8)^2 +
H \left[ (dx^1)^2  +(dx^2)^2 +(dx^9)^2  +(dx^{10})^2\right ] \nonumber \\
&& H(r)=\ln r, \qquad r^2\equiv (x^1)^2+(x^2)^2, \nonumber\\
&& e^{\phi}= H^{1/2} \nonumber\\
&& \widetilde{F}_{r345678}=  \partial_r (1/ H)\, ,
\end{eqnarray}
where $\widetilde{F}_{r345678}$ is the Hodge dual of the 4-form NS field strength $db$. We use the dual because the smearing procedure is always performed in the `electric' description of a  brane.

We first perform the T-duality in the direction 10.  Performing a T-duality on a direction transverse to a NS5 yields a KK5 monopole with the Taub-NUT direction in this transverse direction. Thus the T-duality on the configuration Eqs.(\ref{sns5}) generate a KK5 monopole with Taub-NUT direction 10 and smeared along 9. 
To find from Buscher's rule the transformed configuration, we need the value of the
non-zero $b$ field. Using the Hodge duality and Eqs.(\ref{sns5}), we find that the non-zero component of $b$ is
\begin{equation}
\label{bfield}
b_{9\, 10} =  {\rm arctg}(x^2/x^1) \equiv B\, .
\end{equation}
Using  Eqs.(\ref{busch}), we find  the metric of the smeared KK5 monopole, 
 \begin{equation}
 \label{skk5}
 ds^2_{skk5}=  -(dx^3)^2 +(dx^4)^2+\dots  +(dx^8)^2 +
H \left [ (dx^1)^2  +(dx^2)^2 +(dx^9)^2\right]+H^{-1} \left[(dx^{10}) - B (dx^9)\right]^2
\end{equation}
and $\phi=0$, as it should. We then perform the second T-duality in the direction 9, 
applying Eqs.(\ref{busch}) on Eq.(\ref{skk5}). We get (in the string frame)
\begin{eqnarray}
\label{2tns5}
&&ds^2=  -(dx^3)^2 +(dx^4)^2+\dots  +(dx^8)^2 +
H \left ( (dx^1)^2  +(dx^2)^2\right) + \widetilde{H} \left((dx^9)^2+(dx^{10})^2\right) \nonumber\\
&&e^\phi= \widetilde{H}^{1/2} \\
&& b_{9\, 10} =\widetilde{B}, \nonumber
 \end{eqnarray}
 where $ \widetilde{H}= H/(H^2+B^2)$ and $ \widetilde{B}=-B/(H^2+B^2)$. 
Finally uplifting the configuration Eq.(\ref{2tns5}) back to eleven dimension, we find
 \begin{eqnarray}
 \label{Ml4}
 &&ds^2=  H\widetilde{H}^{-1/3} \left( (dx^1)^2  +(dx^2)^2\right) + \widetilde{H}^{-1/3}\left(-(dx^3)^2 +(dx^4)^2+\dots  
 +(dx^8)^2\right) +\nonumber\\
 &&+\widetilde{H}^{2/3}\left((dx^9)^2+(dx^{10})^2+(dx^{11})^2\right) \nonumber\\
 && A_{9\, 10\, 11}=\widetilde{B}\, .
 \end{eqnarray}
 
 The solution Eq.(\ref{Ml4}) of 11D supergravity is exactly the level 4 solution  Eq.(\ref{metric4}) obtained starting with the level 2 solution (describing the double smeared M5) and performing a Weyl reflexion $W_{\alpha_{11}}$ to go to level 4. This result is in agreement with  the interpretation of the Weyl reflexion  $W_{\alpha_{11}}$ as a double T-duality in the directions 9,10 plus the interchange of the two directions \cite{Elitzur:1997zn, Obers:1998rn, Banks:1998vs, Englert:2003zs}.

 \setcounter{equation}{0}

 \section{Weyl transformations commute with compensations}
 \label{appc}
We will show  that the set of dualitites, compensations and Weyl transformations needed to express, in the M2 and M5 sequences, the Borel representative at a given level  in terms of the level~1 supergravity field  does not depend on the path chosen in Fig.3 and Fig.4. Equivalently we will prove that Weyl transformations and compensations do commute. The same proof can be done for the gravity tower.

We note that the  nature of the compensation matrix, i.e. $SO(2)$ or
$SO(1,1)$,  is unaltered by  the Weyl reflexions $W_{\alpha_{11}}$ or
$W_{-\alpha_{11}+\delta}$. In other words it is the same along a
column in Fig.3 and Fig.4. 
Indeed Eq.(\ref{newinvolve}) shows that the Weyl reflexion mapping the level $k$ generator to the level $k+n$ acts on the involution $\Omega R_k = \epsilon R_{-k}$ where $\epsilon$ is a sign, to yield $\Omega^{\prime} R_{k+n} = U \Omega R_k U^{-1}= \epsilon R_{-k-n}$ with the same sign (irrelevant signs in the Weyl transformed have been dropped). Taking this fact into account, we will analyse simultaneously the $SO(2)$ and $SO(1,1)$ compensations.

We start at a given level from the Borel representative  given by Eqs.(\ref{borel3n}) and  (\ref{borel6n}) for the M2 sequence and by Eqs.(\ref{6Nborel}) and  (\ref{3Nborel}) for the M5 sequence. After a number of dualities and compensations we reach $\boldsymbol{R_i}\equiv R^{[6]}_{1-3n}$ or $\boldsymbol{R_i} \equiv R^{[3]}_{-1-3n}$  defining a Borel representative ${\cal V}_i$. Both cases are shown in Table 8. We will show that ${\cal V}_f$  defined by $\boldsymbol{R_f}$ in the Table is independent of the path joining $\boldsymbol{R_i}$ to $\boldsymbol{R_f}$. Hence compensations and Weyl transformations do commute.

  \begin{table}[h]
\begin{center}
\begin{tabular}{|c|cccccc|}
\hline
levels & &  & & compensation  &&\\
\hline\hline
-1+3p&$R^{[6]}_{-1+ 3p}$& $\dots$&$ \boldsymbol{R_i} \equiv R^{[6]}_{1-3n}$&$\rightarrow $&$R^{[6]}_{-1+3n}$&\dots $\quad$ \\
{} & & & & & &\\
 $ W_{\alpha_{11}}$& & & $\Downarrow$ & & $\downarrow$&\\
{} & & & & &&\\
1+3p&$R^{[3]}_{1+ 3p}$ &$\dots$ &$R^{[3]}_{-1-3n}$&$\Rightarrow$&$\boldsymbol{R_f} \equiv R^{[3]}_{1+3n}$&\dots$\quad$\\
\hline
.& & &. & &. &\\
.& & &. & &.& \\
.& & &. & &. &\\
\hline
1+3r&$R^{[3]}_{1+ 3r}$& $\dots$&$\boldsymbol{R_i}\equiv R^{[3]}_{-1-3n}$&$\rightarrow$&$R^{[3]}_{1+3n}$&\dots$\quad$\\
{} & & & & && \\
$ W_{-\alpha_{11}+\delta}$  & & & $\Downarrow$ & & $\downarrow$&\\
{} & & & & &&\\
-1+3 (r+2) &$R^{[6]}_{-1+ 3(r+2)}$ & $\dots$&$R^{[6]}_{1-3(n+2)}$&$\Rightarrow$&$\boldsymbol{R_f}\equiv R^{[6]}_{-1+3(n+2)}$&\dots$\quad$\\
\hline
\end{tabular}
\caption{\sl \small Commutation of compensations and Weyl transformations: the paths depicted by
  simple arrows $``\rightarrow ''$  and by  double arrows
 $``\Rightarrow ''$ lead to the same result. The table applies to both  M2 and M5 sequences described in Fig.3 and Fig.4. The Weyl transformations are    $ W_{\alpha_{11}}$ and $ W_{-\alpha_{11}+\delta}$, and  $1<p$,  $0<r$, $n< p,r$.}
\end{center}
\label{t2case}
\end{table}

.

\subsection{case 1: $R_i \equiv R^{[6]}_{1-3n}$ and $R_f \equiv R^{[3]}_{1+3n}$}

Let ${\cal V}_a$ be a coset representative along a path joining $\boldsymbol{R_i}$ to $\boldsymbol{R_f}$.  We write the Cartan contribution ${\cal V}_a^{(0)}$ to ${\cal V}_a$ as
\begin{eqnarray}
 \label{reducedC}
\mathcal{V}_a^{(0)}= \exp \left [ \big (\frac{1}{2} \ln \mathcal{R}e \,  \mathcal{E}_a \big) \  \mathcal{C}_a + \Xi\, K^2{}_2 \right ]  \, .
\end{eqnarray}
Here we isolated in $\mathcal{V}_a^{(0)}$ a contribution $\Xi\, K^2{}_2$ left invariant along the paths by both compensations and Weyl transformations. The relevant contribution of the Cartan generators  is the transformed   $[(1/2)\ln\mathcal{R}e \,  \mathcal{E}_a  \  \mathcal{C}_a]$  of $[(1/2)\ln\mathcal{R}e \,  \mathcal{E}_i ) \  \mathcal{C}_i]$ at $\boldsymbol{R_i}$ where
$\mathcal{C}_i$ is the linear combination of the Cartan generators $h_{11}$ and $K^2{}_2$ pertaining to the $SL(2)$ subgroup containing
$R^{[6]}_{1-3n}$. From Eq.(\ref{slb}), we have  $\mathcal{C}_i = \alpha( -h_{11}-n K^2{}_2)$, and $\alpha \,=\, +1\  [-1]$ if the compensation is in $SO(2)$ [$SO(1,1)$]. The Ernst potential $\mathcal{E}_a $ is invariant along a column of Table 8.

We now examine the transformations of $\mathcal{V}_a^{(0)}$ along the paths.

\begin{itemize}
\item{ path ``$\rightarrow$''}

After compensation, both for the $SO(2)$ and $SO(1,1)$ compensations, one gets\footnote{The action of the compensation on $ \mathcal{C}_i$ is obtained by  straightforward generalisation of the compensations performed in Sections 3.2.2 and 3.2.3.} $\mathcal{C}_c = - \, \mathcal{C}_i=  \alpha( h_{11}+n K^2{}_2)$. Performing  the Weyl transformation $W_{\alpha_{11}}$, $K^2{}_2$ is left invariant and the sign of $h_{11}$ changes. The final expression for  $\mathcal{C}_a$ is thus $\mathcal{C}_f= \alpha(- h_{11}+n K^2{}_2)$.

\item{ path ``$\Rightarrow$''}

Performing the Weyl reflexion $ W_{\alpha_{11}}$, we get $\mathcal{C}_{ W_{\alpha_{11}}} = \alpha(h_{11}-n K^2{}_2$). To perform the subsequent compensation, we must identify the Cartan generator of $SL(2)$ subgroup containing $R^{[3]}_{-1-3n}$. From 
Eq.(\ref{sla}) this is precisely $\mathcal{C}_{ W_{\alpha_{11}}}$. After compensation we then obtain $\mathcal{C}_f=- \, \mathcal{C}_{ W_{\alpha_{11}}}=   \alpha (-h_{11}+n K^2{}_2)$.

\end{itemize}

The two different paths yield the same $\mathcal{C}_f$, {\em QED}. 

\noindent
The proof of path equivalence for the contribution of the step operators to ${\cal V}_a$
is immediate except for possible sign shifts in the Weyl transformations. Taking these into account, one easily verifies that this  affects in the same way both columns of Fig.8 connecting $\boldsymbol{R_i}$ to $\boldsymbol{R_f}$, as these columns list generators corresponding to opposite roots (see discussion after Eq.(\ref{permute})). This completes the proof.

\subsection{case 2: $R_i \equiv R^{[3]}_{-1-3n}$ and $R_f \equiv R^{[6]}_{-1+(3n+2)}$ }

Our starting point is Eq.(\ref{reducedC}).  In the relevant contribution of the Cartan generators, $\mathcal{C}_i$ is here the linear combination of  $h_{11}$ and $K^2{}_2$ pertaining to the $SL(2)$ subgroup containing
$R^{[3]}_{-1-3n}$. From Eq.(\ref{sla}), we have  $\mathcal{C}_i = \alpha( h_{11}-n K^2{}_2)$.

\begin{itemize}

\item{ path ``$\rightarrow$''}

After compensation, both for the $SO(2)$ and $SO(1,1)$ compensations, one gets $\mathcal{C}_c = - \, \mathcal{C}_i=  \alpha(- h_{11}+n K^2{}_2)$. Performing  the Weyl transformation $W_{-\alpha_{11} +\delta}$, $K^2{}_2$ is left invariant and  $ h_{11} \rightarrow -h_{11}- 2\, K^2{}_2 $. The final expression for  $\mathcal{C}_a$ is thus $\mathcal{C}_f= \alpha (h_{11}+(n+2)K^2{}_2)$.

\item{ path ``$\Rightarrow$''}

Performing the Weyl reflexion $W_{-\alpha_{11} +\delta}$, we get $\mathcal{C}_{ W_{-\alpha_{11} +\delta}} = \alpha (-h_{11} -(n+2)K^2{}_2 )$. To perform the subsequent compensation, we must identify the Cartan generator of $SL(2)$ subgroup containing $R^{[6]}_{1-3(n+2)}$. From 
Eq.(\ref{slb}) this is precisely $\mathcal{C}_{ W_{-\alpha_{11} +\delta}}$. After compensation we then obtain $\mathcal{C}_f=- \, \mathcal{C}_{ W_{-\alpha_{11} +\delta}}=   \alpha (h_{11}+(n+2)K^2{}_2)$.

\end{itemize}

The two different paths yield the same $\mathcal{C}_f$. It
can also be checked that not only the actions on the generators
commute  but that also the corresponding fields are transformed in the
same way. {\em QED}.

\setcounter{equation}{0}

\section{Redundancy of the two  gravity towers}
\label{apprg}

We will show that there are redundancies at each level 3n of the wave sequence and of the monopole sequence. 

Let us consider first the wave sequence. The metric of the wave sequence solutions follow from the representatives Eqs.(\ref{seqG01}) and (\ref{seqG02}). One has
\begin{eqnarray}
\label{wave6n}
ds^2_{6n}&=& g_{ab} \, dx^a\, dx^b \nonumber \\&=&   {\cal F}_{2n-1}\bar{\cal F}_{2n-1} \Big [(dx^1)^2+(dx^2)^2 \Big ] + (-1)^{n+1} H^{-1}_{2n+1}(dx^3)^2+\Big [(dx^4)^2\dots+(dx^{10})^2 \Big ] \nonumber \\
 &+& (-1)^{n} H_{2n+1} \Big [ dx^{11} -  \Big( (-1)^{n}   H^{-1}_{2n+1}+ (-1)^{n+1} \Big)dx^3 \Big]^2 \, ,  \\
 \nonumber \\
 \label{wave6np}
ds^2_{6 n'}&= &\widetilde{g}_{ab} \, dx^a\, dx^b\nonumber \\&=& {\cal F}_{2n'-1}\bar{\cal F}_{2n'-1} \Big [(dx^1)^2+(dx^2)^2 \Big ] + (-1)^{n'} H^{-1}_{2n'}(dx^3)^2+\Big [(dx^4)^2\dots+(dx^{10})^2 \Big ] \nonumber\\
 &+& (-1)^{n' +1} H_{2n'} \Big [ dx^{11} -  \Big( (-1)^{n' +1}   H^{-1}_{2n'}+ (-1)^{n'} \Big)dx^3 \Big]^2  \, ,
  \end{eqnarray}
where $H_p = {\cal R}e\, {\cal E}_p$.  

We now show that these metric are identical at each level $6n=6n'$ up to interchange of the coordinates $3$ and $11$. Using the relation
\begin{eqnarray} \label{hdn}
{\cal E}(z)_{2n+1}= 2-{\cal E}(z)_{2n} \qquad \Longrightarrow \qquad H_{2n+1}= 2-H_{2n} \, ,
\end{eqnarray}
 in Eq.(\ref{wave6n})  and performing  the following coordinate transformation 
\begin{eqnarray} \label{tsfc}
\left \{ \begin{array}{ll} 
x_3^{\prime} = - x_{11} & \, , \\
x_{11}^{\prime} = x_3& \, ,  \\
x_a^{\prime} = x_a  & a \neq 3, 11 \, ,
 \end{array}  \right.
\end{eqnarray}
we get the transformed metric 
\begin{eqnarray}
ds^{\prime \, 2}_{6n}&=&  {\cal F}_{2n-1}\bar{\cal F}_{2n-1} \Big [(dx^1)^2+(dx^2)^2 \Big ] + (-1)^{n+1} (2- H_{2n})^{-1} (dx^{11})^2+\Big [(dx^4)^2\dots+(dx^{10})^2 \Big ] \nonumber\\
 &+& (-1)^{n}(2- H_{2n}) \Big [ dx^{3} + \Big( (-1)^{n}  (2- H_{2n})^{-1}+ (-1)^{n+1} \Big)dx^{11} \Big]^2 \, .
\end{eqnarray}
We thus conclude that $ds^{\prime \, 2}_{6n} =  ds^2_{6n'} $ for $n = n'$ :
\begin{eqnarray}
\begin{array}{lll}
g^\prime _{11\, 11} &= (-1)^{n+1} H_{2n} &= \widetilde {g} _{11\, 11} \, ,   \\
 g^\prime _{3\, 3}& = (-1)^{n}(2- H_{2n})&= \widetilde {g} _{3\, 3}  \, , \\
g^\prime _{3\, 11} &= -1+ H_{2n} &= \widetilde {g} _{3\, 11}  \, , \\
g^\prime _{a\, a} & = \, \widetilde {g} _{a\, a} \quad a\neq 3, 11 \, .
 & \end{array}
\end{eqnarray}
Let us now  consider the monopole sequence. The metric of the monopole sequence solutions follow from the representatives Eqs.(\ref{seqG31}) and (\ref{seqG32}). One has
\begin{eqnarray}
\label{m6n}
ds^2_{3+ 6n'}&=& g_{ab} \, dx^a\, dx^b \nonumber\\&=& {\cal F}_{2n'-1}\bar{\cal F}_{2n'-1}  H_{2n'+1}\Big [(dx^1)^2+(dx^2)^2 \Big ]\! + H_{2n'+1}(dx^3)^2+(-1)^{n'}  \Big [(dx^4)^2\dots +(dx^{8})^2 \Big ]  \nonumber \\
 &+&\!  \! \!(-1)^{n'+1}  \Big [(dx^9)^2+(dx^{10})^2 \Big ]   + H^{-1}_{2n'+1} \Big [ dx^{11} -  \Big( (-1)^{n'}   B_{2n'+1}\Big)dx^3 \Big]^2  \\
 \nonumber \\
 \label{m63n}
ds^2_{- 3+ 6n}&=& \widetilde{g}_{ab} \, dx^a\, dx^b\nonumber\\  &=& {\cal F}_{2n-1}\bar{\cal F}_{2n-1}  H_{2n}\Big [(dx^1)^2+(dx^2)^2 \Big ] + H_{2n}(dx^3)^2+(-1)^{n+1}  \Big [(dx^4)^2\dots+(dx^{8})^2 \Big ]  \nonumber \\
 &+& \!  \! \! (-1)^{n}  \Big [(dx^9)^2+(dx^{10})^2 \Big ]   + H^{-1}_{2n} \Big [ dx^{11} -  \Big( (-1)^{n+1}   B_{2n}\Big)dx^3 \Big]^2 \,. 
 \end{eqnarray}
 where $B_p = {\cal I}m\, {\cal E}_p$.
 
We now show that these metric are identical at each level $3+ 6n'= - 3+ 6n$, i.e. $n= n'+1$, up to interchange of the coordinates $3$ and $11$. Replacing in Eq.(\ref{m63n}) $n$ by $n'+1$, we get 
\begin{eqnarray}
\label{m63n2}
ds^2_{-3+ 6(n'+1)} &=&   {\cal F}_{2n'-1}\bar{\cal F}_{2n'-1} {\cal E}_{2n'+1}\bar{\cal E}_{2n'+1}   H_{2n'+2}\Big [(dx^1)^2+(dx^2)^2 \Big ] \nonumber\\&+& H_{2n'+2}(dx^3)^2+(-1)^{n'}  \Big [(dx^4)^2\dots+(dx^{8})^2 \Big ]  \nonumber \\
 &+& (-1)^{n'+1}  \Big [(dx^9)^2+(dx^{10})^2 \Big ]   + H^{-1}_{2n'+2} \Big [ dx^{11} -  \Big( (-1)^{n'}   B_{2n'+2}\Big)dx^3 \Big]^2 \, .
\end{eqnarray}
Using the relation
\begin{eqnarray} \label{hdn2}
{\cal E}(z)_{2n'+2}= ({\cal E}(z)_{2n'+1} )^{-1}\qquad &\Longrightarrow& \qquad H_{2n'+2}= \frac{H_{2n'+1}}{ {\cal E}_{2n'+1}\bar{\cal E}_{2n'+1} } \, ,\\
&\Longrightarrow& \qquad B_{2n'+2}= -\frac{ B_{2n'+1}}{ {\cal E}_{2n'+1}\bar{\cal E}_{2n'+1} } \, ,
\end{eqnarray}
 in Eq.(\ref{m63n2})  and performing  the  transformation of coordinates Eq.(\ref{tsfc}), we get the transformed metric 
\begin{eqnarray}
ds^{2 \prime}_{-3+ 6(n'+1)} &=&   {\cal F}_{2n'-1}\bar{\cal F}_{2n'-1}   H_{2n'+1}\Big [(dx^1)^2+(dx^2)^2 \Big ] + \frac{H_{2n'+1}}{ {\cal E}_{2n'+1}\bar{\cal E}_{2n'+1} }(dx^{11})^2\nonumber \\&+&(-1)^{n'}  \Big [(dx^4)^2\dots+(dx^{8})^2 \Big ]  
 + (-1)^{n'+1}  \Big [(dx^9)^2+(dx^{10})^2 \Big ] \nonumber \\  &+ &(\frac{H_{2n'+1}}{ {\cal E}_{2n'+1}\bar{\cal E}_{2n'+1} })^{-1} \Big [ dx^{3} +  \Big( (-1)^{n'+1}   \frac{B_{2n'+1}}{ {\cal E}_{2n'+1}\bar{\cal E}_{2n'+1} }\Big)dx^{11} \Big]^2 \, . 
\end{eqnarray}
We thus conclude that $ds^{2 \prime}_{-3+ 6(n'+1)} =  ds^2_{3+6n'}$:
\begin{eqnarray}
\begin{array}{lll}
g^\prime _{11\, 11} &=  H^{-1}_{2n'+1} &= g _{11\, 11} \, ,   \\
 g^\prime _{3\, 3}& = H^{-1}_{2n'+1}  (  {\cal E}_{2n'+1}\bar{\cal E}_{2n'+1} )&= g _{3\, 3}  \, ,\\
g^\prime _{3\, 11} &= (-1)^{n'+1}B_{2n'+1} H^{-1}_{2n'+1}  &= g _{3\, 11}  \, , \\
g^\prime _{a\, a} & =g _{a\, a} \quad a\neq 3, 11 .& 
 \end{array}
\end{eqnarray}

We see that there is only one gravity tower, the left and the right tower of Fig.5b being equivalent (except for the level 0 KK-wave) as each of them contains the full wave and monopole sequences.

\section{Structure of the $A_1^+$ U-duality group}
\label{affapp}

Our U-duality group in two non-compact dimensions is the infinite
order Weyl group $\cW$ of an affine group. The structure of such Weyl
groups is known in terms of translations and the finite Weyl group of
the underlying finite group of rank $r$ \cite{Kac:book}. The affine Weyl
group is the semi-direct product of translations $\ints$ and the
finite Weyl group. For the case of affine
$A_1^+$, which features prominently in this
paper, the affine Weyl group is simply
\begin{eqnarray}
{\cal{W}} = \mathbb{Z}_2 \ltimes \mathbb{Z}.
\end{eqnarray}

To derive this fact it is useful to denote the two simple roots of 
$A_1^+$ by $\alpha_1$ and $\alpha_2$. These can be identified to
$\alpha_{11}$ and $-\alpha_{11} +\delta$   for the M2 and the M5
sequences,  and to $\lambda$ and $-\lambda + \delta$ for the KK-wave
and the KK-monopole sequences.  The simple roots $\alpha_1$ and 
$\alpha_2$ are a basis of the root lattice (they span the ladder
diagrams of Fig.2 and Fig.5). To describe the fundamental reflexions
$W_1,W_2$ in these two  roots, it is sufficient to give their action
on a basis: 
\begin{align}
W_1(\alpha_1) &= -\alpha_1,&\quad W_1(\alpha_2)&= \alpha_2 + 2
\alpha_1,&\nonumber\\ 
W_2(\alpha_1) &= \alpha_1+2\alpha_2,&\quad W_2(\alpha_2)&= -\alpha_2.&
\end{align}
We take the $\ints_2$ to be generated by the horizontal
Matzner--Misner reflexion $W_2$. The Coxeter relations for this Weyl
group are $(W_1 W_2)^\infty = \text{id}$ and $(W_2 W_1)^\infty =
\text{id}$, in other words there are no mixed relations. Therefore all
Weyl group elements are of the form
\be
\label{cox}
(W_1 W_2)^n, \quad\quad (W_2 W_1)^n, \quad\quad W_1 (W_2 W_1)^n,\quad\quad W_2 (W_1 W_2)^n, \quad\quad\text{id}\, ,
\ee
for some $m,n\ge 0$. Defining $T^n = (W_1W_2)^n$
for $n\ge 0$ and $T^n=(W_2W_1)^{(-n)}$ for $n\le 0$ one deduces for the
$\mathbb{Z}_2$ generated by $W_2$ the structure
\begin{eqnarray}
W_2 T^n = T^{-n} W_2\, ,
\end{eqnarray}
illustrating the semi-directness of the product in this case. The set
of all elements of the infinite order Weyl group is thus 
\begin{eqnarray}
{\cal{W}}=\{ T^n : n\in\mathbb{Z}\} \cup \{ W_2 T^n : n\in\mathbb{Z}\}\, ,
\end{eqnarray}
with relations
\begin{eqnarray}
W_2W_2 &=& 1\nn\\
W_2 T^n &=& T^{-n} W_2\nn\\
T^nT^m &=& T^{n+m}.
\end{eqnarray} 
The translations $T$ act vertically in the diagrams  of Fig.2a, Fig.3, Fig. 4, Fig.5a and Fig.5b. They connect the points lying both on the same tower and on the same sequence.

We can act with the Weyl group on any integrable representation $\rho:A_1^+\to End(V)$, by letting \cite{Kac:book}
\begin{eqnarray}\label{weyl}
U_i = \exp(\rho (f_i)) \exp(-\rho (e_i)) \exp(\rho (f_i))\in GL(V).
\end{eqnarray}
Here $ e_i$ and $f_i$ are the simple Chevalley generators. It is not hard to see that this definition implies that $U_i$ is
actually an element of $SO(V)$ in the sense that
$U_iU_i^T=\text{id}_V$ for $i=1,2$ where the {\em Chevalley}
transposed element is
$U_i^T=\exp(\rho(e_i))\exp(-\rho (f_i))\exp(\rho (e_i))$.\footnote{The
  definition 
  Eq.(\ref{weyl}) does not necessarily imply $U_i U_i=\text{id}_V$. In
  order to arrive at the proper Weyl group one has to factor out the
  subgroup generated by $U_i U_i$ from the $GL(V)$
  subgroup generated by the $U_i$, see \S3.8 in~\cite{Kac:book}.}

From Eq.(\ref{weyl}) it is straightforward to show that Weyl reflexions
are always elements of the compact subgroup of the split real form of
the associated group, so in our case this means $K_{10}^+$.

\newpage

 \addcontentsline{toc}{section}{References}
\bibliographystyle{hunsrt}
\bibliography{Refdata}
\end{document}